%% file: jpsipipi_JHEP.tex
\pdfoutput=1 
\documentclass[a4paper,11pt]{article}
\usepackage{atlasphysics_modified}
\usepackage{subfigure}
\usepackage{multirow}
\usepackage{afterpage}
\usepackage{mathrsfs}
\usepackage{rotating}
\usepackage{epstopdf}
\usepackage{caption} 
\usepackage{mcite}
\usepackage{placeins} 

\usepackage{jheppub} 

\usepackage[T1]{fontenc} 

\RequirePackage{ifpdf}

\def\Jpipi{\ensuremath{\Jpsi\,\pi^{+}\pi^{-}}}
\def\Jmumupipi{\ensuremath{\Jpsi(\to\!\!\mu^+\mu^-)\pi^+\pi^-}}

\usepackage{preprintcover}  
\PreprintCoverPaperTitle{\boldmath Measurement of the production cross-section of
 $\psi(2\mathrm{S})\to\Jmumupipi$\\
  in $pp$ collisions at $\sqrt{s}=7$~TeV at ATLAS}
\PreprintIdNumber{CERN-PH-EP-2014-095}  
\PreprintCoverAbstract{
The prompt and non-prompt production cross-sections for $\psi(2\mathrm{S})$ mesons are measured using 
2.1~fb$^{-1}$ of $pp$ collision data at a centre-of-mass energy of 7 TeV recorded by the ATLAS experiment at the LHC.
The measurement exploits the $\psi(2\mathrm{S})\rightarrow J/\psi(\to\mu^+\mu^-)\pi^+\pi^-$ decay mode, and probes
$\psi(2\mathrm{S})$ mesons with transverse momenta in the range $10\leq p_{\rm T}<100$~GeV and rapidity $|y|<2.0$.
The results are compared to other measurements of $\psi(2\mathrm{S})$ production at the LHC 
and to various theoretical models for prompt and non-prompt quarkonium production.
}
\PreprintJournalName{JHEP}  

\title{\boldmath Measurement of the production cross-section of
  $\psi(2\mathrm{S})\to\Jmumupipi$ in $pp$ collisions at $\sqrt{s}=7$~TeV at ATLAS}

\author{The ATLAS Collaboration}

\arxivnumber{}

\abstract{The prompt and non-prompt production cross-sections for $\psi(2\mathrm{S})$ mesons are measured using 
2.1~fb$^{-1}$ of $pp$ collision data at a centre-of-mass energy of 7 TeV recorded by the ATLAS experiment at the LHC.
The measurement exploits the $\psi(2\mathrm{S})\rightarrow J/\psi(\to\mu^+\mu^-)\pi^+\pi^-$ decay mode, and probes
$\psi(2\mathrm{S})$ mesons with transverse momenta in the range $10\leq p_{\rm T}<100$~GeV and rapidity $|y|<2.0$.
The results are compared to other measurements of $\psi(2\mathrm{S})$ production at the LHC 
and to various theoretical models for prompt and non-prompt quarkonium production.}

\begin{document} 

\maketitle
\flushbottom

\section{Introduction}

The production of quarkonium states in hadronic collisions has been the subject of intense theoretical and experimental study 
for many decades, especially since measurements of prompt $J/\psi$ and $\Upsilon$ production at the Tevatron\,\cite{CDFjpsianomaly1,CDFjpsianomaly2,CDFupsilonanomaly1,CDFjpsianomaly3,D0jpsi1,D0jpsi2,D0upsilon1}
exposed order-of-magnitude differences between data and theoretical expectations\,\cite{CDFupsilonanomaly2}.
Despite these being among the most studied heavy-quark bound states, there is still no satisfactory understanding of the mechanisms of their formation.
Quarkonium production acts as a unique and important testing ground for quantum chromodynamics (QCD) in its own right. 
While the production of a heavy quark pair occurs at a hard scale and is generally well-described by QCD, 
its subsequent evolution into a bound state includes many non-perturbative effects at much softer scales that pose a challenge to current theoretical methods.
With the data obtained from the Large Hadron Collider (LHC), it is possible to perform stringent tests of theoretical models across a large range of momentum transfer.

Studies of heavy quarkonia were conducted previously by ATLAS in the $J/\psi\rightarrow\mu^+\mu^-$\,\cite{JPsi} and $\Upsilon(nS)\rightarrow\mu^+\mu^-$\,\cite{Upsilon2010,Upsilon} decay modes. 
The measurements described here are based on an analysis of 2.1~fb$^{-1}$ of $pp$ collision data at $\sqrt{s}=7$~TeV, 
and study the prompt and non-prompt production of the $\psi(2\mathrm{S})$ meson through its decay to $\Jmumupipi$.
The prompt production arises from direct QCD production mechanisms and the non-prompt production arises from weak decays of $b$-hadrons.
The $\Jmumupipi$ final state offers improvements in $\psi(2\mathrm{S})$ mass resolution and background discrimination over exclusive dilepton channels.
Unlike prompt $J/\psi$ production, which can occur through either direct QCD production of $J/\psi$ or the production of excited 
states that subsequently decay into $J/\psi+X$ final states, 
no appreciable prompt production of excited states decaying into $\psi(2\mathrm{S})$ has been established in hadron collisions.
In this respect the $\psi(2\mathrm{S})$ is a unique state with no significant feed-down from
higher quarkonium resonances, which decay predominantly to $D\overline{D}$ pairs. 

The measurement presented here, when combined with a concurrent measurement of the prompt and non-prompt production of $P$-wave $\chi_{cJ}$ states\,\cite{ATLAS_chic} and existing measurements of the production cross-section
of the $J/\psi$\,\cite{JPsi}, provides a rather comprehensive picture of the production of both prompt and non-prompt charmonia.
These $\psi(2\mathrm{S})$ cross-sections are compared with the results from LHCb\,\cite{LHCb} and CMS\,\cite{CMS} and with a variety of theoretical
models for both prompt and non-prompt production, and complement recent measurements from ALICE\,\cite{Abelev:2014qha} at low $p_{\rm T}$.

\section{The ATLAS detector}
\label{sec:Detector}
The ATLAS detector\,\cite{ATLAS:det} is composed of an inner tracking
system, calorimeters, and a muon spectrometer.
The inner detector (ID) surrounds the proton--proton collision point and
consists of a silicon pixel detector, a silicon microstrip detector, and a
transition radiation tracker, all of which are immersed in a 2\,T axial magnetic
field. The inner detector spans the pseudorapidity\footnote{ATLAS uses a right-handed coordinate system with its origin at the nominal interaction point (IP) in the centre of the detector and the $z$-axis along the beam pipe. The $x$-axis points from the IP to the centre of the LHC ring, and the $y$-axis points upward. Cylindrical coordinates $(r,\phi)$ are used in the transverse plane, $\phi$ being the azimuthal angle around the $z$-axis. The pseudorapidity $\eta$ is defined in terms of the polar angle $\theta$ as $\eta=-\ln\tan(\theta/2)$ and the transverse momentum $p_{\rm T}$ is defined as $p_{\rm T}=p\sin\theta$. The rapidity is defined as $y=0.5\ln\left(\left( E + p_{\rm z} \right)/ \left( E - p_{\rm z} \right)\right)$, where $E$ and $p_{\rm z}$ refer to energy and longitudinal momentum, respectively. The $\eta$--$\phi$ distance between two particles is defined as 
 $\Delta R=\sqrt{(\Delta\eta)^2 + (\Delta\phi)^2}$. } range $\abseta<2.5$ and is
enclosed by a system of electromagnetic and hadronic calorimeters.
Surrounding the calorimeters is the muon spectrometer (MS)
consisting of three large air-core superconducting magnets (each with eight coils) providing a toroidal field, a system of
precision tracking chambers, and fast detectors for triggering.
This spectrometer is equipped with monitored drift tubes and cathode-strip chambers that provide precision measurements in the bending plane of muons
within the pseudorapidity range $\abseta<2.7$. Resistive-plate and thin-gap chambers with fast response are 
primarily used to make fast trigger decisions in the ranges $\abseta<1.05$ and $1.05<\abseta<2.4$ respectively,
and also provide position measurements in the non-bending
plane and improve overall pattern recognition and track reconstruction.
Momentum measurements in the muon spectrometer are based on track segments formed in at least two of the three precision chamber planes.

The ATLAS detector employs a three-level trigger system\,\cite{ATLAS:trig}, which reduces the 20\,MHz proton bunch collision rate
to the several-hundred Hz transfer rate to mass storage.
The level-1 muon trigger searches for hit coincidences between
different muon trigger detector layers inside pre-programmed
geometrical windows that bound the path of muon candidates over a given transverse momentum ($p_{\rm T}$) threshold and provide a rough estimate of its position within the
pseudorapidity range $\abseta<2.4$. 
At level-1, muon candidates are reported in ``regions of interest'' (RoI). 
Only a single muon can be associated with a given RoI of spatial extent 
$\Delta\phi \times \Delta\eta \approx 0.1 \times 0.1$. This limitation has a small 
effect on the trigger efficiency for $\psi(2\mathrm{S})$ mesons, which is corrected in the analysis using a 
data-driven method based on analysis of $J/\psi\to\mu^+\mu^-$ and $\varUpsilon\to\mu^+\mu^-$ decays.
There are two subsequent higher-level, software-based trigger selection
stages. Muon candidates reconstructed at these higher levels incorporate, with increasing precision, information from both
the muon spectrometer and the inner detector, reaching position and momentum resolutions close to those provided by the offline muon reconstruction.

In this analysis, muon candidates are reconstructed using algorithms reliant on the combination of both an MS track and an ID track. 
Because of this ID coverage requirement, muon reconstruction is possible only within $|\eta|<2.5$.  
The muons selected for this analysis are further restricted to $|\eta|<2.3$.
This ensures high-quality tracking and triggering, and reduces the number of fake muon candidates. 
It also removes regions of strongly varying efficiency and acceptance.

\section{Data and event selection}
\label{sec:selection}

Data for this analysis were collected in 2011, during LHC proton--proton collisions at a centre-of-mass energy of $7$~TeV.  
The data sample was collected using a trigger requiring two oppositely charged muon candidates with no explicit requirement on the transverse momentum at level-1 of the trigger.
The higher-level trigger stage subsequently requires each muon to have transverse momentum satisfying $p_{\rm T}>4$~GeV. Muon candidates are also required to 
fulfil additional quality criteria and the dimuon pair must be consistent with having originated from a common vertex, and have invariant mass $2.5<m_{\mu^+\mu^-}<4.3\textrm{ GeV}$.
The data collected with this trigger configuration corresponded to a total integrated luminosity of $2.09\pm 0.04$~fb$^{-1}$\,\cite{lumi} in the full 7~TeV dataset.

The $\psi(2\mathrm{S})\to \Jpipi$ candidates are reconstructed with a technique similar to the one used by ATLAS for $B_s\rightarrow J/\psi \phi$\,\cite{BJpsiPhi} candidates.
The selected events 
contain at least two oppositely charged muons, identified by the muon spectrometer and with associated tracks reconstructed in the inner detector.
The two muon tracks are considered a $J/\psi\rightarrow\mu^+\mu^-$ candidate if they can be fitted to a 
common vertex with a dimuon invariant mass between $2.8$~GeV and $3.4$~GeV. 
The muon track parameters are taken from the ID measurement alone, since the MS does not improve the precision in the momentum range relevant for 
the $\psi(2\mathrm{S})$ measurements presented here.
To ensure accurate inner detector measurements, each muon track must contain at least six hits in the silicon microstrip detector and at least one hit in the pixel detector.
Muon candidates satisfying these criteria are required to have $p_{\rm T}>4\GeV$, $\abseta < 2.3$, and a successful fit to a common vertex.
Good spatial matching, $\Delta R < 0.01$, between each reconstructed muon candidate and a trigger 
identified candidate is required to accurately correct for trigger inefficiencies.
The dimuon pair is further required to satisfy $p_{\rm T}>8$~GeV and $|y|<2.0$ to ensure that the $J/\psi$ candidates are reconstructed in a fiducial region where acceptance and efficiency
corrections do not vary too rapidly. An additional requirement on the dimuon vertex-fit $\chi^2$ helps to remove spurious dimuon combinations.

The two pions in the
$\psi(2\mathrm{S})\to \Jpipi$ decay are reconstructed by taking all pairs of the remaining oppositely charged tracks with $p_{\rm T}>0.5$~GeV and $|\eta|<2.5$ and assigning the pion mass hypothesis to each reconstructed track.
A constrained four-particle vertex fit is performed to all $\psi(2\mathrm{S})$ candidates, where the $J/\psi\rightarrow\mu^+\mu^-$ candidates have their invariant mass constrained to the world average value for 
the $J/\psi$ mass~(3096.916 MeV)\,\cite{PDG}. 
A $\chi^2$ probability requirement of ${\mathcal P}(\chi^2)>0.005$ is applied to the vertex fit quality, which considerably reduces combinatorial background from incorrect dipion candidate assignment.
The constrained vertex fit also provides significantly improved invariant mass resolution for the $\Jpipi$ system over that attainable from momentum resolution alone.
Corrections are made for signal selection inefficiencies ($\sim 5\%$--$8\%$) arising from the dimuon invariant mass, $p_{\rm T}$, and rapidity selections, the vertex requirements on dimuon candidates, and the 
constrained-fit quality criterion of the four-particle vertex.

\begin{figure}[tbp]
  \begin{center}
\includegraphics[width=0.71\textwidth]{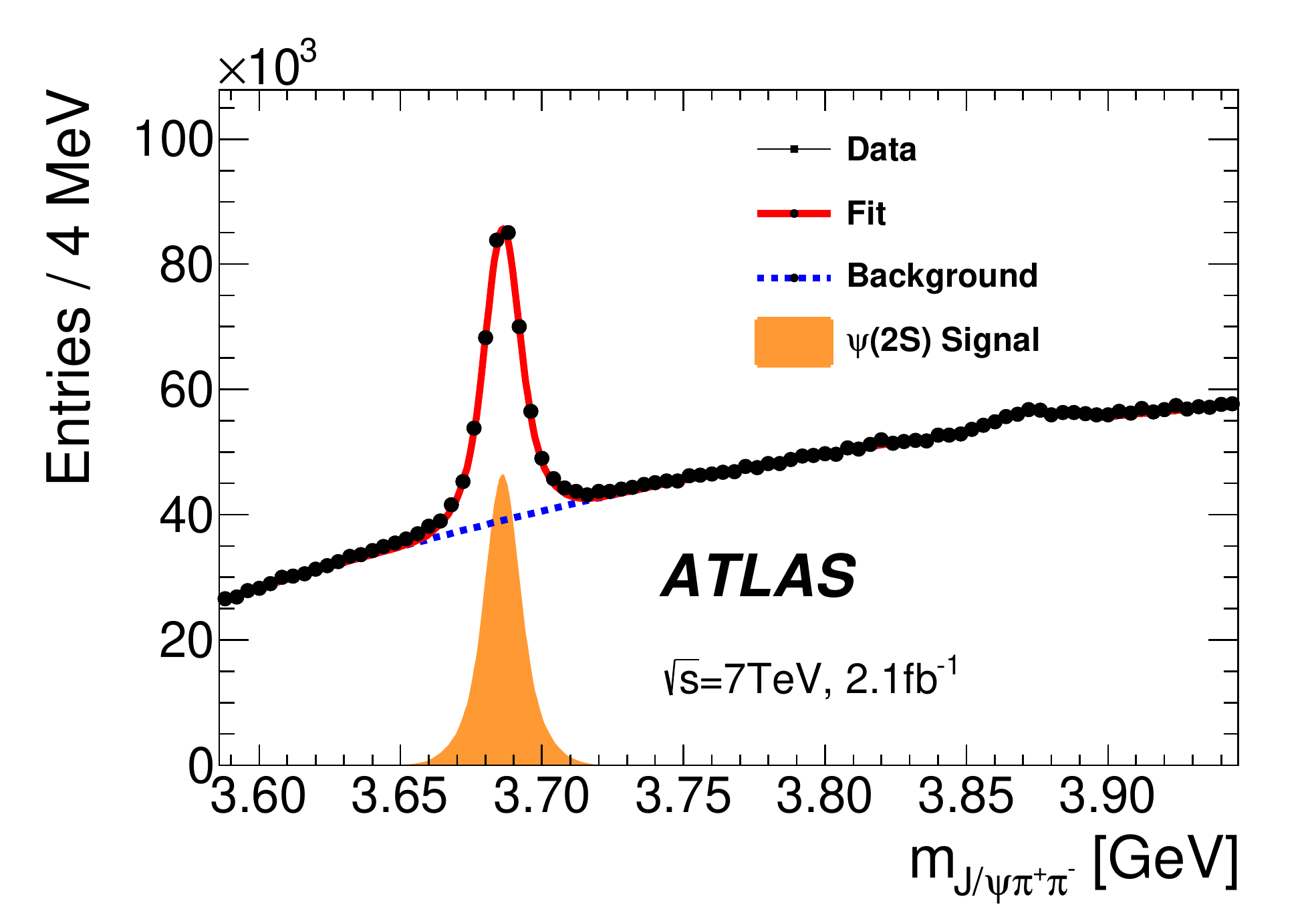}
    \caption{The uncorrected $\Jpipi$ mass spectrum between $3.586$~GeV and $3.946$~GeV.
Superimposed on the data points is the result of a fit using a double Gaussian distribution to describe the $\Jpipi$ signal peak, 
and a second-order Chebyshev polynomial to model the background, where the region within $\pm 25\MeV$ of the $X(3872)$ ($m_{\Jpipi}=3.872\,\GeV$) is excluded from the fit.\label{fig:massSpectrum}}
  \end{center}
\end{figure}

Figure~\ref{fig:massSpectrum} shows the $\Jpipi$ invariant mass distribution after the above selection criteria are applied.
A clear peak of the $\psi(2\mathrm{S})$ is observed near 3.69~GeV. At larger invariant mass, a further structure is also observed, identified as the $X(3872)$.

\begin{figure}[thp]
  \begin{center}
    \subfigure[$|y| < 0.75$]{\includegraphics[width=0.49\textwidth]{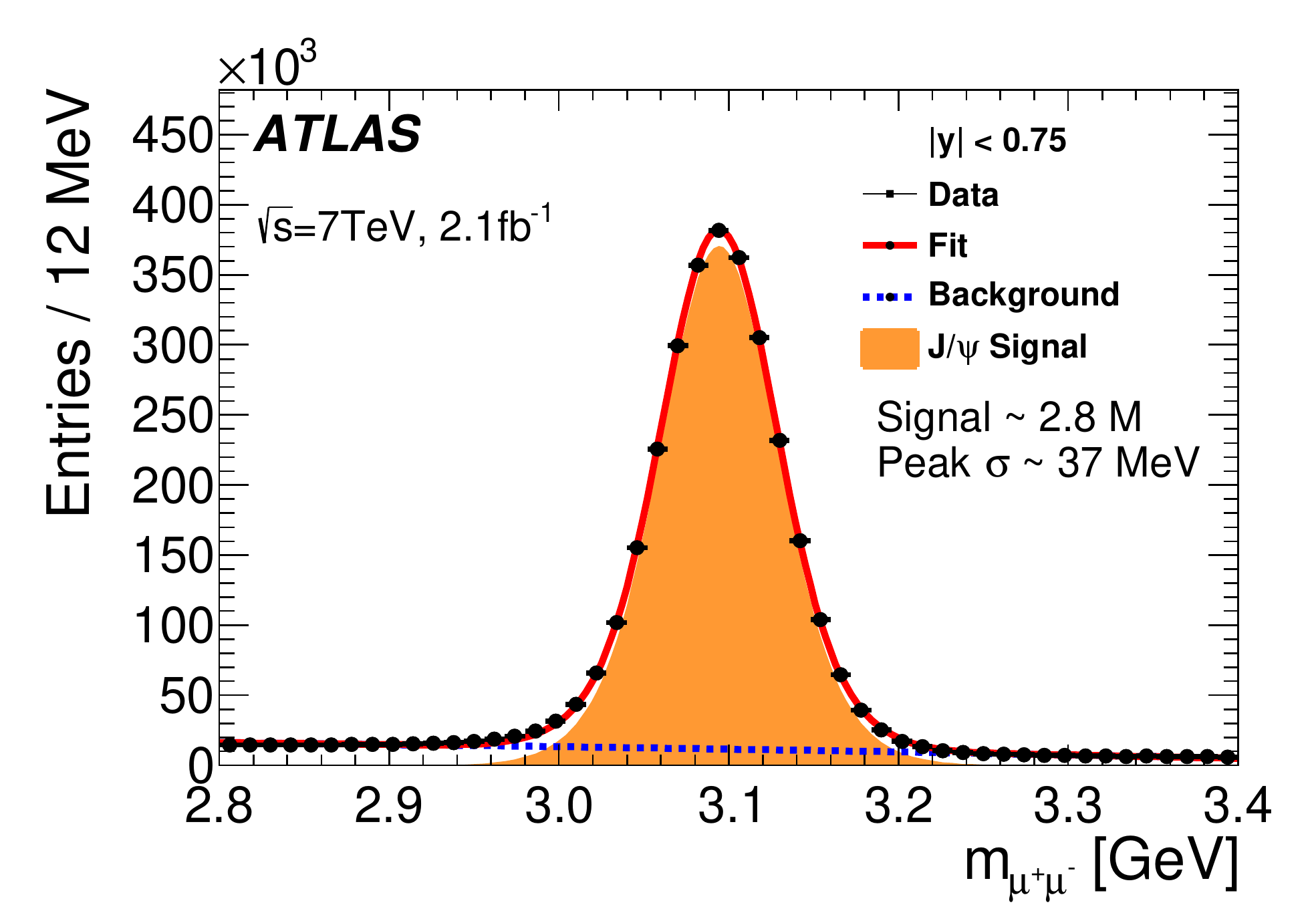}}
    \subfigure[$|y| < 0.75$]{\includegraphics[width=0.49\textwidth]{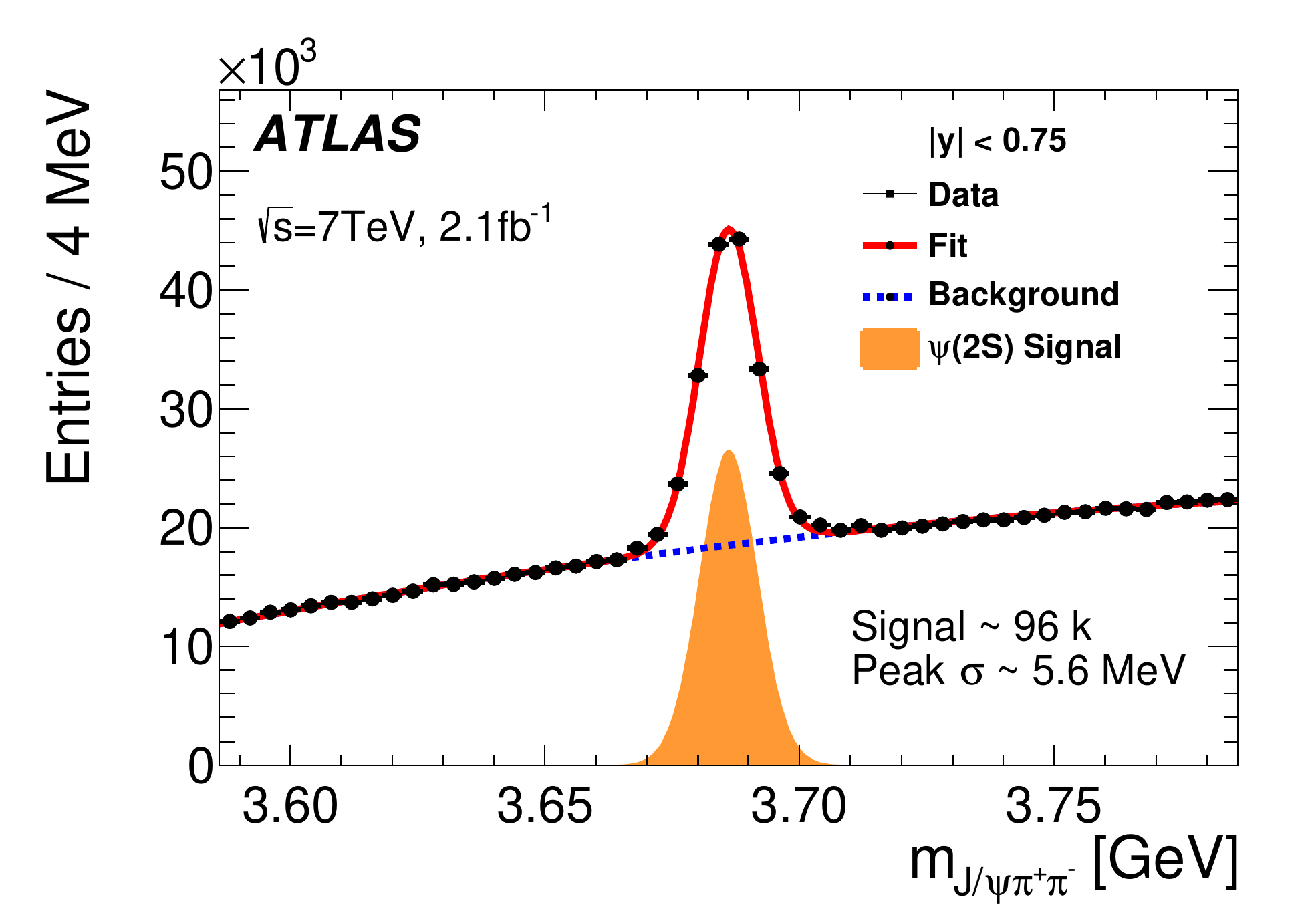}}
    \subfigure[$0.75 \leq |y| < 1.5$]{\includegraphics[width=0.49\textwidth]{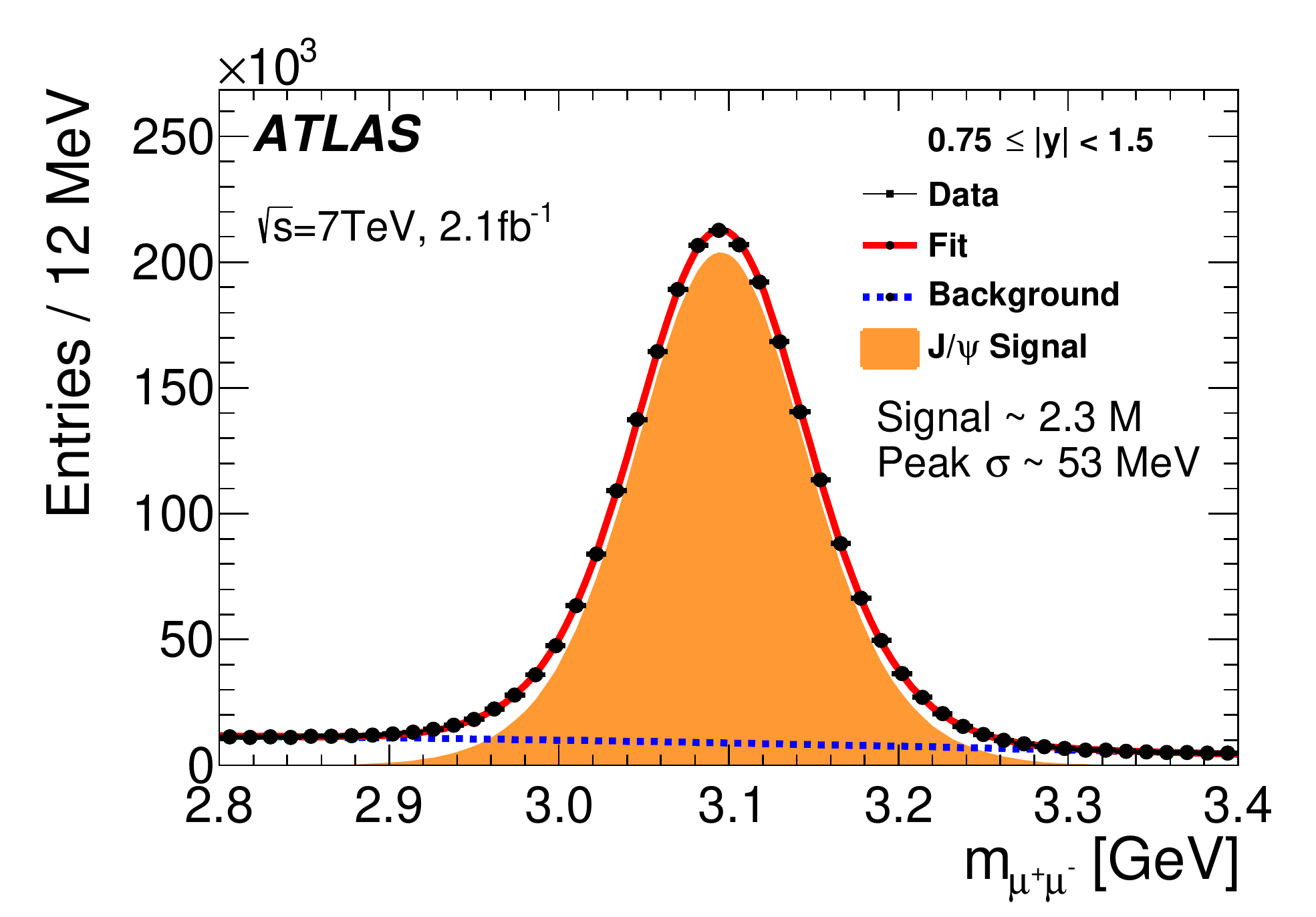}}
    \subfigure[$0.75 \leq |y| < 1.5$]{\includegraphics[width=0.49\textwidth]{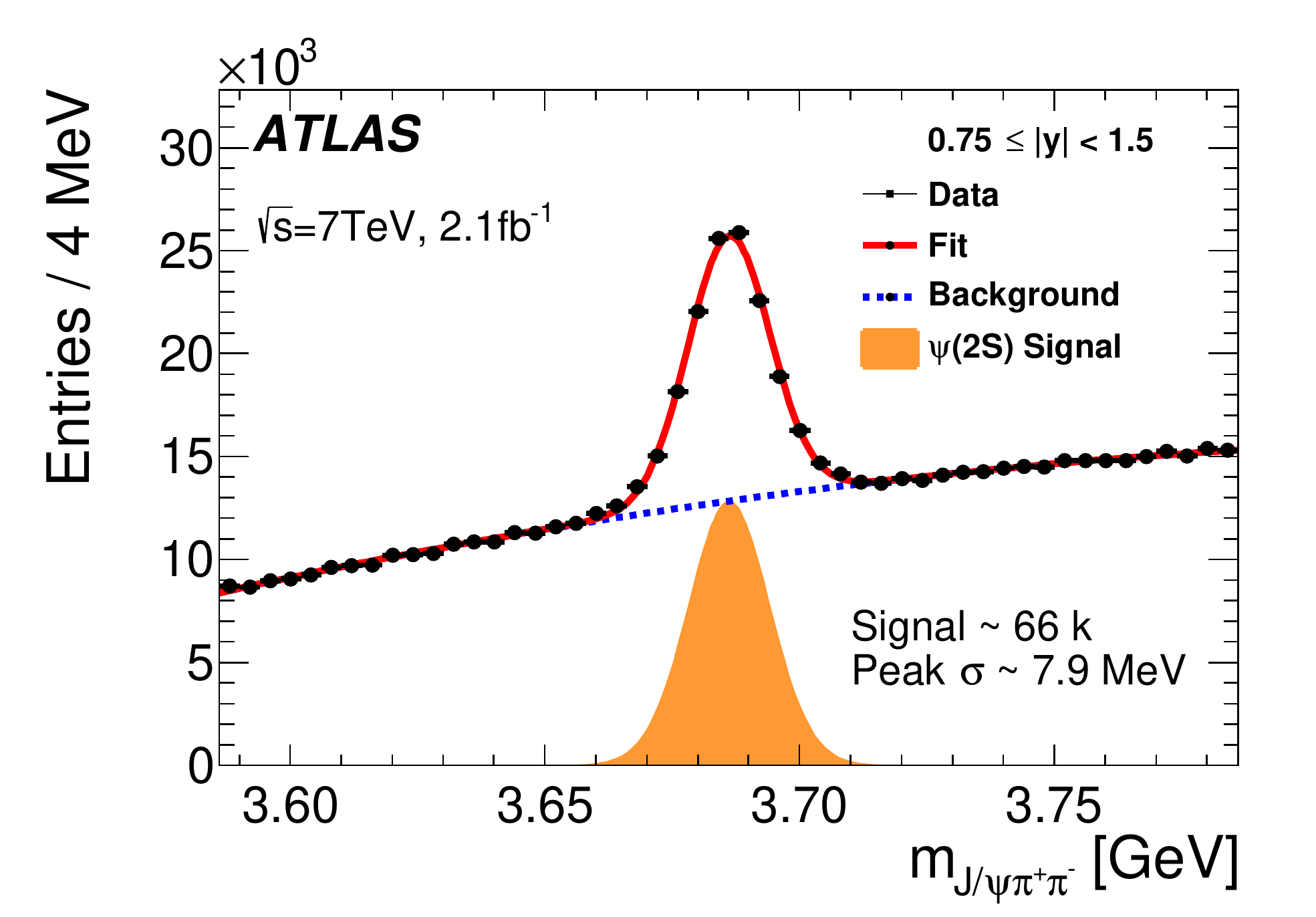}}
    \subfigure[$1.5 \leq |y| < 2.0$]{\includegraphics[width=0.49\textwidth]{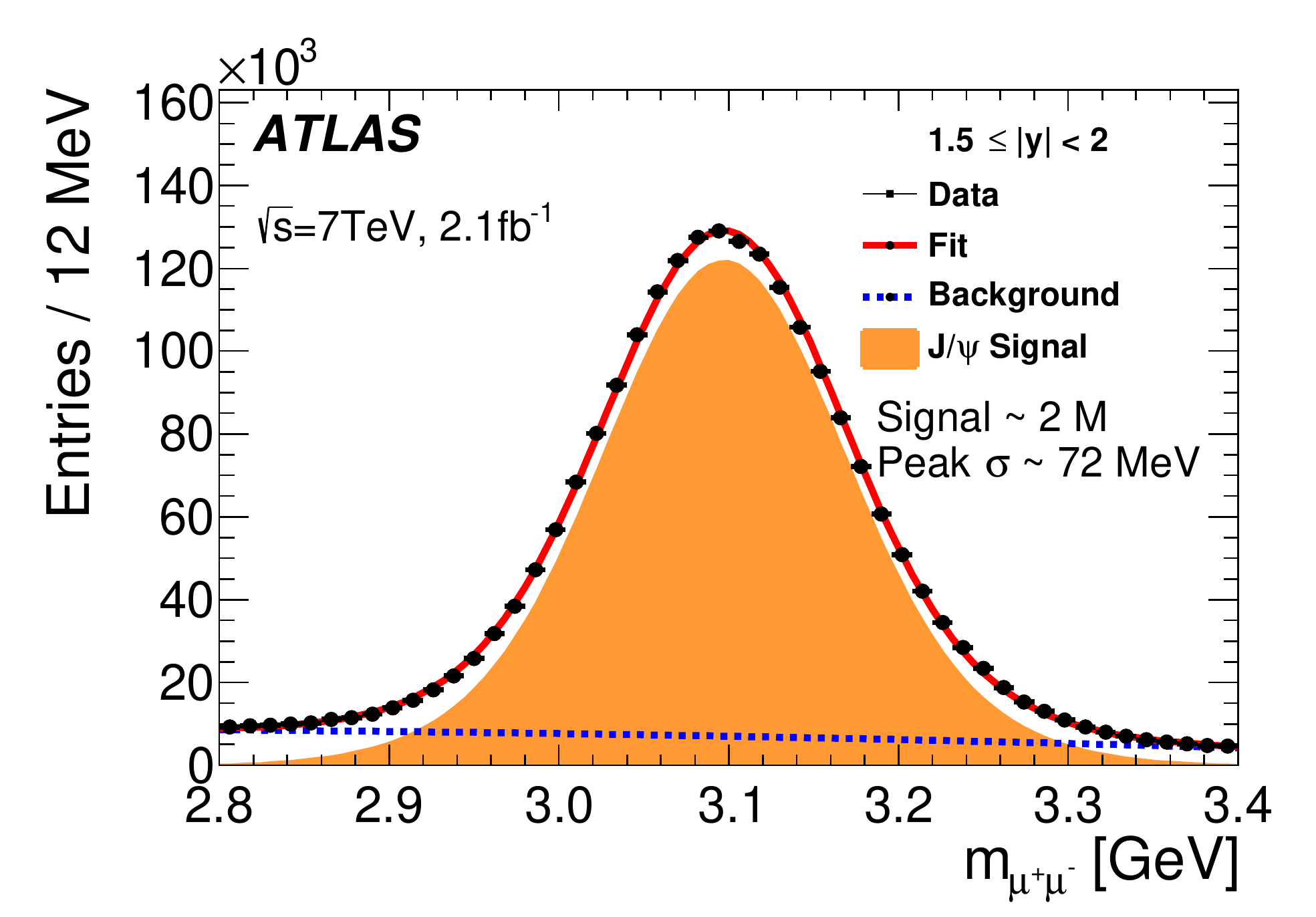}}
    \subfigure[$1.5 \leq |y| < 2.0$]{\includegraphics[width=0.49\textwidth]{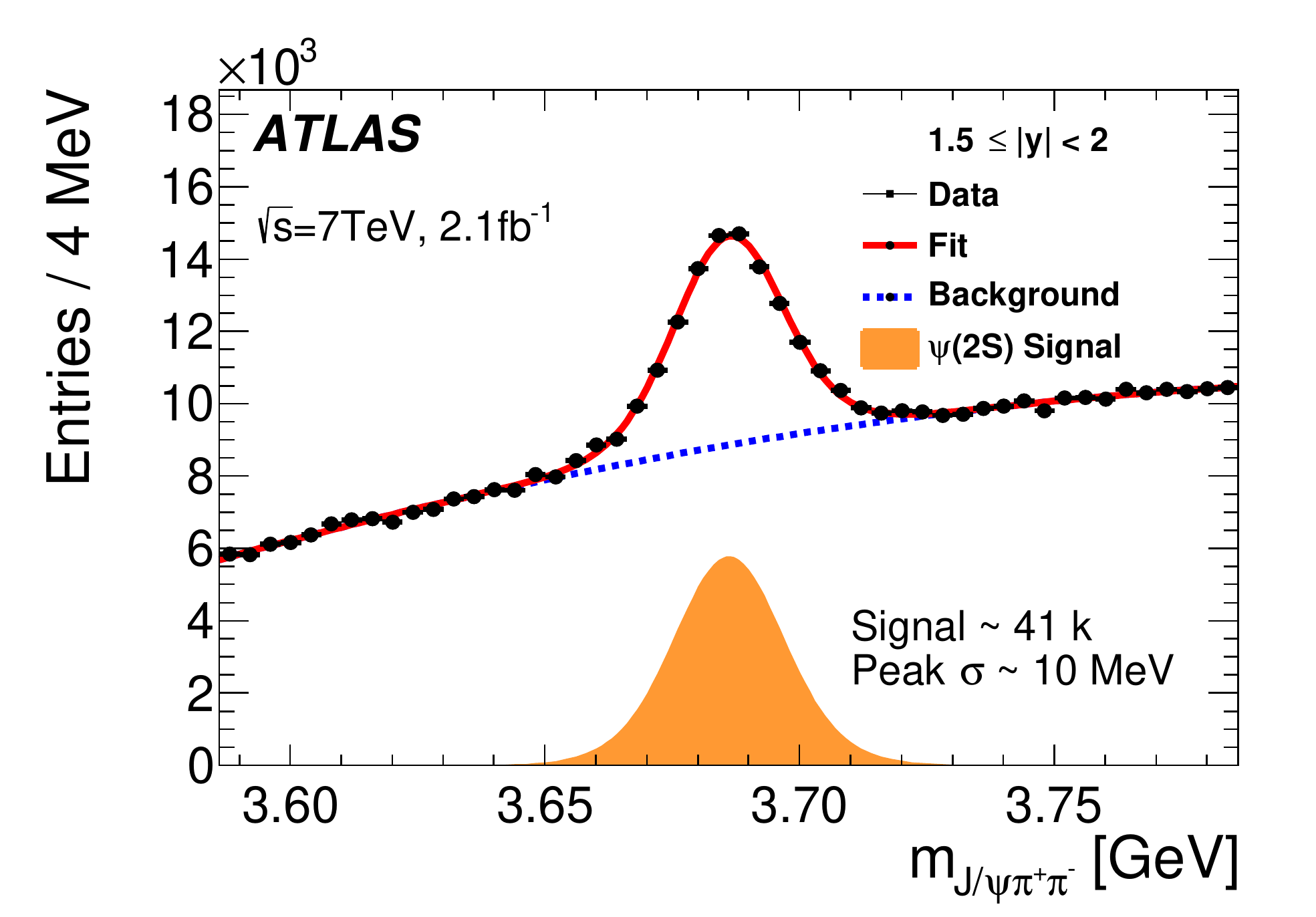}}
    \caption{
      Invariant mass distributions for the dimuon
      (left) and $\Jpipi$ system after the dimuon mass-constrained fit (right) in the three rapidity ranges of the measurement. 
      The data distributions are fitted with a combination a double Gaussian distribution (for the signals)
       and a second-order Chebyshev polynomial (for backgrounds).
    \label{fig:Yields}
    }
  \end{center}
\end{figure}

The cross-section measurements are presented in three $\psi(2\mathrm{S})$ rapidity intervals: $|y| < 0.75$, $0.75 \leq |y| < 1.5$, and $1.5 \leq |y| < 2.0$, and in ten $p_{\rm T}$ intervals
for each of the rapidity intervals, spanning $10\leq p_{\rm T}<100$~GeV.
Figure~\ref{fig:Yields} illustrates the uncorrected yields and the invariant mass resolutions of the dimuon and $\Jpipi$ systems in the three rapidity regions, 
which comprise about $96\, 000$, $66\, 000$ and $41\, 000$ $\psi(2\mathrm{S})$ candidates respectively.
For both the dimuon and the $\Jpipi$ invariant mass fits, a double Gaussian is used to describe the signal shape, and a second-order Chebyshev polynomial to model the background.

\section{Cross-section determination}

The differential production cross-section for $\psi(2\mathrm{S})$ can be apportioned between prompt production and non-prompt production.
Non-prompt $\psi(2\mathrm{S})$ production processes are distinguished from prompt processes by their longer apparent lifetimes, with production occurring 
through the decay of a $b$-hadron.
To distinguish between these prompt and non-prompt processes, a parameter called the pseudo-proper lifetime $\tau$ is constructed using the $\Jpipi$ transverse momentum:
\begin{equation}
\tau = \frac{L_{xy}\, m_{\Jpipi}}{p_{\rm T}},
\end{equation}
with \noindent$L_{xy}$ defined by the equation:
\begin{equation}
L_{xy} \equiv {\vec{L}}\cdot {\vec{p}_{\rm T}} / p_{\rm T},
\end{equation}
where $\vec{L}$ is the vector from the primary vertex to the $\Jpipi$ decay vertex and $\vec{p}_{\rm T}$ is the transverse
momentum vector of the $\Jpipi$ system. The primary vertex is defined as the vertex with the largest 
scalar sum of associated charged-particle track $p^2_{\rm T}$,
and identified as the location of the primary proton--proton interaction. The presence of additional simultaneous proton--proton collisions, and the effect 
of associating the final-state particles with the wrong collision was found\,\cite{BJpsiPhi}
to have a negligible impact on the discrimination and extraction of short and long-lived components of the signal.

To obtain a measurement of the production cross-sections, the reconstructed candidates are individually weighted to correct for detector effects, 
such as acceptance, muon reconstruction efficiency, pion reconstruction efficiency and trigger efficiency, which are discussed below in detail.
The distribution of candidates in $\psi(2\mathrm{S})$ $p_{\rm T}$ and $|y|$ intervals are then fitted using a weighted two-dimensional unbinned maximum likelihood method, 
performed on the invariant mass and pseudo-proper lifetime distributions to isolate signal candidates from the backgrounds and separate the prompt signal from the non-prompt signal. 
The corrected prompt and non-prompt signal yields ($N_{\mathrm{P}}^{\psi(2\mathrm{S})}$, $N_{\mathrm{NP}}^{\psi(2\mathrm{S})}$) 
are then used to calculate the prompt and non-prompt differential cross-section ($\sigma_\mathrm{P}$, $\sigma_\mathrm{NP}$) times branching ratio, using the equation:

\begin{equation}
\mathcal{B}\left(\psi(2\mathrm{S})\rightarrow \Jmumupipi\right) 
\times \frac{\mathrm{d}^{2}\sigma_\mathrm{P, NP}^{\psi(2\mathrm{S})}}{\mathrm{d}p_{\rm T}\mathrm{d}y} 
= \frac{N_{\mathrm{P, NP}}^{\psi(2\mathrm{S})}}{\Delta p_{\rm T}\Delta y \int \mathcal{L}\mathrm{d}t},
\end{equation}
where $\int \mathcal{L}\mathrm{d}t$ is the total integrated luminosity, $\Delta p_{\rm T}$ and $\Delta y$ represent the 
intervals in $\psi(2\mathrm{S})$ transverse momentum and rapidity, respectively, and $\mathcal{B}\left(\psi(2\mathrm{S})\rightarrow \Jmumupipi\right)$
is the total branching ratio of the signal decay, taken to be $(2.02\pm 0.03)\%$, obtained by combining the world average values for $\mathcal{B}\left(J/\psi\to\mu^{+}\mu^{-}\right)$
and $\mathcal{B}\left(\psi(2\mathrm{S})\rightarrow \Jpipi\right)$\,\cite{PDG}.

In addition to the prompt and non-prompt production cross-sections, the non-prompt $\psi(2\mathrm{S})$ production fraction $f_{B}^{\psi(2\mathrm{S})}$
is simultaneously extracted from the maximum likelihood fits in the same kinematic intervals. This fraction is defined as 
the corrected yield of non-prompt $\psi(2\mathrm{S})$ divided by the corrected total yield of produced $\psi(2\mathrm{S})$, as given in the equation:
\begin{eqnarray}
f_{B}^{\psi(2\mathrm{S})} &\equiv &
\frac
{N_{\mathrm{NP}}^{\psi(2\mathrm{S})}}
{N_{\mathrm{P}}^{\psi(2\mathrm{S})} + N_{\mathrm{NP}}^{\psi(2\mathrm{S})}}.
\label{eqn:npfrac}
\end{eqnarray}
Measurement of this fraction benefits from improved precision over absolute cross-section measurements 
through cancellation or reduction of overall acceptance and efficiency corrections in the ratio.

\subsection*{Acceptance\label{sec:accept}}

The acceptance $\mathcal{A}(p_{\rm T},y,m_{\pi\pi})$ is defined as the probability that the decay products in $\psi(2\mathrm{S})\to \Jmumupipi$
fall within the fiducial volume
($p_{\rm T}(\mu^\pm)>4$~GeV, $|\eta(\mu^\pm)|<2.3$, $p_{\rm T}(\pi^\pm)>0.5$~GeV, $|\eta(\pi^\pm)| < 2.5$). 
The acceptance depends on the spin-alignment of $\psi(2\mathrm{S})$. For the central results obtained in this analysis, the $\psi(2\mathrm{S})$ 
decay was assumed to be isotropic, with variations corresponding to a number of extreme spin-alignment scenarios described below.

\begin{figure}[htbp]
\centering 
\subfigure[
\label{fig:accMap:1}]{\includegraphics[width=0.49\textwidth]{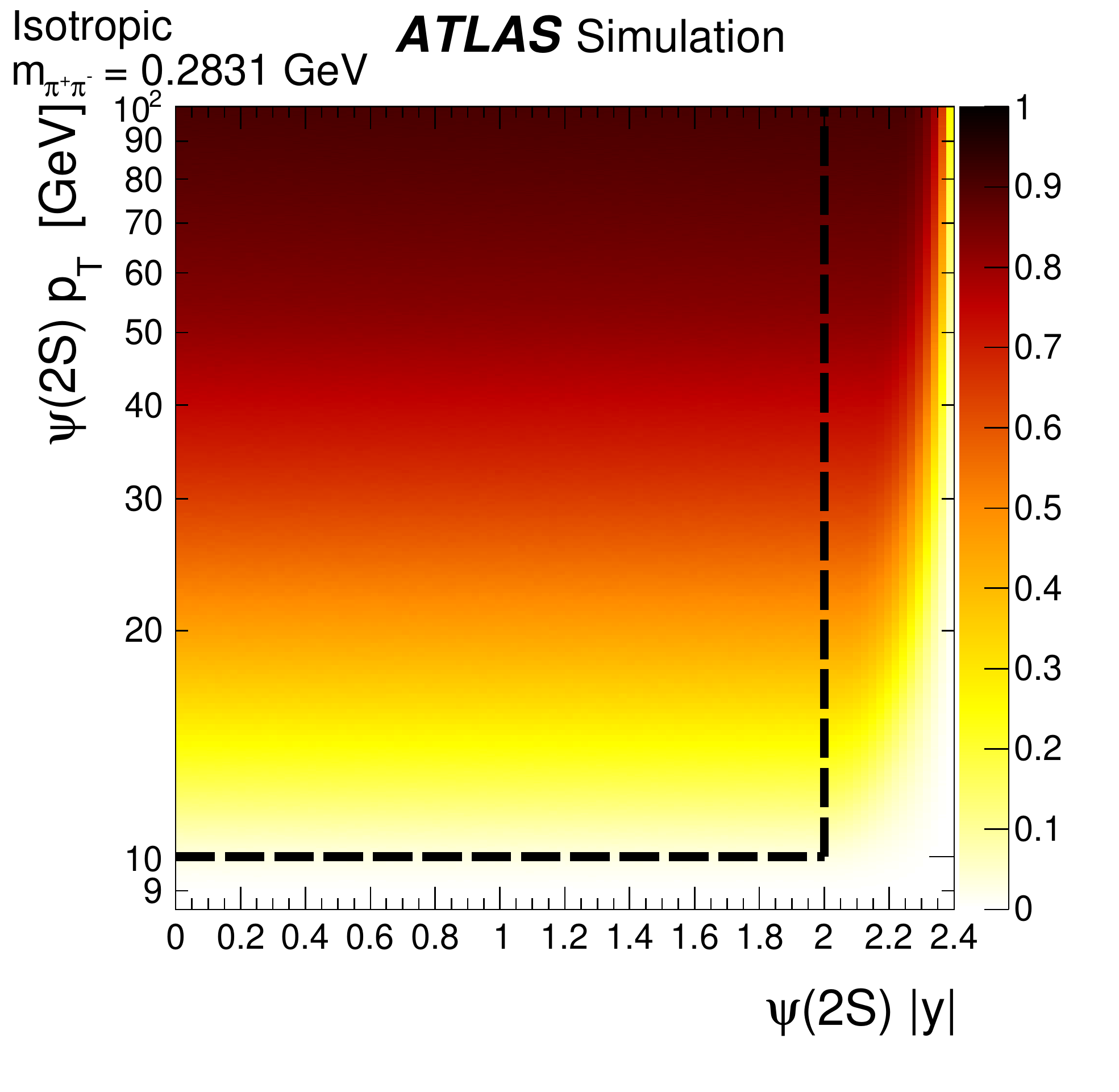}}
\subfigure[
\label{fig:accMap:3}]{\includegraphics[width=0.49\textwidth]{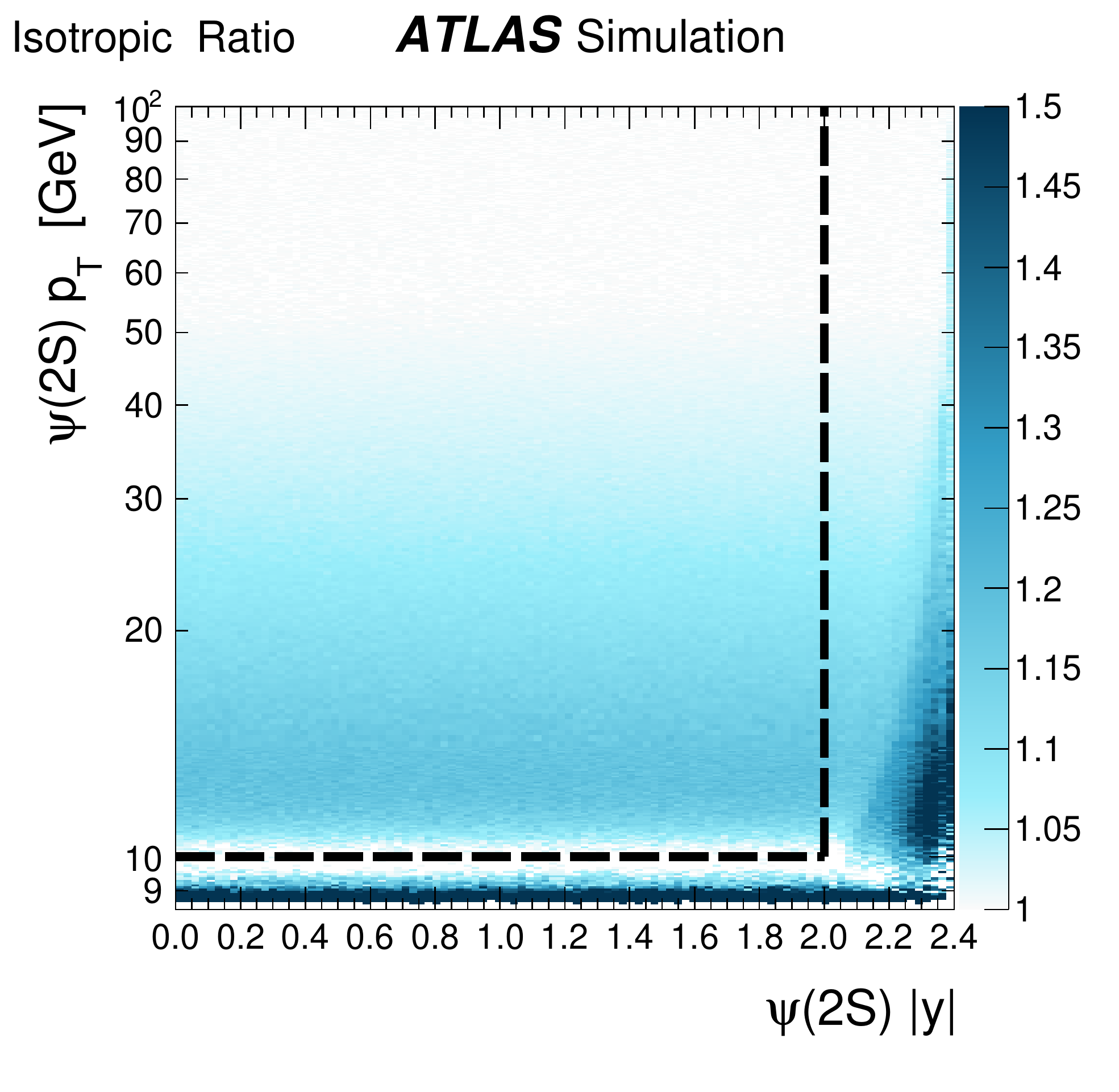}}
\caption{(a) Example of $\psi(2\mathrm{S})\to\Jmumupipi$ acceptance for isotropic $\psi(2\mathrm{S})$ production for the lowest dipion mass, $m_{\pi^+\pi^-}= 2m_\pi$,
and (b) the ratio of the acceptance at the lowest dipion masses to that at the highest dipion masses.
Dashed lines show the $p_{\rm T}$--$y$ bounds of the measurement.
  \label{fig:accMap}
}
\end{figure}

Acceptance maps are created
using a large sample of generator-level Monte Carlo (MC) simulation, which randomly creates and decays $\psi(2\mathrm{S})\to\Jmumupipi$, 
as a function of the $\psi(2\mathrm{S})$ transverse momentum and rapidity, in finely binned intervals of the dipion invariant mass $m_{\pi^+\pi^-}$ 
covering the allowed range, $2m_\pi < m_{\pi^+\pi^-} <m_{\psi(2\mathrm{S})} - m_{J/\psi}$. 
An example of the acceptance map for the lowest dipion mass ($m_{\pi^+\pi^-} \simeq 2m_{\pi}$) is shown in figure~\ref{fig:accMap:1} for the isotropic $\psi(2\mathrm{S})$ assumption. 
The variation of acceptance with dipion mass is illustrated
by the ratio of the acceptance 
at the lowest dipion mass $m_{\pi^+\pi^-} \simeq 2m_{\pi}$  to the acceptance at the highest dipion mass 
$m_{\pi^+\pi^-} \simeq m_{\psi(2\mathrm{S})} - m_{J/\psi}$ shown in figure~\ref{fig:accMap:3}.
The largest variations are observed at low $p_{\rm T}$ and at high rapidity, reaching $\pm 20\%$ within the $p_{\rm T}$--$y$ range of this measurement ($\psi(2\mathrm{S})$ rapidity $|y|<2.0$
and transverse momentum between 10~GeV and 100~GeV).

It has been shown\,\cite{BES} that the dipion state is largely dominated by the angular momentum configuration where
the two pions are in a relative $S$-wave state, and the $J/\psi$ and dipion system are in an $S$-wave state as well. 
The spin-alignment of $J/\psi$ from $\psi(2\mathrm{S})$ decay is thus
assumed to be fully transferred from the spin-alignment of $\psi(2\mathrm{S})$ and hence, in its decay frame, the angular dependence of the decay $J/\psi\to\mu^+\mu^-$ is given by
\begin{equation}
\frac{\mathrm{d}^2 N}{\mathrm{d}\cos \theta^* \mathrm{d}\phi^*} \propto \left(\frac{1}{3+\lambda_\theta}\right) \left( 1 + \lambda_{\theta} \cos^2 \theta^* + \lambda_{\phi} \sin^2 \theta^* \cos 2\phi^* + \lambda_{\theta \phi} \sin 2\theta^* \cos \phi^* \right),
\label{eqn:spin}
\end{equation}
where the $\lambda_i$ are coefficients related to the spin density matrix elements of the $\psi(2\mathrm{S})$ wavefunction\,\cite{Faccioli:2010kd}.
The polar angle $\theta^*$ and the azimuthal angle $\phi^*$ are defined by the momentum of the positive muon in the $J/\psi\to\mu^+\mu^-$ decay frame with respect to the direction
of the $\psi(2\mathrm{S})$ momentum in the lab frame.
In the default case of isotropic $\psi(2\mathrm{S})$ decay, all three 
$\lambda_i$ coefficients in eq.~(\ref{eqn:spin}) are equal to zero. 
This assumption is compatible with measurements for prompt\,\cite{CMSPol,Aaij:2014qea} and non-prompt\,\cite{Abulencia:2007us} production.

In certain areas of the phase space, the acceptance $\mathcal{A}$ may depend quite strongly on the values of the $\lambda_i$ coefficients in eq.~(\ref{eqn:spin}).
Seven extreme cases that lead to the largest possible variations of acceptance within the phase space of this measurement are identified.
These cases, described in table~\ref{tab:spin}, are used to define a range in which the results may vary under any physically allowed spin-alignment assumptions.

\begin{table}[htbp]
\begin{center}
\caption{Values of angular coefficients describing spin-alignment scenarios 
with maximal effect on the measured rate for a given total production cross-section.}
\vspace{2mm}
\begin{tabular}[h]{r|ccc}
\hline\hline
      & \multicolumn{3}{c}{Angular coefficients} \\ 
      & $\lambda_{\theta}$ & $\lambda_{\phi}$ & $\lambda_{\theta\phi}$ \\ \hline
Isotropic {\em (central value)}            & $0$ & $0$ & $0$ \\
Longitudinal         & $-1$ & $0$ & $0$ \\
Transverse positive  & $+1$ & $+1$ & $0$ \\
Transverse zero      & $+1$                & $0$ & $0$ \\
Transverse negative  & $+1$            & $-1$ & $0$ \\
Off-($\lambda_{\theta}$--$\lambda_{\phi}$)-plane positive   & $0$ & $0$ & $+0.5$ \\
Off-($\lambda_{\theta}$--$\lambda_{\phi}$)-plane negative   & $0$ & $0$ & $-0.5$ \\
\hline\hline
\end{tabular}
\label{tab:spin}
\end{center}
\end{table}

\begin{figure}[tb]
  \begin{center}
    \subfigure[$|y| < 0.75$\label{fig:polRatio1}]{\includegraphics[width=0.49\textwidth]{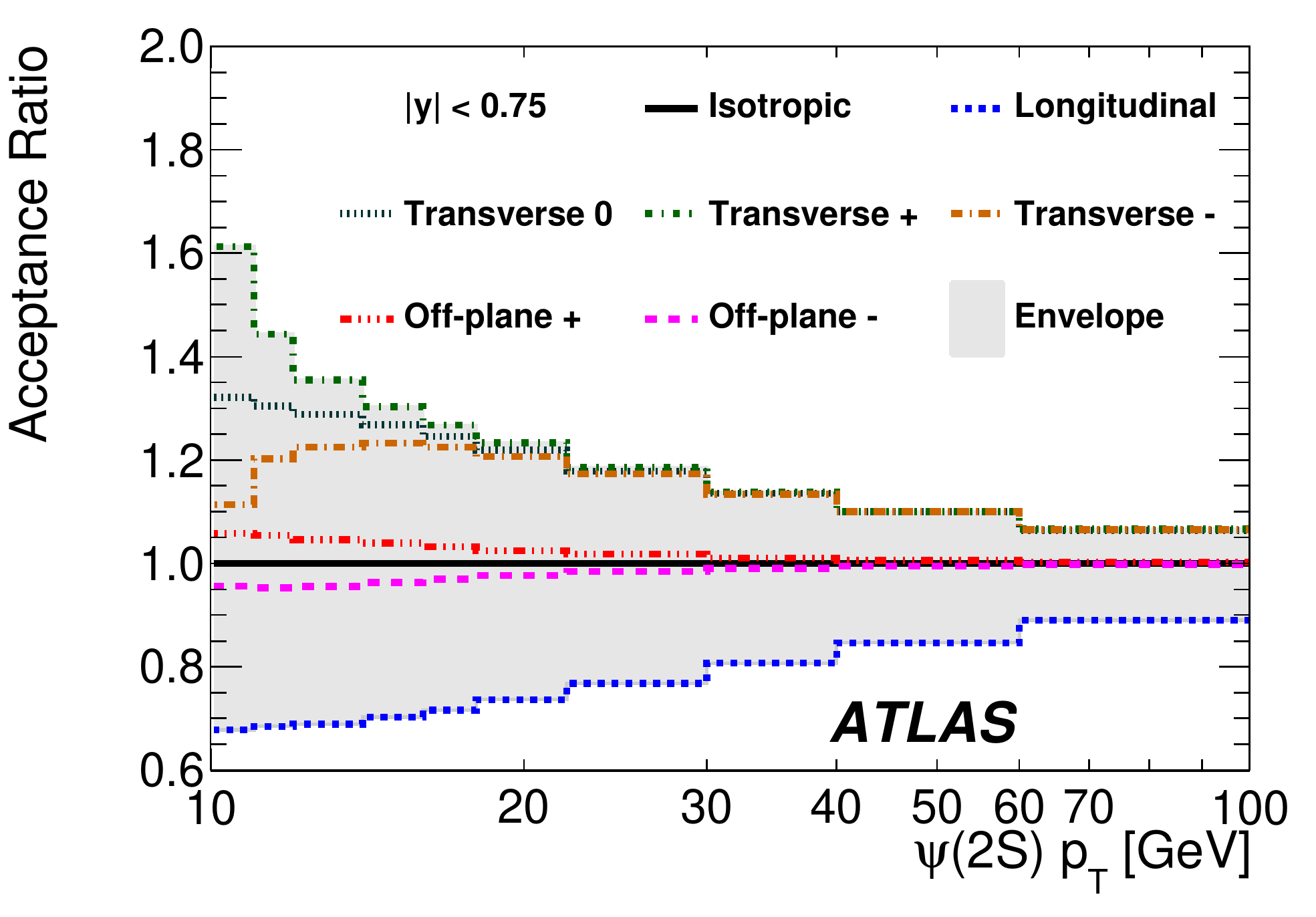}}
    \subfigure[$0.75 \leq |y| < 1.5$\label{fig:polRatio2}]{\includegraphics[width=0.49\textwidth]{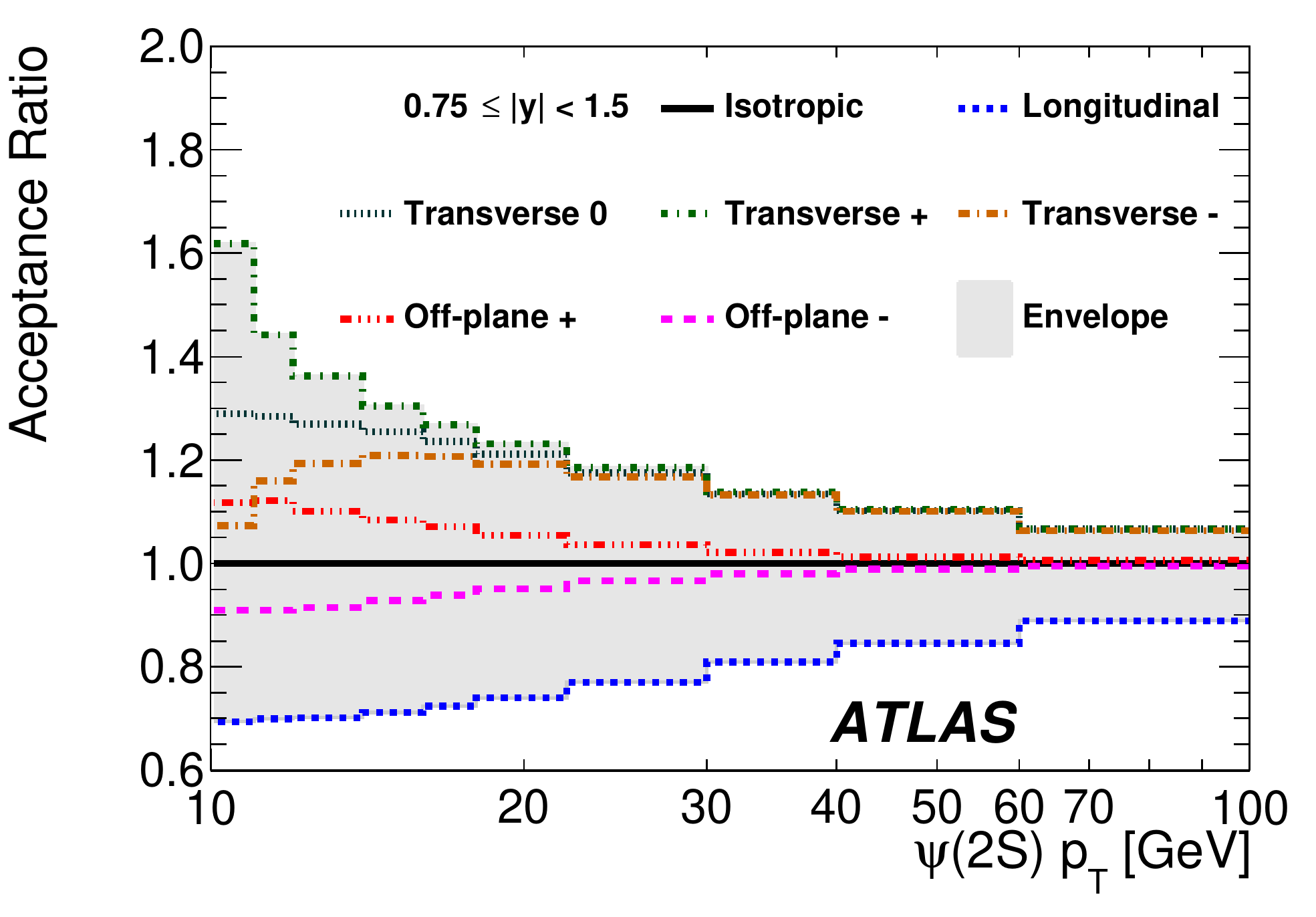}}
    \subfigure[$1.5 \leq |y| < 2.0$\label{fig:polRatio3}]{\includegraphics[width=0.49\textwidth]{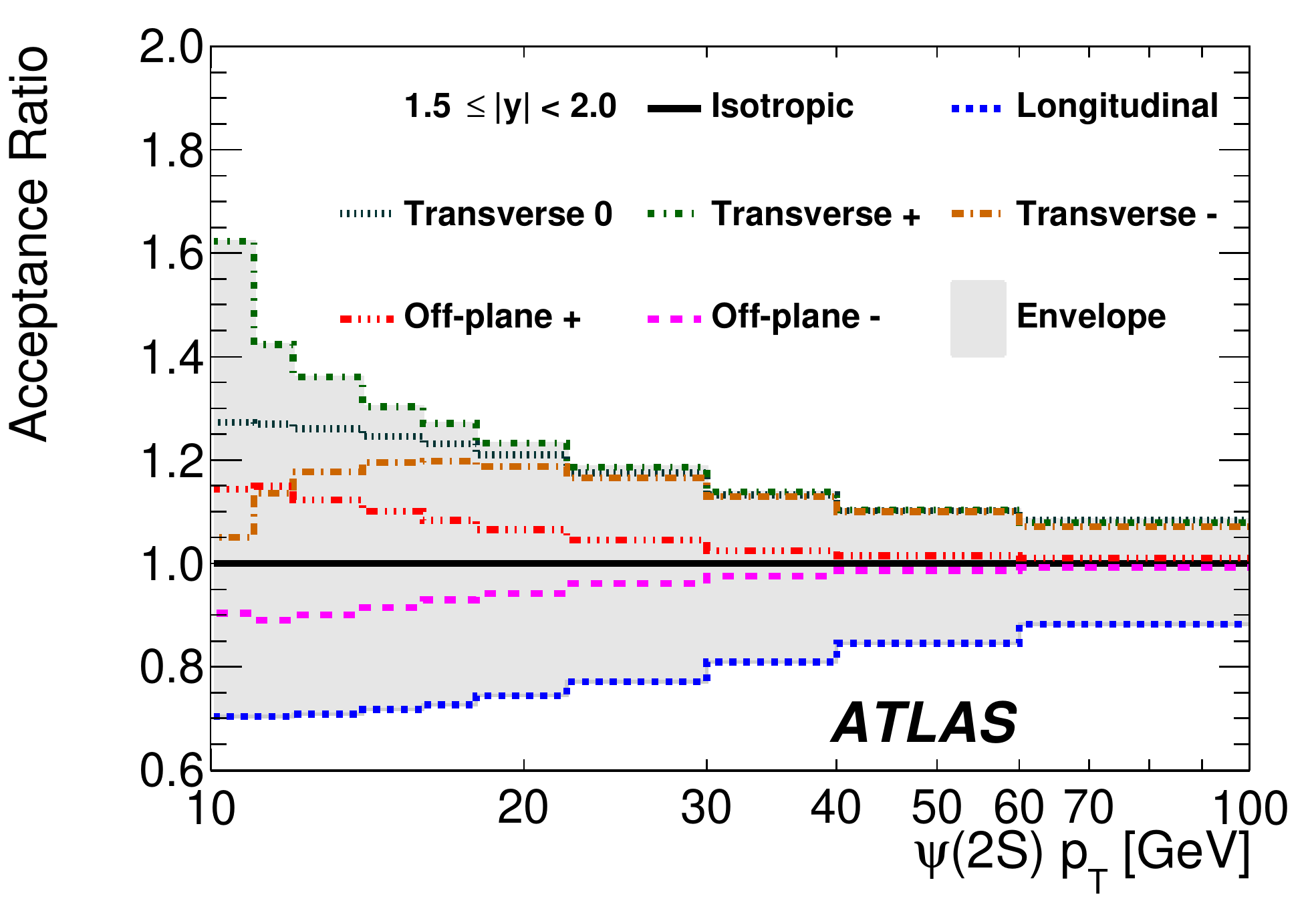}}
    \subfigure[$10 \leq p_{\rm T} < 100$ {GeV}\label{fig:polRatioRap}]{\includegraphics[width=0.49\textwidth]{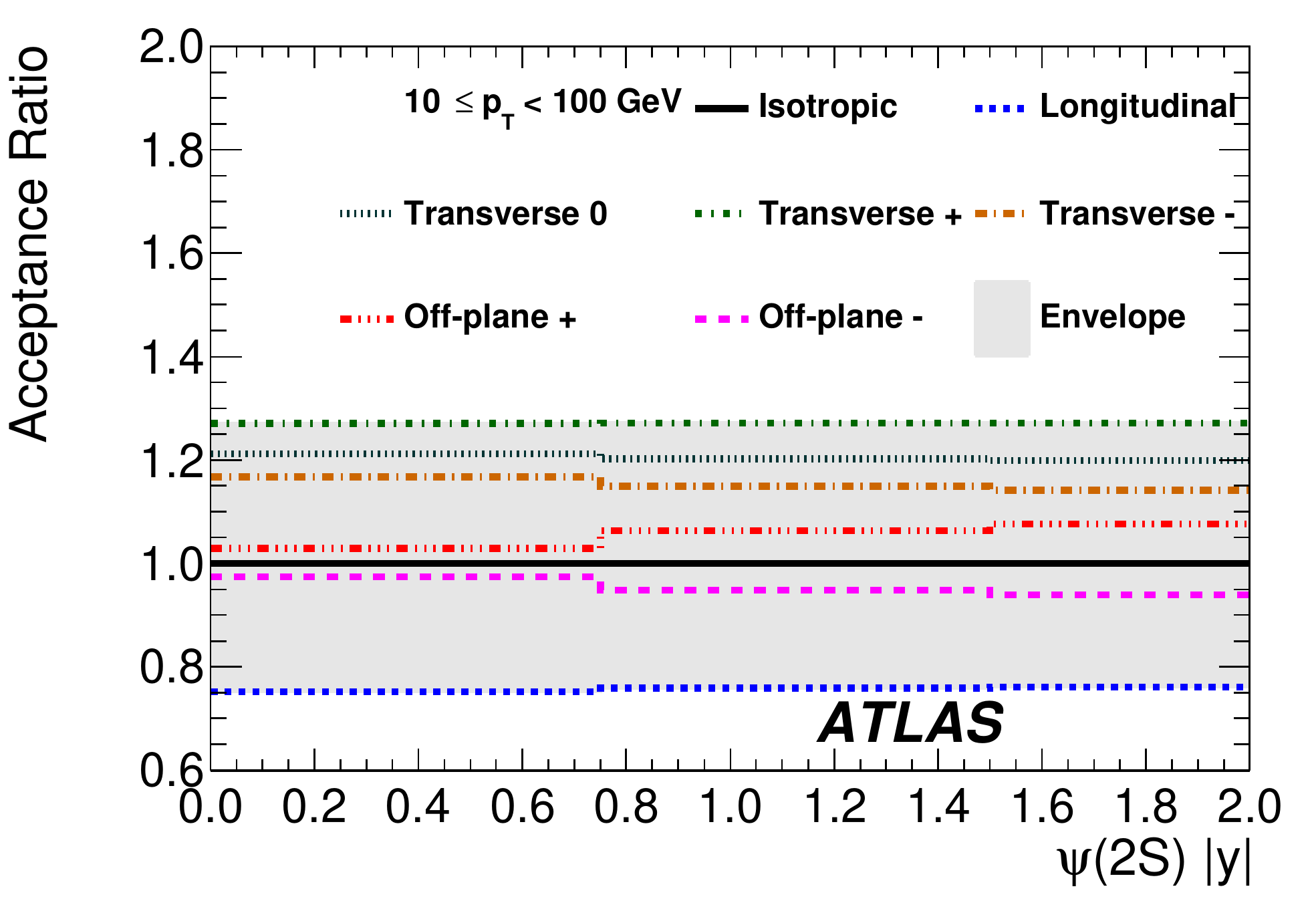}}
    \caption{Average acceptance correction relative to the isotropic scenario for the six extreme spin-alignment scenarios described in the text, 
       (a)-(c) as a function of $\psi(2\mathrm{S})$ transverse momentum in the three rapidity regions,
      and (d) versus $\psi(2\mathrm{S})$  rapidity for $10<p_{\rm T}<100$~GeV.
    \label{fig:polRatio}
    }
  \end{center}
\end{figure}

\begin{figure}[tb]
  \begin{center}
    \subfigure[$|y| < 0.75$\label{fig:polRatio4}]{\includegraphics[width=0.49\textwidth]{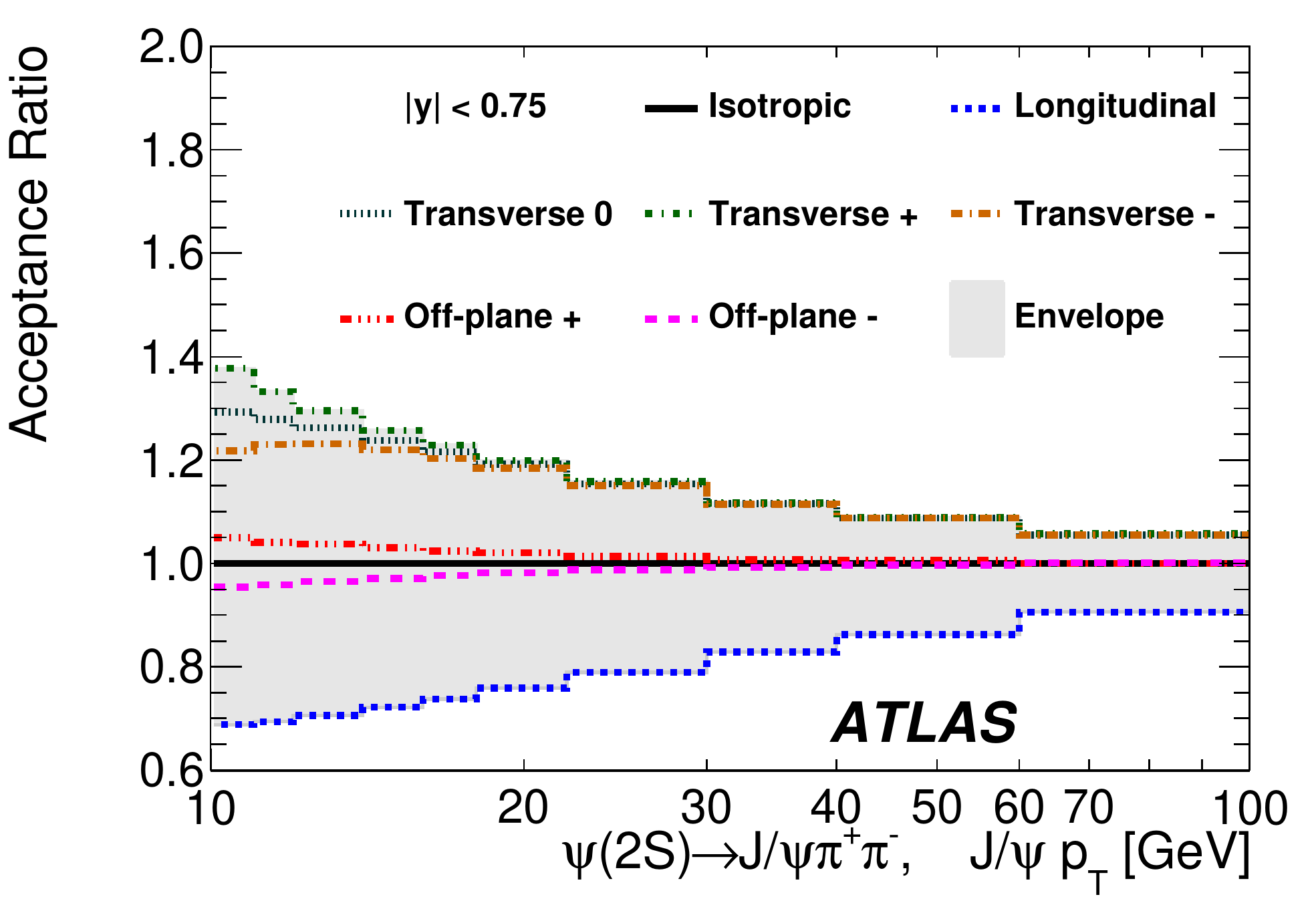}}
    \subfigure[$0.75 \leq |y| < 1.5$\label{fig:polRatio5}]{\includegraphics[width=0.49\textwidth]{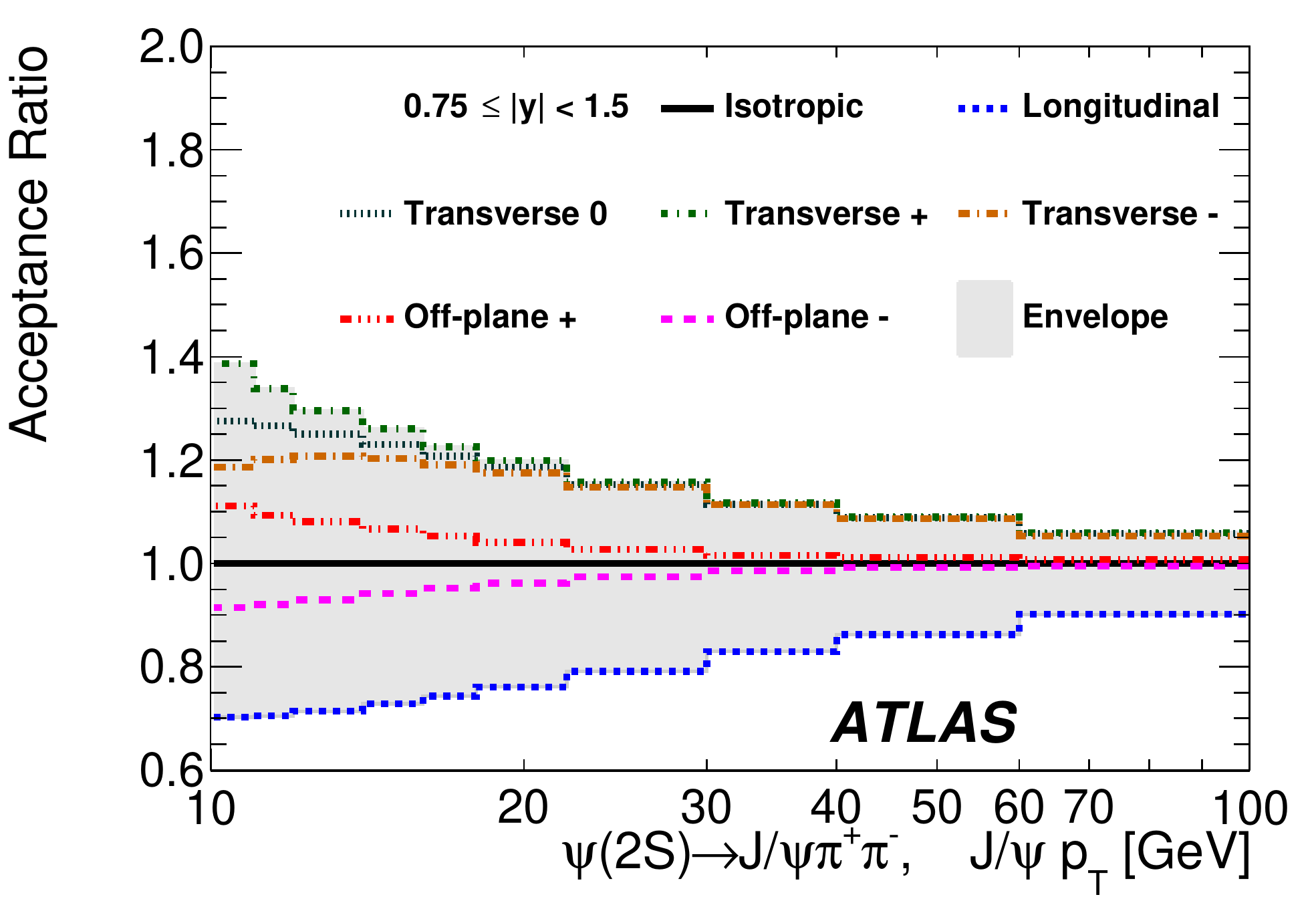}}
    \subfigure[$1.5 \leq |y| < 2.0$\label{fig:polRatio6}]{\includegraphics[width=0.49\textwidth]{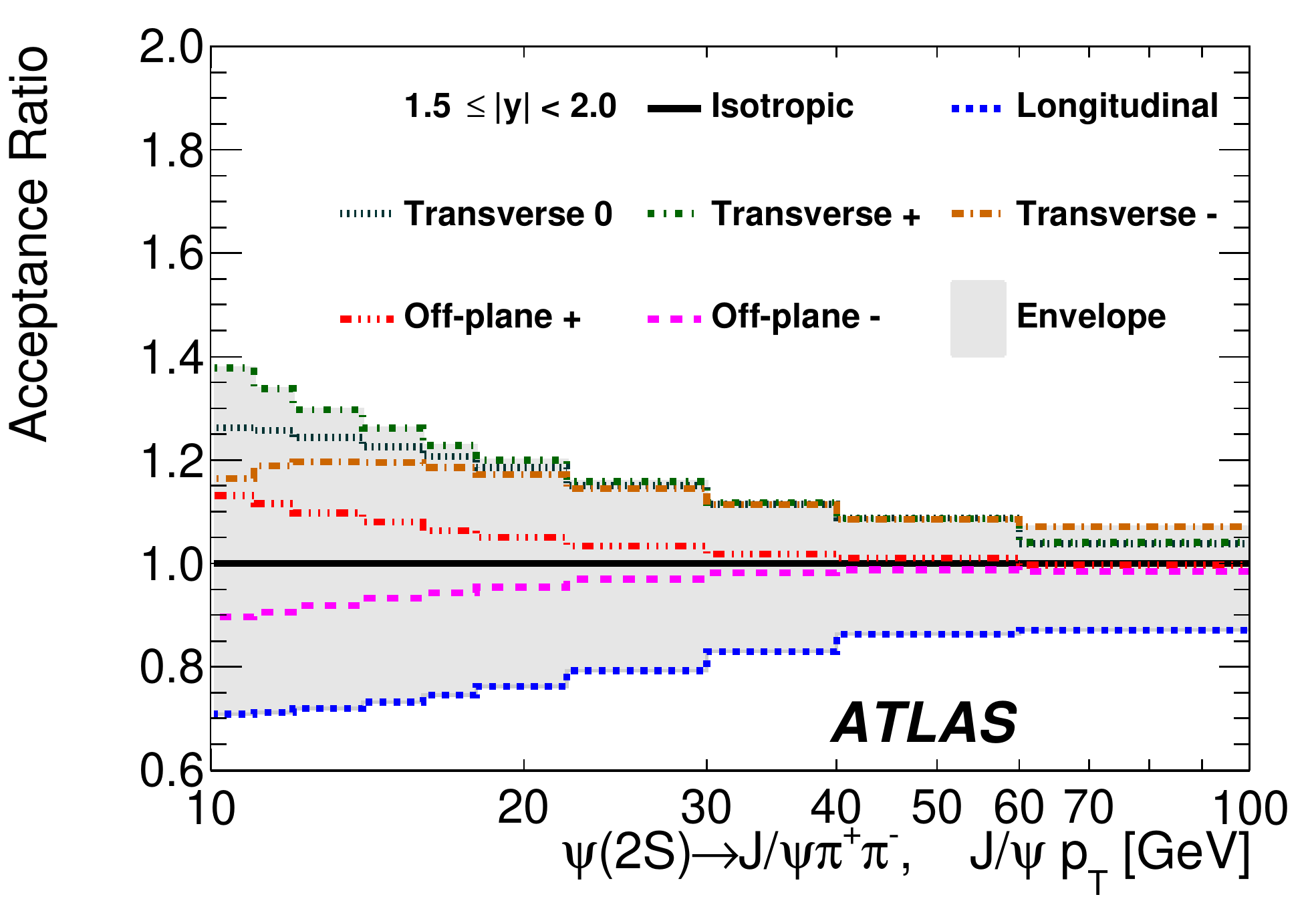}}
    \subfigure[$10 \leq p_{\rm T} < 100$ {GeV}\label{fig:polRatioRap2}]{\includegraphics[width=0.49\textwidth]{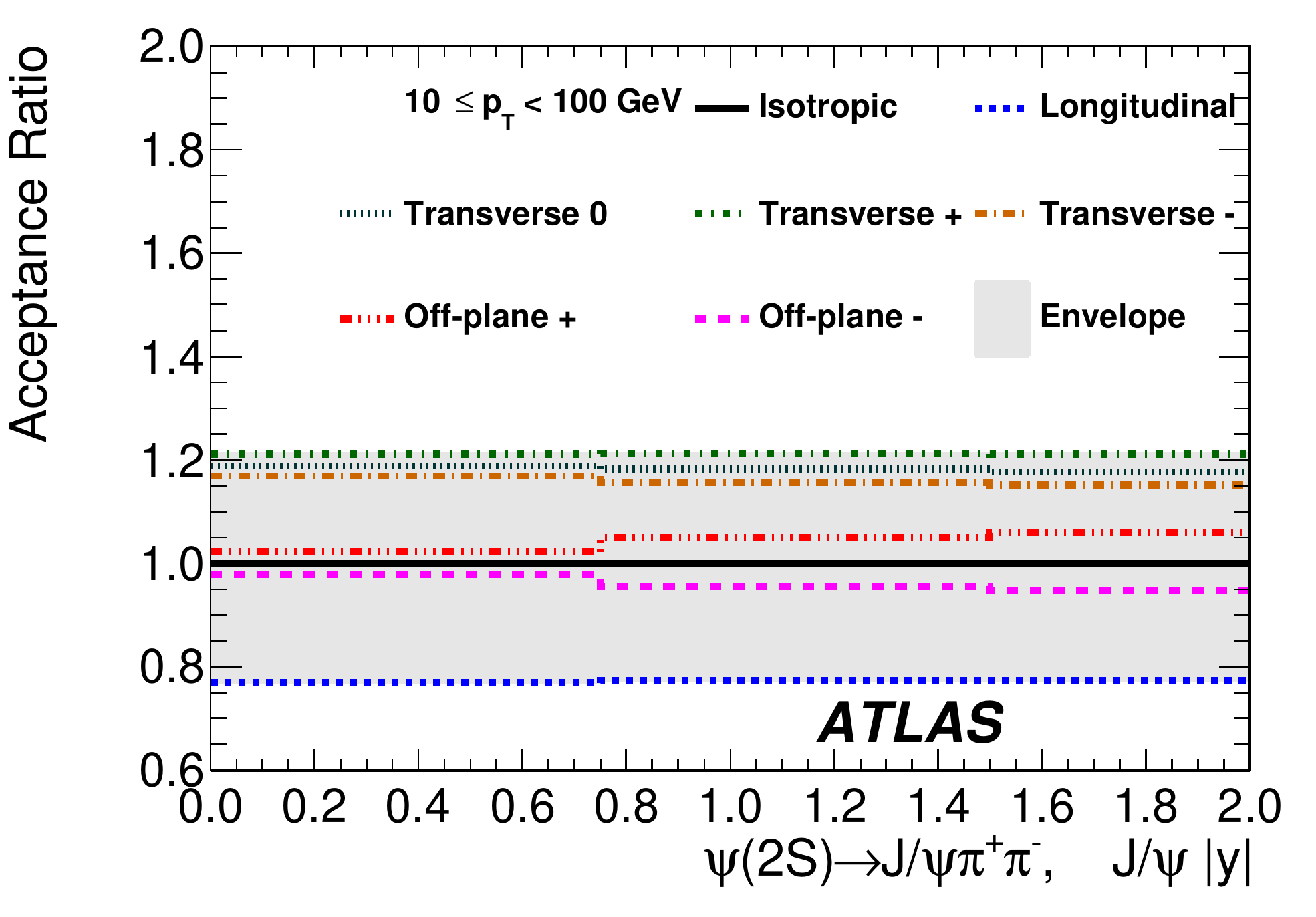}}
    \caption{Average acceptance correction relative to the isotropic scenario for the six extreme spin-alignment scenarios described in the text, 
       as a function of the transverse momentum of the $J/\psi$ in $\psi(2\mathrm{S})\to\Jmumupipi$ decays in (a)-(c) the three rapidity regions,
       and (d) versus $J/\psi$ rapidity for $10\leq p_{\rm T}<100$~GeV.
    \label{fig:polRatioJpsi}
    }
  \end{center}
\end{figure}

Figures~\ref{fig:polRatio} and~\ref{fig:polRatioJpsi} illustrate
the variation of the acceptance correction weights with $p_{\rm T}$ and rapidity of $\psi(2\mathrm{S})$ and $J/\psi$ from the $\psi(2\mathrm{S})\to\Jmumupipi$ decay,
for the six anisotropic spin-alignment scenarios described above, relative to the isotropic case.
There is a clear dependence on the spin-alignment scenario. This can be as large as $(+62\%,-32\%)$ for strong polarisations at the lowest $p_{\rm T}$ probed, but the effect is limited to $(+8\%,-12\%)$
at the highest $p_{\rm T}$ probed.
Since spin-alignment is regarded as an ultimately resolvable model-dependence issue rather than an intrinsic experimental shortcoming,
the associated uncertainties are handled here differently from purely experimental systematic uncertainties.
The range of variation of our cross-section results due to possible spin-alignment scenarios is documented in appendix~\ref{sec:acccorr}.

\subsection*{Dimuon reconstruction efficiency}

The dimuon reconstruction efficiency, determined via a data-driven tag-and-probe method\,\cite{Upsilon} from $J/\psi\to\mu^+\mu^-$ decays, is given by:
\begin{equation}
  \epsilon^{\mu}_{\mathrm{reco}} = \epsilon_{\rm trk}(p^{\mu 1}_{\rm T},\eta^{\mu 1}) \cdot \epsilon_{\rm trk}(p^{\mu 2}_{\rm T}, \eta^{\mu 2}) \cdot \epsilon_{\mu} (p^{\mu 1}_{\rm T},\, q^{\mu 1}\cdot \eta^{\mu 1}) \cdot \epsilon_{\mu} (p^{\mu 2}_{\rm T},\, q^{\mu 2}\cdot \eta^{\mu 2}),
\end{equation}
where $q$ is the charge of the muon, $\epsilon_{\rm trk}$ is the muon track reconstruction efficiency in the ID, while $\epsilon_{\mu}$
is the efficiency of the muon reconstruction algorithm given that the muon track has been reconstructed in the ID. 
The dependence on charge is due to the effect of the toroidal field bending particles into or out of the detector at low momenta and high rapidities.
The muon track reconstruction efficiency $\epsilon_{\rm trk}$ is determined\,\cite{Upsilon} to be $(99\pm 1)\%$ per muon candidate within the kinematic range of interest.
Possible correlation effects were found to be negligible due to the large spatial separation of the two reconstructed muon candidates relative to the spatial resolution of the detector.

\subsection*{Dipion reconstruction efficiency}

The dipion reconstruction efficiency $\epsilon^{\pi}_{\mathrm{reco}}$ is given by:
\begin{equation}
\epsilon^{\pi}_{\mathrm{reco}} =
\epsilon_{\pi} (p^{\pi 1}_{\rm T},\, \eta^{\pi 1}) \cdot \epsilon_{\pi} (p^{\pi 2}_{\rm T},\, \eta^{\pi 2}),
\end{equation}
where the two $\epsilon_{\pi}$ are individual pion reconstruction efficiencies. These are 
determined using techniques derived for tracking-efficiency measurements\,\cite{TrkEff}.
Pions produced in MC event simulation using a PYTHIA6\,\cite{pythia} sample of $\psi(2\mathrm{S})\to\Jmumupipi$ decays
were used to determine the efficiencies in the interval $p_{\rm T}>0.5$~GeV and $|\eta|<2.5$.
The MC sample was produced using the ATLAS 2011 MC tuning\,\cite{pythia_tune} and simulated using
the ATLAS GEANT4\,\cite{Agostinelli:2002hh} detector simulation\,\cite{:2010wqa}.
The pion track reconstruction efficiencies are calculated in intervals of 
track pseudorapidity 
and transverse momentum.
In addition to the statistical uncertainties on the efficiency due to the size of the MC sample,
each efficiency value also contains an additional uncertainty to account for any possible mismodelling in the simulations.
Pion candidates were found to have spatial separations sufficient to not require additional corrections for correlations in reconstruction efficiency.

\subsection*{Trigger efficiency}

The efficiency of the dimuon trigger used in this analysis was measured in a previous analysis\,\cite{Upsilon}
from $J/\psi\to\mu^+\mu^-$ and $\Upsilon\to\mu^+\mu^-$ decays using a data-driven method. 
The trigger efficiency is the efficiency for the trigger system to select signal events that also pass the reconstruction-level analysis selection, and is parameterised as:
\begin{equation}
\epsilon_{\mathrm{trig}} = \epsilon_{\mathrm{RoI}}(p^{\mu 1}_{\rm T},\, q^{\mu 1},\,\eta^{\mu 1})\cdot \epsilon_{\mathrm{RoI}}(p^{\mu 2}_{\rm T},\, q^{\mu 2},\,\eta^{\mu 2})\cdot c_{\mu\mu}(\Delta R, |y_{\mu\mu}|),
\end{equation}
where $\epsilon_\mathrm{RoI}$ is the efficiency of the trigger system to find an RoI for a single muon 
and $c_{\mu\mu}$ is a correction term taking into account muon--muon correlations, 
dependent on the angular separation $\Delta R$ between the two muons, and the absolute rapidity of the dimuon system, $|y_{\mu\mu}|$.
The invariant mass requirement of the trigger was found to be fully efficient, with a correction for an efficiency of $(99.7\pm 0.3)\%$ 
applied to account for possible signal loss as determined from MC simulation.

\subsection*{Total weight}

The total weight $w$ for each $\Jpipi$ candidate was calculated as the inverse of the product of acceptance and efficiency corrections, as described by: 
\begin{equation}
w^{-1} = \mathcal{A}(p_{\rm T},y,m_{\pi\pi}) \cdot 
\epsilon^{\mu}_{\mathrm{reco}} \cdot
\epsilon^{\pi}_{\mathrm{reco}} \cdot
\epsilon_{\mathrm{trig}}.
\label{eqn:weight}
\end{equation} 

\begin{figure}[tbp]
\centering 
\subfigure[$|y| < 0.75$\label{fig:weights1}]{\includegraphics[width=0.49\textwidth]{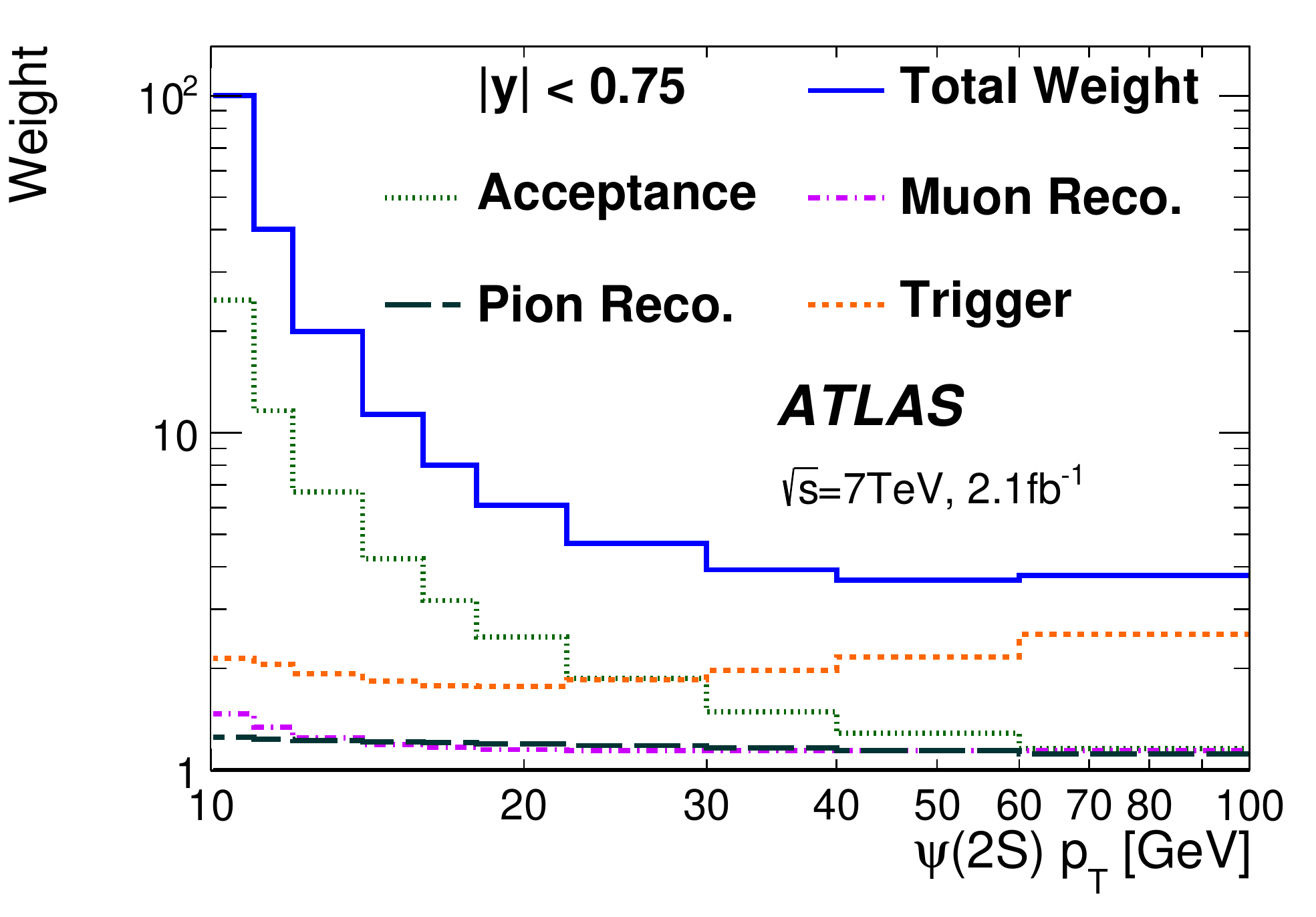}}
\subfigure[$0.75 \leq |y| < 1.5$\label{fig:weights2}]{\includegraphics[width=0.49\textwidth]{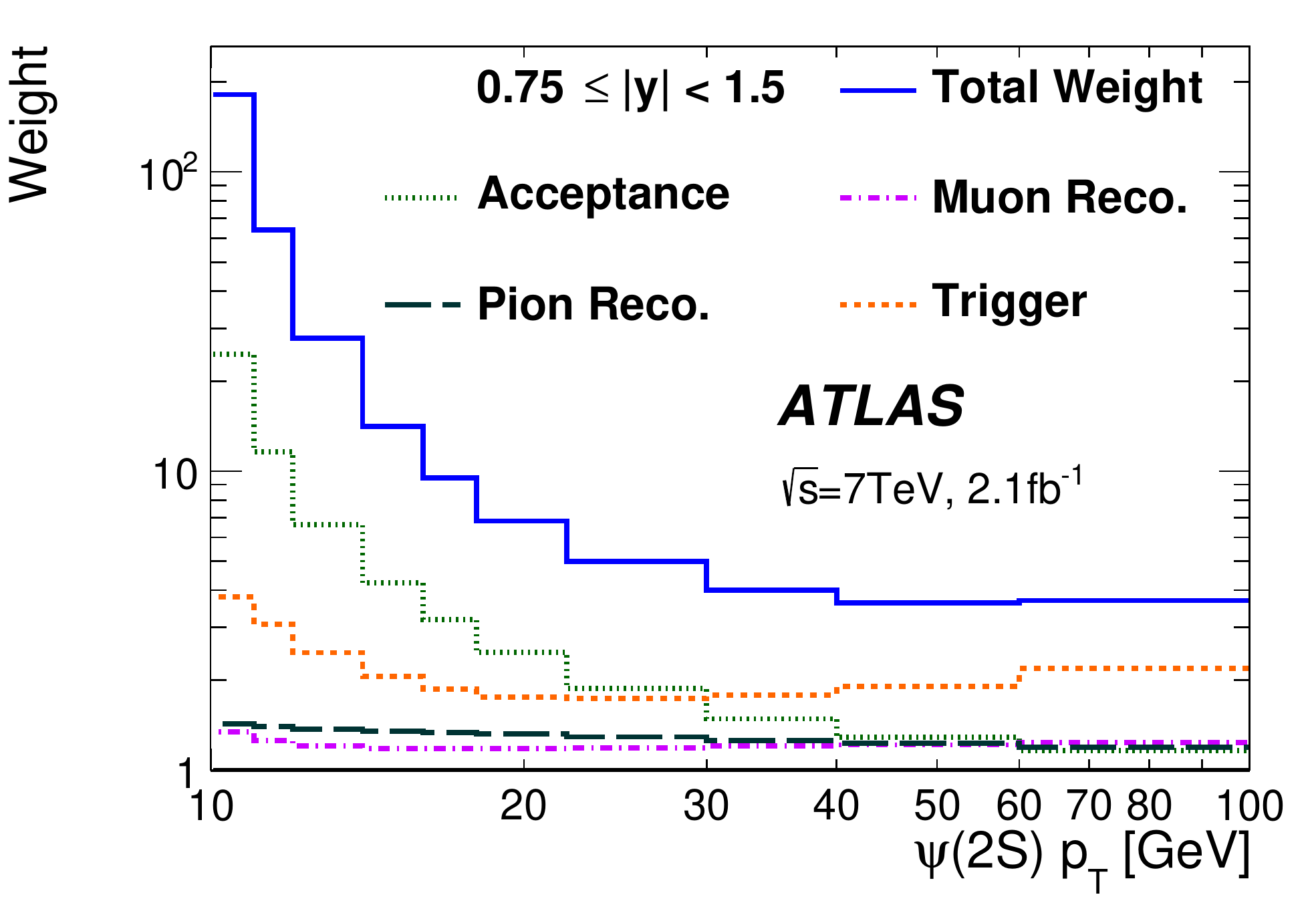}}
\subfigure[$1.5 \leq |y| < 2.0$\label{fig:weights3}]{\includegraphics[width=0.49\textwidth]{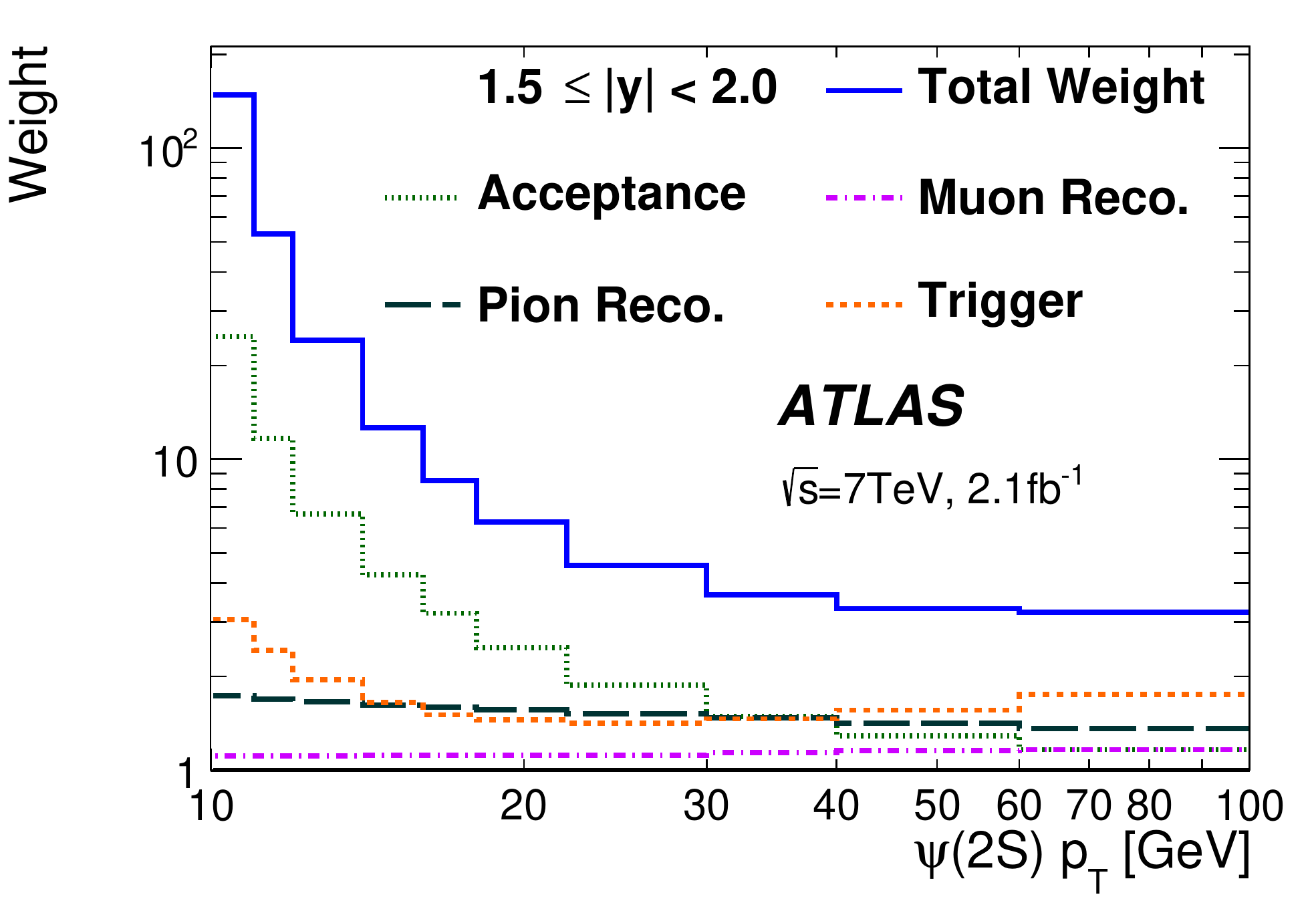}}
\subfigure[$10 \leq p_{\rm T} < 100$ {GeV}\label{fig:weightsRap}]{\includegraphics[width=0.49\textwidth]{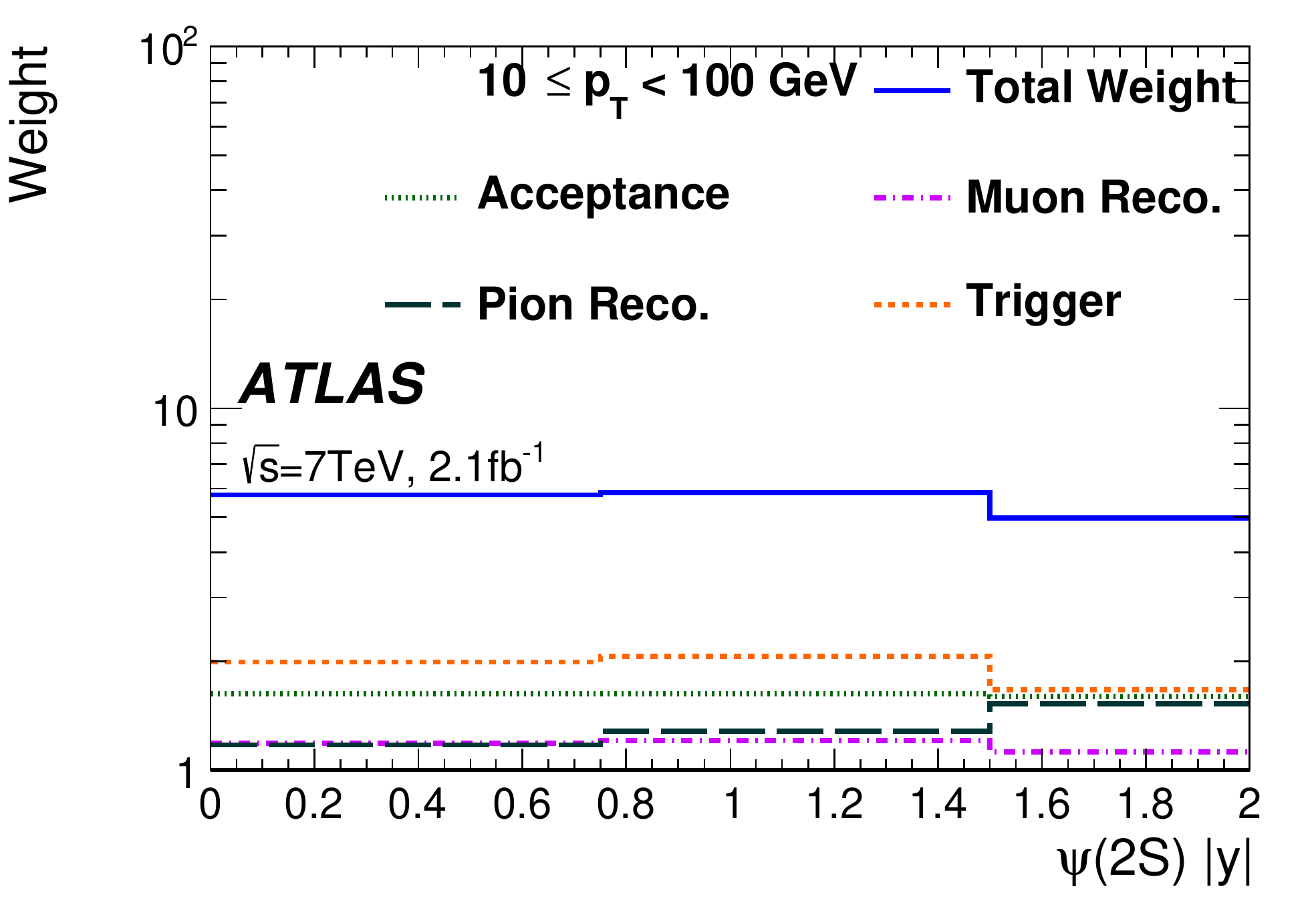}}
\caption{Average correction weights (a)-(c) for the three rapidity regions versus $p_{\rm T}$ and (d) for the full $p_{\rm T}$ region versus $|y|$.}
\label{fig:weights}
\end{figure}

No lifetime dependence was observed in any of the efficiency corrections.
While weights are applied to the data on a candidate-by-candidate basis, 
the average of the total weight and its breakdown into individual sources is shown in figure~\ref{fig:weights} for the three rapidity regions and in each $p_{\rm T}$ bin of the measurement,
and as an average over the full transverse momentum range ($10\leq p_{\rm T}<100$~GeV) versus rapidity. The inverse of these weights illustrate a representative average efficiency correction in each measurement interval.

\subsection*{Fitting procedure}
\label{sec:Fitting}
The corrected  prompt and non-prompt $\psi(2\mathrm{S})$ signal yields are
extracted from two-dimensional weighted unbinned maximum likelihood fits performed on the
$\Jpipi$ invariant mass ($m$) and pseudo-proper lifetime ($\tau$) in each $p_{\rm T}$--$|y|$ interval.
The probability density function (PDF) for the fit is defined as a
normalised sum, where each term is factorised into mass-  and lifetime-dependent functions.
The PDF can be written in a compact form as
\begin{equation}
\mathrm{PDF}(m,\tau) = \sum_{i=1}^{5} \kappa_i f_i(m) \cdot h_i(\tau)
\otimes G(\tau),
\label{eqn:fitModel}
\end{equation}
where $\kappa_i$ represents the relative normalisation of the $i^{\rm th}$ term (such that $\sum_i \kappa_i \equiv 1$),
$f_i(m)$ is the mass-dependent term, and $\otimes$ represents the convolution of the lifetime-dependent function $h_i(\tau)$ with
the lifetime resolution term, $G(\tau)$.
The latter is modelled by a Gaussian distribution with mean fixed to zero and resolution determined from the fit.

\begin{table}[htbp]
\begin{center}
\caption{Components of the probability density function used to extract
the prompt (P) and non-prompt (NP) contributions for signal (S) and background (B). 
}
\vspace{2mm}
\begin{tabular}[h]{ccccc}
\hline\hline
i & Type               & Source           &  $f_i(m)$ & $h_i(\tau)$ \\ \hline 
1 & S           & P          & $\omega G_{1}(m) + (1-\omega) G_{2} (m) $ & $\delta(\tau)$ \\
2 & S           & NP  & $\omega G_{1}(m) + (1-\omega) G_{2} (m) $  & $E_{1}(\tau)$ \\ \hline
3 & B & P          & $C_{1}(m)$                   & $\delta(\tau)$ \\
4 & B & NP & $C_{2}(m)$                   & $\rho E_{2}(\tau) + (1-\rho) E_{3}(\tau)$ \\
5 & B & NP & $C_{3}(m)$                   & $E_{4}(|\tau|)$ \\ \hline\hline
\end{tabular}
\label{tab:fitModel}
\end{center}
\end{table}
Table~\ref{tab:fitModel} shows the five contributions to the overall PDF with the corresponding  $f_i$ and $h_i$ functions. 
Here $G_1$ and $G_2$ are Gaussian distributions with the same mean, but different width parameters (see below), while $C_1$, $C_2$ and $C_3$ 
are different linear combinations of Chebyshev polynomials up to second order. The exponential functions $E_1$, $E_2$, $E_3$ and $E_4$ have different slope parameters, 
where $E_1(\tau)$, $E_2(\tau)$ and $E_3(\tau)$ are required to vanish for $\tau<0$, whereas $E_4(|\tau|)$ is a double-sided exponential
with the same slope parameter on either side of $\tau=0$.
The parameters $\omega$ and $\rho$ represent the fractional contributions of 
the components shown, while $\delta(\tau)$ is the Dirac delta function
modelling the lifetime distribution of prompt candidates.

\begin{figure}[htbp]
\centering
\subfigure[$|y| < 0.75$, $11\leq p_{\rm T} <12$~GeV]{\includegraphics[width=0.7\textwidth]{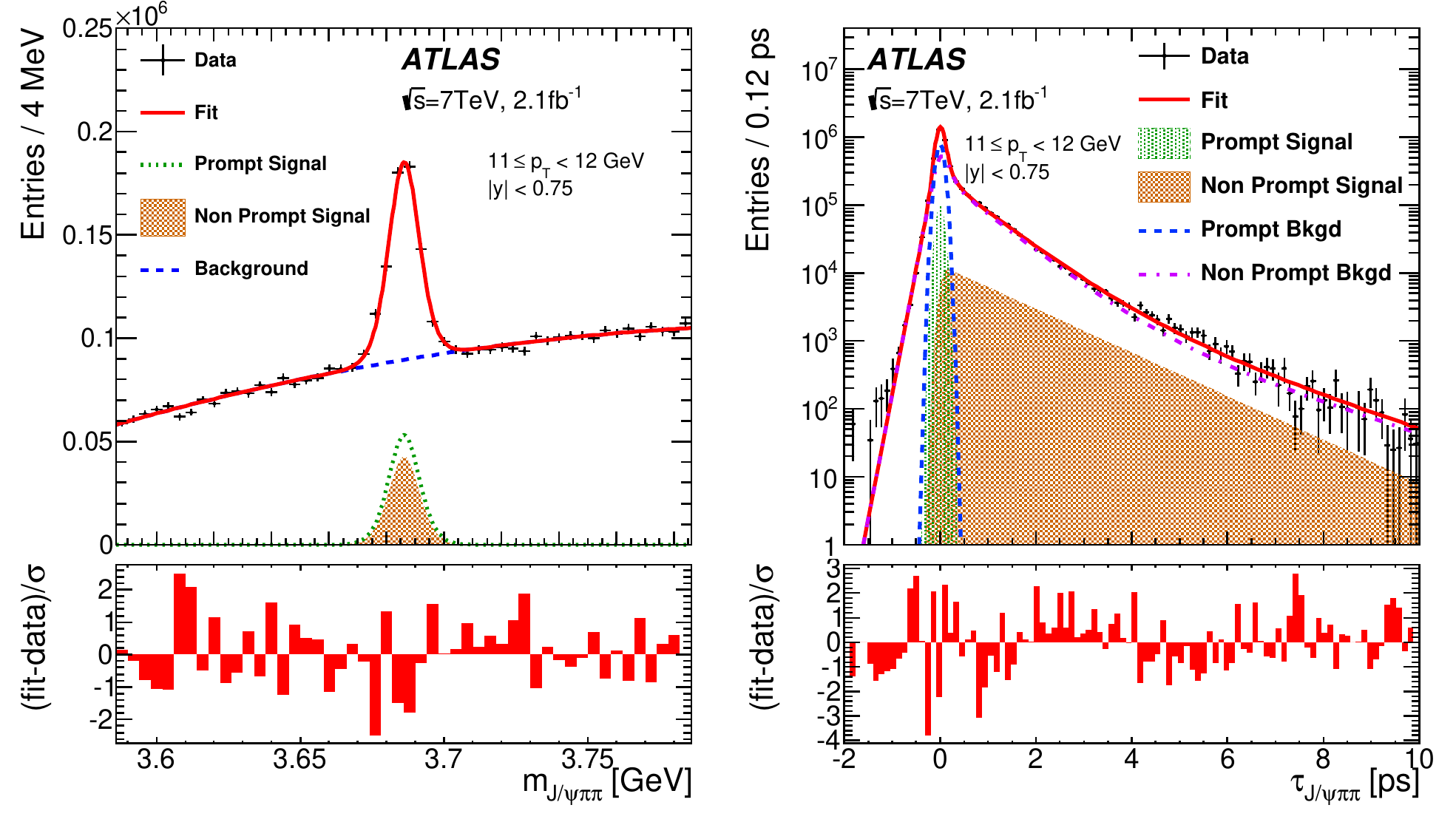}}
\subfigure[$0.75\leq |y| <1.5$, $16\leq p_{\rm T} <18$~GeV]{\includegraphics[width=0.7\textwidth]{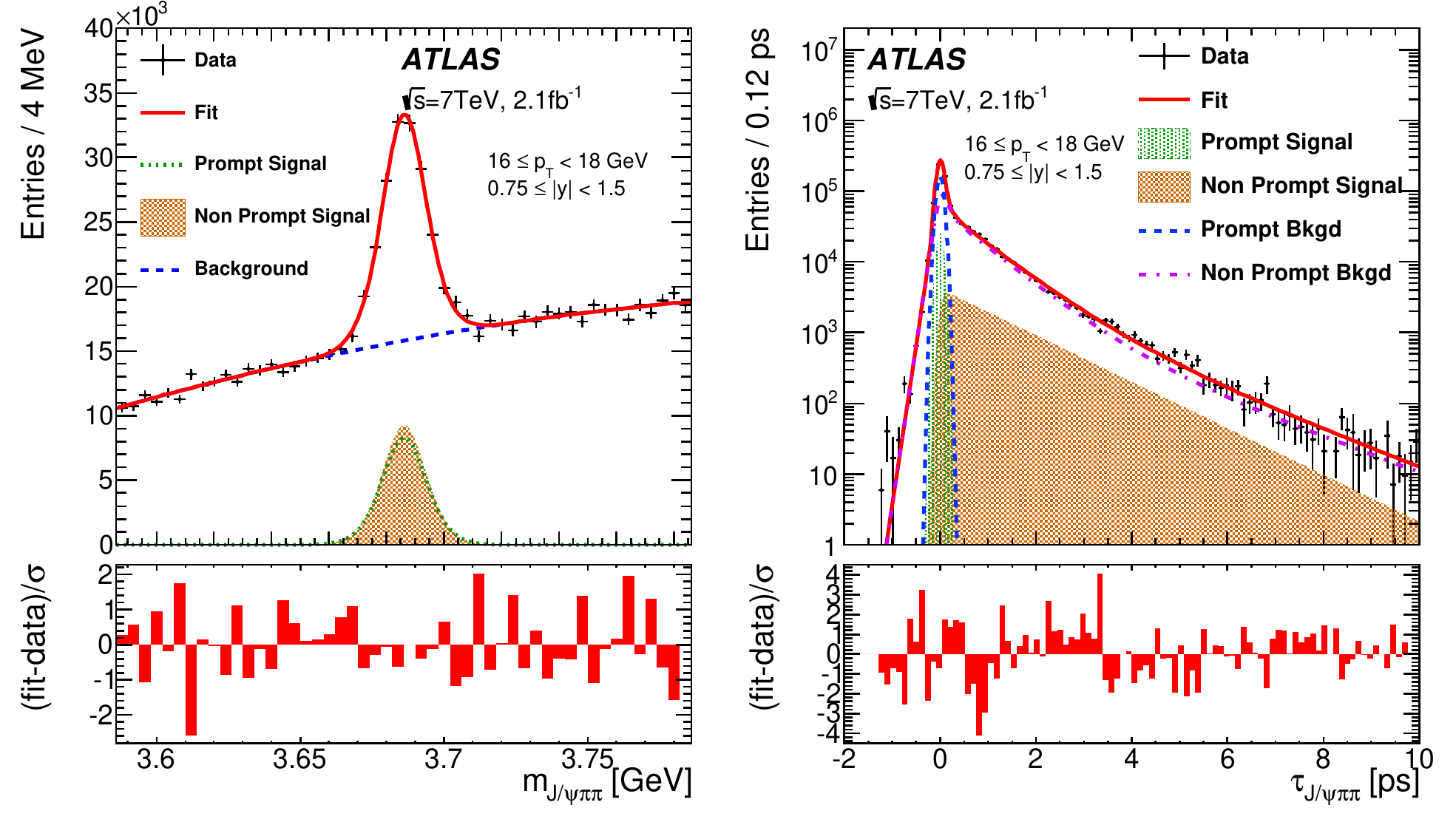}}
\subfigure[$1.5\leq |y| <2.0$, $40\leq p_{\rm T} <60$~GeV]{\includegraphics[width=0.7\textwidth]{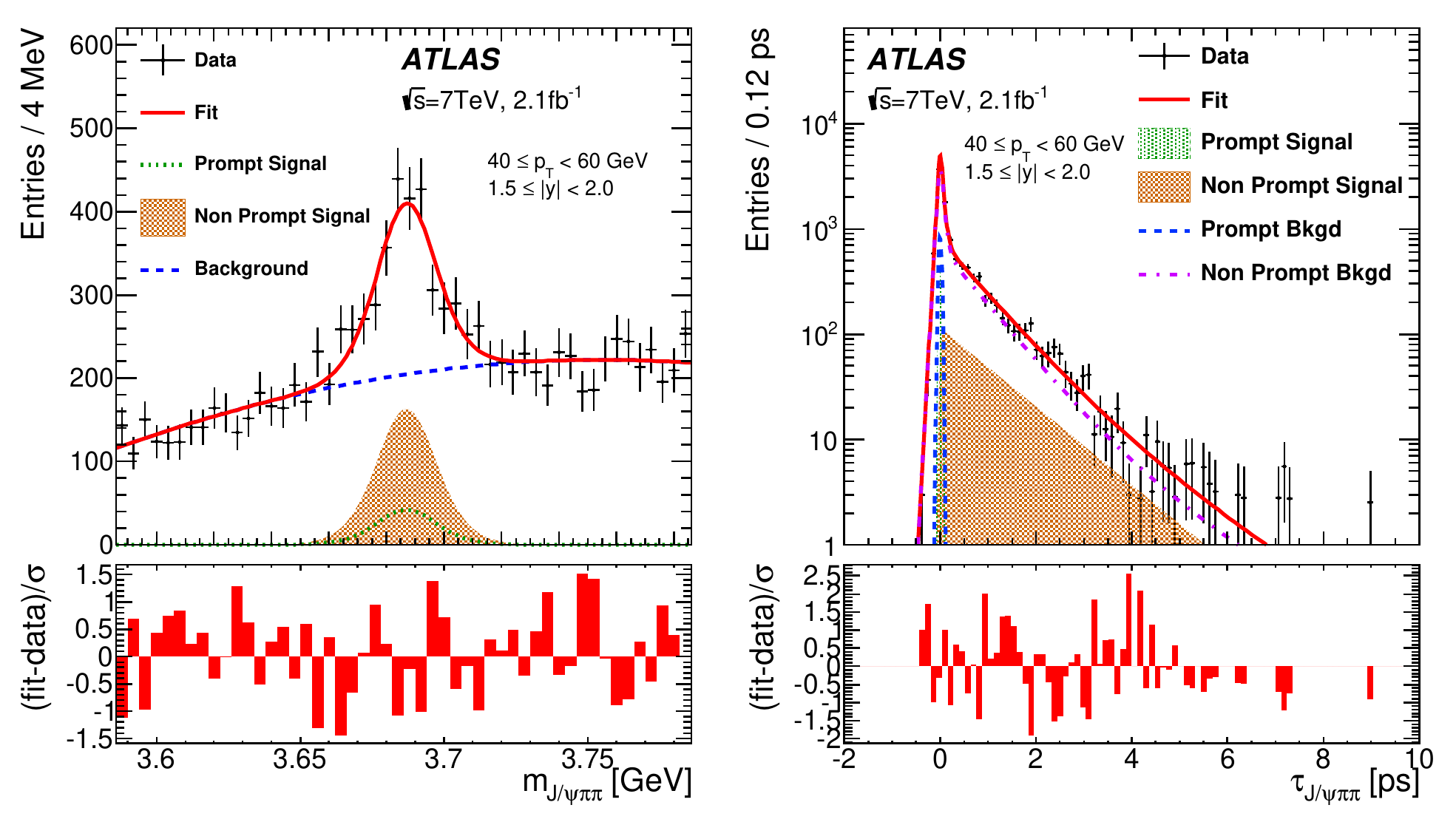}}
\caption{Unbinned maximum likelihood fit and data projections onto the invariant mass and pseudo-proper lifetimes of the $\psi(2\mathrm{S})$ candidates
  for three representative kinematic intervals studied in this measurement. Total signal-plus-background fits to the data are shown, along with the breakdown
  by prompt/non-prompt production for the $\psi(2\mathrm{S})$ signal. The bottom panel shows the pull distribution between the fit and the data.
  \label{fig:ExampleFits}
}
\end{figure}

To better constrain the fit model at high $p_{\rm T}$, the widths of
the Gaussian distributions $G_1$ and $G_2$ are required to satisfy the relation $\sigma_2 = s\sigma_1$.
The values of $\sigma_1$ and $s$ are obtained as a function of $p_{\rm T}$, for each  $|y|$ range, from separate one-dimensional mass fits.
A value of $s=1.5$ is used for the central fit results, and its variation considered within the systematics.
The relative normalisations, $\kappa_i$, $\rho$, and $\omega$, are kept free in all fits, and any autocorrelation effects
are accounted for as part of the systematic uncertainties in the fit procedure.
Projections of the fit results, for three representative  $p_{\rm T}$--$|y|$ intervals, are presented in figure~\ref{fig:ExampleFits}.

\afterpage{\clearpage}

\section{Systematic uncertainties}

Various sources of systematic uncertainties in the measurement are considered and are outlined below.

\subsection*{Acceptance corrections}
The acceptance maps were generated using large event samples from MC simulation. Statistical uncertainties in the maps are assigned as a systematic effect on the acceptance correction (a sub-$1\%$ effect). 
Possible deviation of the spin-alignment from an isotropic configuration is accounted for separately (see figures~\ref{fig:polRatio} and~\ref{fig:polRatioJpsi}).
Other effects, such as smearing of the primary vertex position and momentum resolution causing migrations between particle-level and
reconstruction-level kinematic intervals were studied using methods discussed in previous publications\,\cite{JPsi,Upsilon}.
Corrections due to migration effects were found to be negligible $(<1\%)$, largely because of improved momentum
resolution due to the vertex-constrained and mass-constrained fits.

\subsection*{Fit model variations}
The uncertainty due to the fit procedure was determined by changing one component at a time in the fit model described in section~\ref{sec:Fitting}, 
creating a set of new fit models. For each new fit model, the cross-section was recalculated, and in each $p_{\rm T}$ and $|y|$ interval the maximal variation from the central fit model was used as its systematic uncertainty. 
Table~\ref{tab:tableModel} shows the changes made to the mass and lifetime PDFs in the central fit model, 
as defined in eq.~(\ref{eqn:fitModel}) and table~\ref{tab:fitModel},
where CB is a Crystal Ball function~\mcite{CB1,CB2,CB3}, with parameters $\alpha$ and $n$ fixed, $\alpha=2.0, n=2.0$, 
as determined from test fits. 
\begin{table}[htbp]
\begin{center}
\caption{Fit models used to test the variation from the central model,
where the changes made are highlighted in bold. Definitions of the symbols
are described in the text.
}
\vspace{2mm}
\begin{tabular}[h]{cl}
\hline \hline
 Variation & \\ \hline
 & Mass PDF variation [$f_i(m)$] \\ \hline
1 & $\omega G_1 + (1-\omega) G_2$ (fit $\sigma_1$, $\sigma_2 =
1.5\times\sigma_1$) $\rightarrow$ (fit $\sigma_1$,
$\sigma_2=\mathbf{1.2}\times\sigma_1$ ) \\
2 & $\omega G_1 + (1-\omega) G_2$ (fit $\sigma_1$, $\sigma_2 =
1.5\times\sigma_1$) $\rightarrow$ (fit $\sigma_1$,
$\sigma_2=\mathbf{2.0}\times\sigma_1$  )\\
3 & $\omega G_1 + (1-\omega) G_2$ (fit $\sigma_1$, $\sigma_2 =
1.5\times\sigma_1$) $\rightarrow$ (\textbf{free} $\sigma_1$,
$\sigma_2=1.5\times\sigma_1$ ) \\
4 & $\omega G_1 + (1-\omega) G_2 \rightarrow$  \textbf{CB}(fit $\sigma$,
fixed $\alpha$, n)  \\ \hline
5 & $C_{1,2,3}$  second-order $\rightarrow$ {\bf {third-order}} \\
6 & $C_{1,2,3}$ $\rightarrow$ {\bf Gaussian} \\ \hline
 & Lifetime PDF variation [$h_i(\tau)$ and $G(\tau)$] \\ \hline
7 & Resolution $G(\tau) \rightarrow $ {\bf Double Gaussian}
($\sigma_2=2.0\times\sigma_1$ ) \\
8 & $E_{1}\rightarrow \mathbf{\rho' E_5+ (1-\rho') E_6}$ \\ \hline
9 & $\rho E_2 + (1-\rho) E_3 \rightarrow \mathbf{E_7}$ \\
\hline \hline
\end{tabular}
\label{tab:tableModel}
\end{center}
\end{table}
In table~\ref{tab:tableModel}, ``fit $\sigma$'' means that the result is obtained using the fitted $\sigma$ 
(defined in section~\ref{sec:Fitting})
while ``free $\sigma$'' means that the width $\sigma$ is completely free in the fit.
Fit model changes cause signal yield variations of up to $5\%$--$10\%$ and form one of the dominant uncertainties 
in the cross-section measurement, however
no single variation was found to dominate the total systematic variation in the whole kinematic range.


\subsection*{ID tracking efficiency for muons}

The ID tracking efficiency for muon tracks varies as a function of track transverse momentum and pseudorapidity in the kinematic intervals studied in this analysis.
The tracking efficiency also has a small dependence on the number of proton--proton collisions that contribute to the event. These variations
are contained within a band of $\pm1.0\%$ around the nominal value of $99.0\%$ determined for the efficiency per-track, and this band is directly assigned as a systematic uncertainty in measured cross-sections.

\subsection*{Muon reconstruction and trigger efficiencies}

Uncertainties in the muon reconstruction and muon trigger efficiencies arise predominantly from statistical uncertainties due to the size of the data samples
used to determine the efficiencies. The uncertainties in the $\psi(2\mathrm{S})$ yields are determined for each efficiency map independently by fluctuating each entry in the efficiency maps according to their uncertainty
independently from bin-to-bin to create a series of toy efficiency maps through many such trials. These fluctuated maps are used to recalculate the corrected signal yields in each kinematic bin of the measurement. 
A fit of a Gaussian distribution to the resultant yields (relative to the nominal extraction) allows determination of the $\pm 1\sigma$ variations of these yields up and down 
due to the uncertainties in the individual efficiencies, that affect the measurement at the $3\%$--$5\%$ level.

\subsection*{Pion track reconstruction efficiency}

The pion track reconstruction uncertainty contains the contributions from statistical
uncertainties in the pion efficiency maps, which are estimated using the same procedure as for the muon efficiency maps. 
Systematic uncertainties in the efficiencies are assigned based on tracking efficiency variations observed in alternative detector material and geometry simulations\,\cite{TrkEff}.
The total uncertainties are determined to be $2\%$--$3\%$ per pion in the $p_{\rm T}$ ranges considered,
varying with rapidity, with an additional $1\%$ contribution per pion from the hadronic track reconstruction uncertainties.

\subsection*{Selection criteria}

The efficiency of the constrained $\Jpipi$ vertex-fit quality criterion, ${\mathcal P}(\chi^2)>0.005$, was estimated from data and MC studies to vary
between $93\%$ and $97\%$ as a function of rapidity and $p_{\rm T}$, 
with an uncertainty of about $2\%$, 
determined from data/MC comparison
and the variation of the efficiency with transverse momentum.

Additional inefficiencies from the other selection 
criteria described in section~\ref{sec:selection} 
and their corresponding uncertainties were estimated 
using simulations, and were found to be less than $1\%$ 
in the first two rapidity regions and less than $2\%$ in the highest 
rapidity region. 
These were combined with the efficiencies of the constrained-fit quality requirement 
to calculate the total selection efficiency, which was found to
vary between $92\%$ and $95\%$ with a $2\%$ uncertainty.

\subsection*{Luminosity}

The uncertainty in the integrated luminosity for the dataset used in this analysis was determined\,\cite{lumi} to be $\pm 1.8\%$.
This systematic uncertainty does not affect the measurement of the non-prompt production fraction.

\subsection*{Total uncertainties}

\begin{figure}[h!bpt]
\centering
\includegraphics[width=0.98\textwidth]{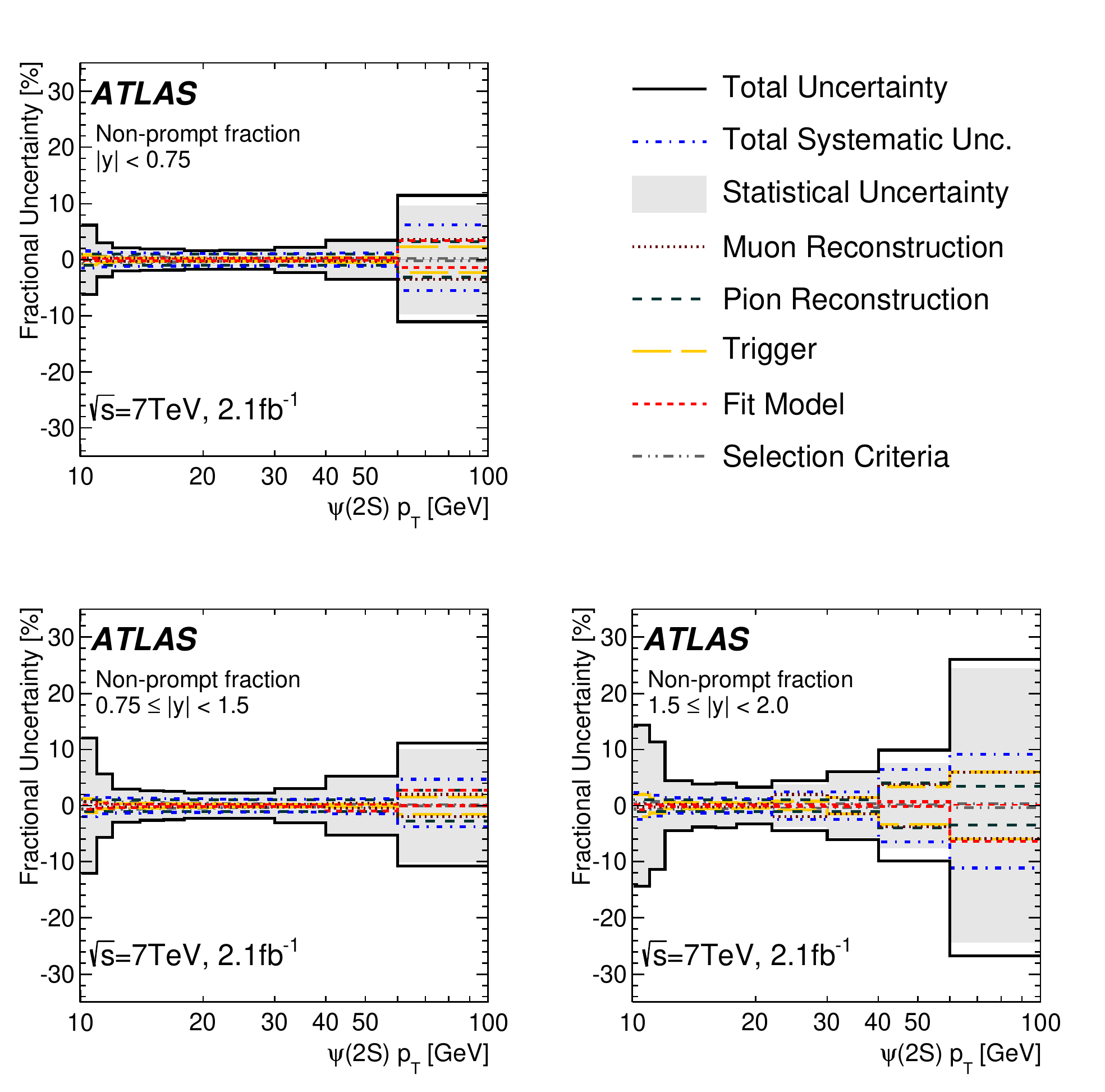}
\caption{Summary of the positive and negative uncertainties for the non-prompt fraction measurement in three $\psi(2\mathrm{S})$ rapidity intervals. The plots do not include the spin-alignment uncertainty.}
\label{fig:Uncertainty:npf}
\end{figure}

\begin{figure}[tbp]
\centering
\includegraphics[width=0.98\textwidth]{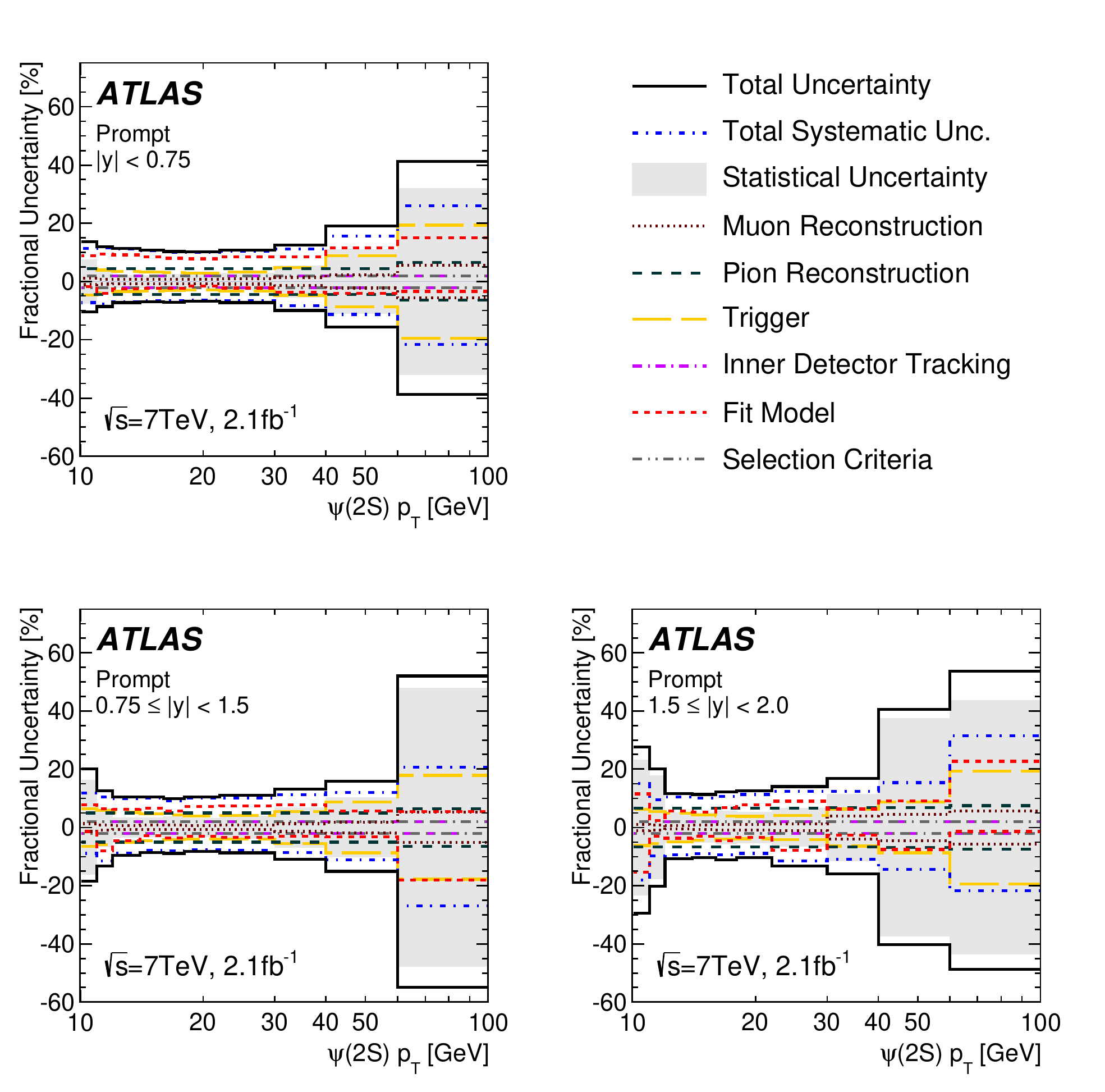}
\caption{Summary of the positive and negative uncertainties for the prompt cross-section measurement in three $\psi(2\mathrm{S})$ rapidity intervals. The plots do not include the constant 1.8$\%$ luminosity uncertainty or the spin-alignment uncertainty.}
\label{fig:Uncertainty:p}
\end{figure}

\begin{figure}[tbp]
\centering
\includegraphics[width=0.98\textwidth]{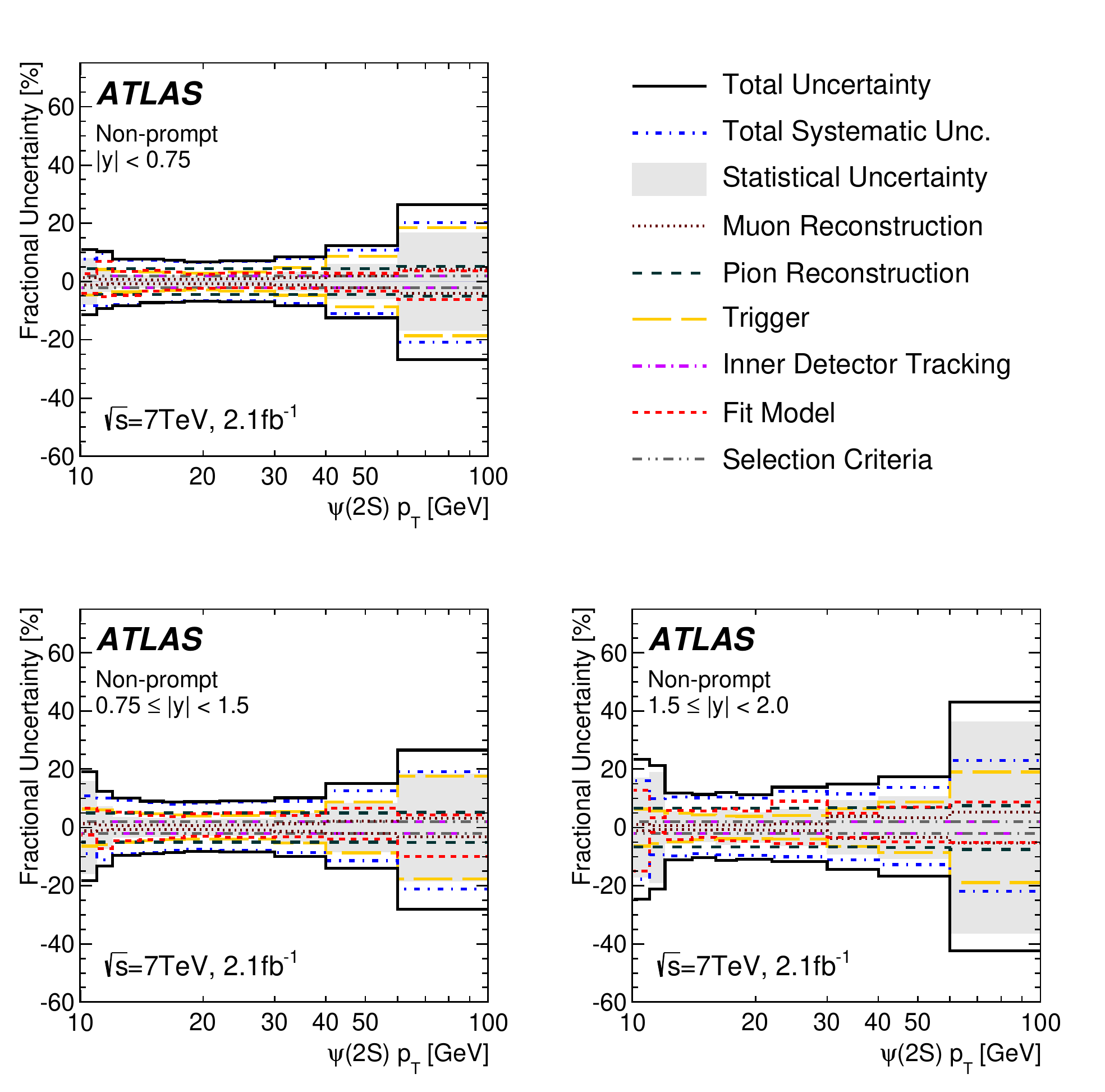}
\caption{Summary of the positive and negative uncertainties for the non-prompt cross-section measurement in three $\psi(2\mathrm{S})$ rapidity intervals. The plots do not include the constant 1.8$\%$ luminosity uncertainty or the spin-alignment uncertainty.}
\label{fig:Uncertainty:np}
\end{figure}

Figures~\ref{fig:Uncertainty:npf}--\ref{fig:Uncertainty:np} 
summarise the total systematic and statistical uncertainties in the measurement of the non-prompt production fraction and the prompt/non-prompt cross-sections.


\section{Production of {\boldmath $\psi(2\mathrm{S})$} as a function of {\boldmath $\Jpsi$ $p_{\rm T}$} and rapidity}
In order to better understand the various feed-down contributions to $\Jpsi$ production
it is important to measure the differential cross-section of the production of $\Jpsi$ mesons 
from prompt and non-prompt $\psi(2\mathrm{S})\to \Jpipi$ decays, as a function of the transverse
momentum of the $\Jpsi$.
The procedure is very similar to the measurement of $\psi(2\mathrm{S})$ production:
the invariant mass distributions of all $\psi(2\mathrm{S})\to \Jpipi$  candidates (selected and
fully corrected for acceptance and efficiency, according to eq.~(\ref{eqn:weight})), are fitted again to
extract the yield
of $\psi(2\mathrm{S})$ mesons, but this time in bins of $\Jpsi$ $p_{\rm T}$ and rapidity. Fitting and
uncertainty estimation procedures remain the same.
As the fiducial volume from which $\Jpipi$ candidates are reconstructed extends well beyond the kinematic range over which the measurements are presented, no additional corrections are needed
to present the data as a function of $J/\psi$ kinematic variables. The absence of any need for additional corrections was cross-checked using MC simulations.

\clearpage
\section{Results and discussion}

\begin{figure}[h!tbp]
  \begin{center}
    \includegraphics[width=0.99\textwidth]{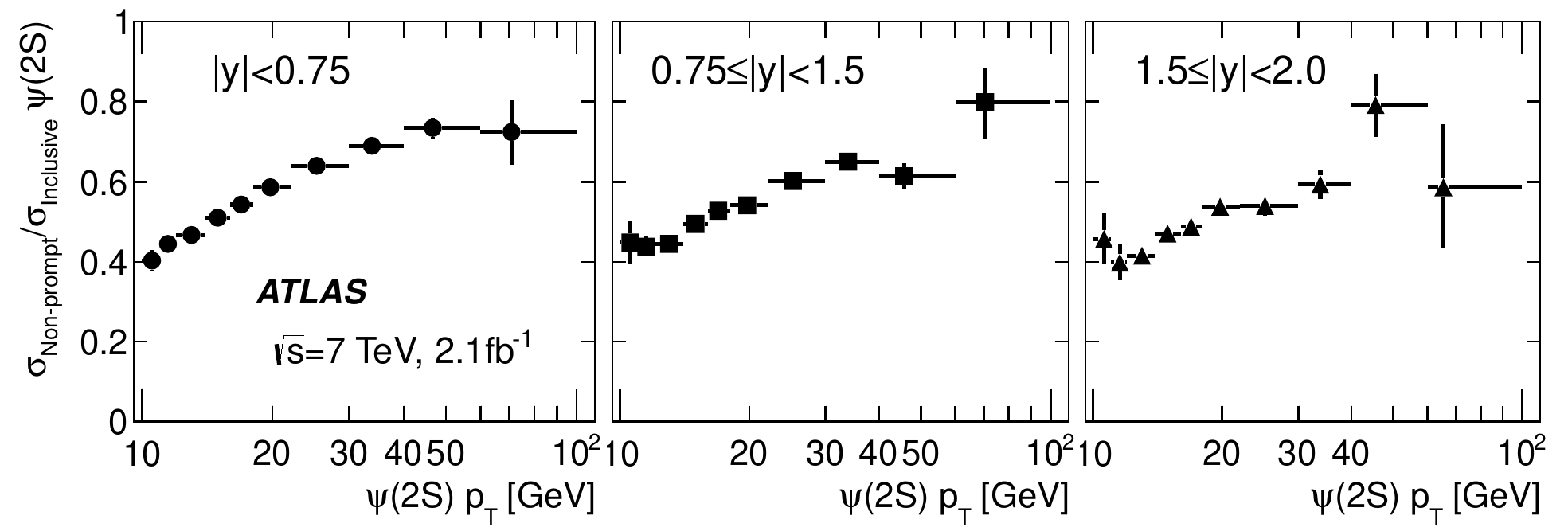}
    \caption{Non-prompt $\psi(2\mathrm{S})$ production fraction is calculated using eq.~(\ref{eqn:npfrac}), and is shown here as a function of $\psi(2\mathrm{S})$ transverse momentum in three intervals of $\psi(2\mathrm{S})$ rapidity.
The data points are at the mean of the efficiency and acceptance corrected $p_{\rm T}$ distribution in each $p_{\rm T}$ interval, indicated by the horizontal error bars, and the vertical error bars represent the total statistical and systematic uncertainty (see figure~\ref{fig:Uncertainty:npf}).
\label{fig:results_fraction}
    }
  \end{center}
\end{figure}

The corrected non-prompt $\psi(2\mathrm{S})$ production fraction, and the prompt and non-prompt $\psi(2\mathrm{S})$ production cross-sections are measured
in intervals of $\psi(2\mathrm{S})$ transverse momentum and three ranges of $\psi(2\mathrm{S})$ rapidity. All measurements are presented assuming the $\psi(2\mathrm{S})$
decays isotropically.
Figure~\ref{fig:results_fraction} shows the fully corrected measured non-prompt production fraction $f_{B}^{\psi(2\mathrm{S})}$ as a function of $p_{\rm T}$. 
A rise in the relative non-prompt production rate is observed with increasing $p_{\rm T}$ for all three rapidity intervals. 
This behaviour is similar to that seen for the non-prompt $J/\psi$ production fraction\,\cite{JPsi}.
Whereas at large $p_{\rm T}(>50~{\rm GeV})$ the non-prompt $\psi(2\mathrm{S})$ fraction approaches that of the $J/\psi$, at low $p_{\rm T}$
the non-prompt fraction for $\psi(2\mathrm{S})$ is somewhat larger than is observed for $J/\psi$.
The data shows no significant dependence on rapidity at the lowest transverse momenta probed, 
but a systematic reduction in the non-prompt fraction with increasing rapidity is observed as the $\psi(2\mathrm{S})$ transverse momentum increases.
The data are tabulated in table~\ref{tab:NPfraction}.

\begin{table}[htbp]
  \begin{center}
\footnotesize
\caption{\label{tab:NPfraction}Non-prompt $\psi(2\mathrm{S})$ production fraction as a function of $\psi(2\mathrm{S})$ $p_{\rm T}$ for three $\psi(2\mathrm{S})$ rapidity intervals.
  The first uncertainty is statistical, the second is systematic. Spin-alignment uncertainties are not included.}
\vspace{2mm}
\begin{tabular}{cc r@{\hspace{1mm}$\pm$\hspace{1mm}}l@{\hspace{0.5mm}$\pm$\hspace{0.5mm}}l}
\hline\hline
\multicolumn{5}{c}{$0\leq |y|<0.75$} \\
$p_{\rm T}$ interval [GeV]  & $\langle p_{\rm T}\rangle $ [GeV]  & \multicolumn{3}{c}{Non-prompt fraction} \\[-0.5ex]
$10.0$--$11.0$ & $ 10.6$ & $ 0.404$ & $0.024$ & $0.006$ \\
$11.0$--$12.0$ & $ 11.5$ & $ 0.445$ & $0.012$ & $0.006$ \\
$12.0$--$14.0$ & $ 13.0$ & $ 0.466$ & $0.007$ & $0.006$ \\
$14.0$--$16.0$ & $ 15.0$ & $ 0.509$ & $0.007$ & $0.006$ \\
$16.0$--$18.0$ & $ 17.0$ & $ 0.543$ & $0.008$ & $0.006$ \\
$18.0$--$22.0$ & $ 19.8$ & $ 0.586$ & $0.007$ & $0.007$ \\
$22.0$--$30.0$ & $ 25.2$ & $ 0.639$ & $0.008$ & $0.007$ \\
$30.0$--$40.0$ & $ 33.8$ & $ 0.690$ & $0.014$ & $0.008$ \\
$40.0$--$60.0$ & $ 46.6$ & $ 0.735$ & $0.024$ & $0.009$ \\
$60.0$--$100.0$ & $ 70.8$ & $ 0.724$ & $0.070$ & $0.042$ \\
\hline\hline
\multicolumn{5}{c}{$0.75\leq |y|<1.5$} \\
$p_{\rm T}$ interval [GeV]  & $\langle p_{\rm T}\rangle $ [GeV]  & \multicolumn{3}{c}{Non-prompt fraction} \\[-0.5ex]
$10.0$--$11.0$ & $ 10.6$ & $ 0.448$ & $0.053$ & $0.009$ \\
$11.0$--$12.0$ & $ 11.5$ & $ 0.438$ & $0.024$ & $0.006$ \\
$12.0$--$14.0$ & $ 13.0$ & $ 0.445$ & $0.012$ & $0.006$ \\
$14.0$--$16.0$ & $ 15.0$ & $ 0.495$ & $0.011$ & $0.006$ \\
$16.0$--$18.0$ & $ 16.9$ & $ 0.527$ & $0.012$ & $0.006$ \\
$18.0$--$22.0$ & $ 19.8$ & $ 0.542$ & $0.010$ & $0.006$ \\
$22.0$--$30.0$ & $ 25.2$ & $ 0.602$ & $0.012$ & $0.007$ \\
$30.0$--$40.0$ & $ 33.8$ & $ 0.649$ & $0.018$ & $0.007$ \\
$40.0$--$60.0$ & $ 45.6$ & $ 0.614$ & $0.031$ & $0.008$ \\
$60.0$--$100.0$ & $ 70.4$ & $ 0.798$ & $0.081$ & $0.034$ \\
\hline\hline
\multicolumn{5}{c}{$1.5\leq |y|<2$} \\
$p_{\rm T}$ interval [GeV]  & $\langle p_{\rm T}\rangle $ [GeV]  & \multicolumn{3}{c}{Non-prompt fraction} \\[-0.5ex]
$10.0$--$11.0$ & $ 10.6$ & $ 0.457$ & $0.064$ & $0.011$ \\
$11.0$--$12.0$ & $ 11.5$ & $ 0.398$ & $0.045$ & $0.007$ \\
$12.0$--$14.0$ & $ 13.0$ & $ 0.414$ & $0.018$ & $0.006$ \\
$14.0$--$16.0$ & $ 14.9$ & $ 0.471$ & $0.017$ & $0.006$ \\
$16.0$--$18.0$ & $ 16.9$ & $ 0.488$ & $0.018$ & $0.006$ \\
$18.0$--$22.0$ & $ 19.8$ & $ 0.537$ & $0.016$ & $0.007$ \\
$22.0$--$30.0$ & $ 25.1$ & $ 0.539$ & $0.020$ & $0.013$ \\
$30.0$--$40.0$ & $ 33.9$ & $ 0.593$ & $0.033$ & $0.014$ \\
$40.0$--$60.0$ & $ 45.5$ & $ 0.791$ & $0.059$ & $0.051$ \\
$60.0$--$100.0$ & $ 65.3$ & $ 0.587$ & $0.143$ & $0.069$ \\
\hline\hline
\end{tabular}
  \end{center}
\end{table}

\begin{figure}[htbp]
  \begin{center}
    \subfigure[Prompt production vs.\ $\psi(2\mathrm{S})$ $p_{\rm T}$\label{fig:promptData}]{\includegraphics[width=0.49\textwidth]{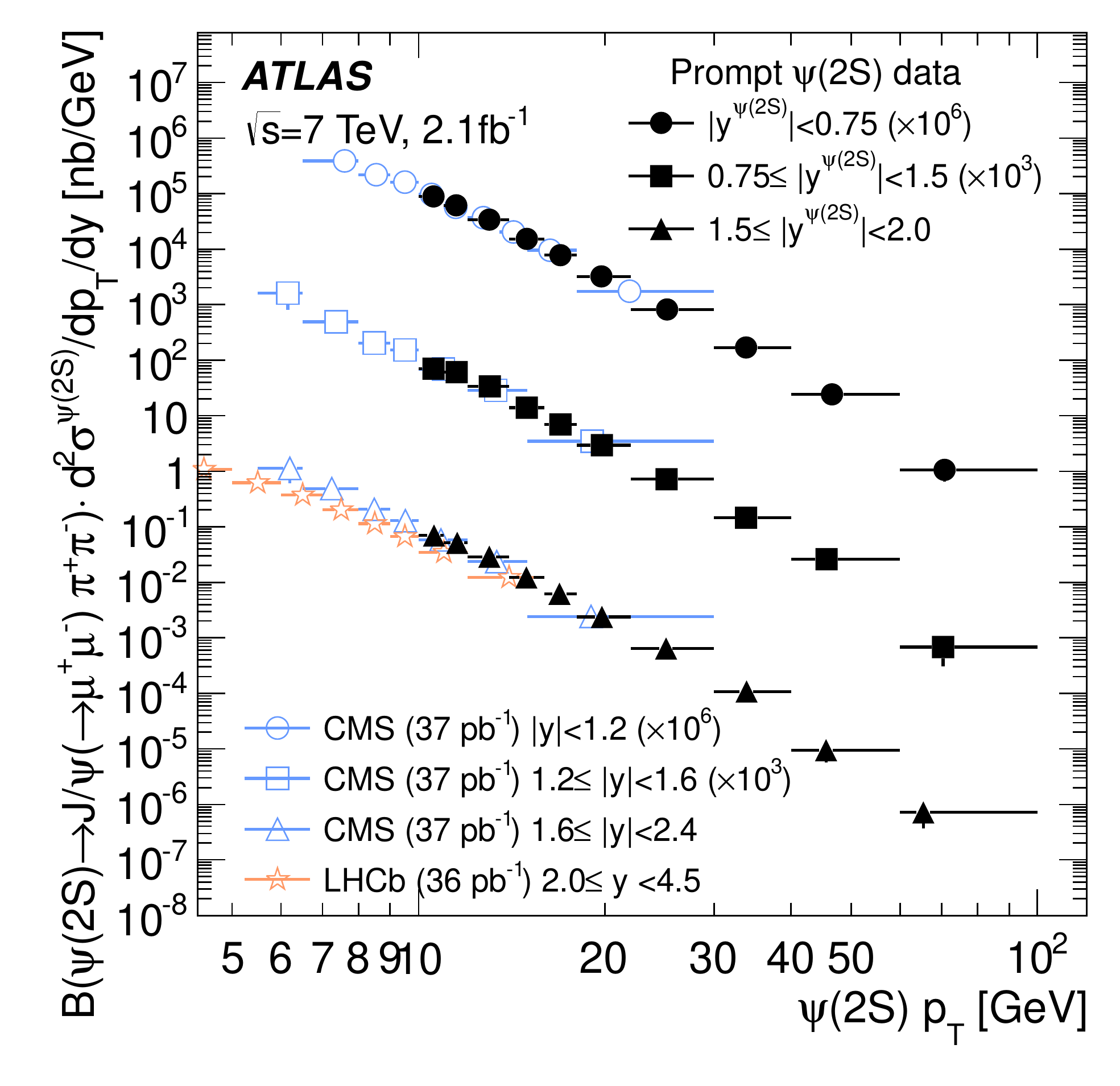}}
    \subfigure[Non-prompt production vs.\ $\psi(2\mathrm{S})$ $p_{\rm T}$\label{fig:nonpromptData}]{\includegraphics[width=0.49\textwidth]{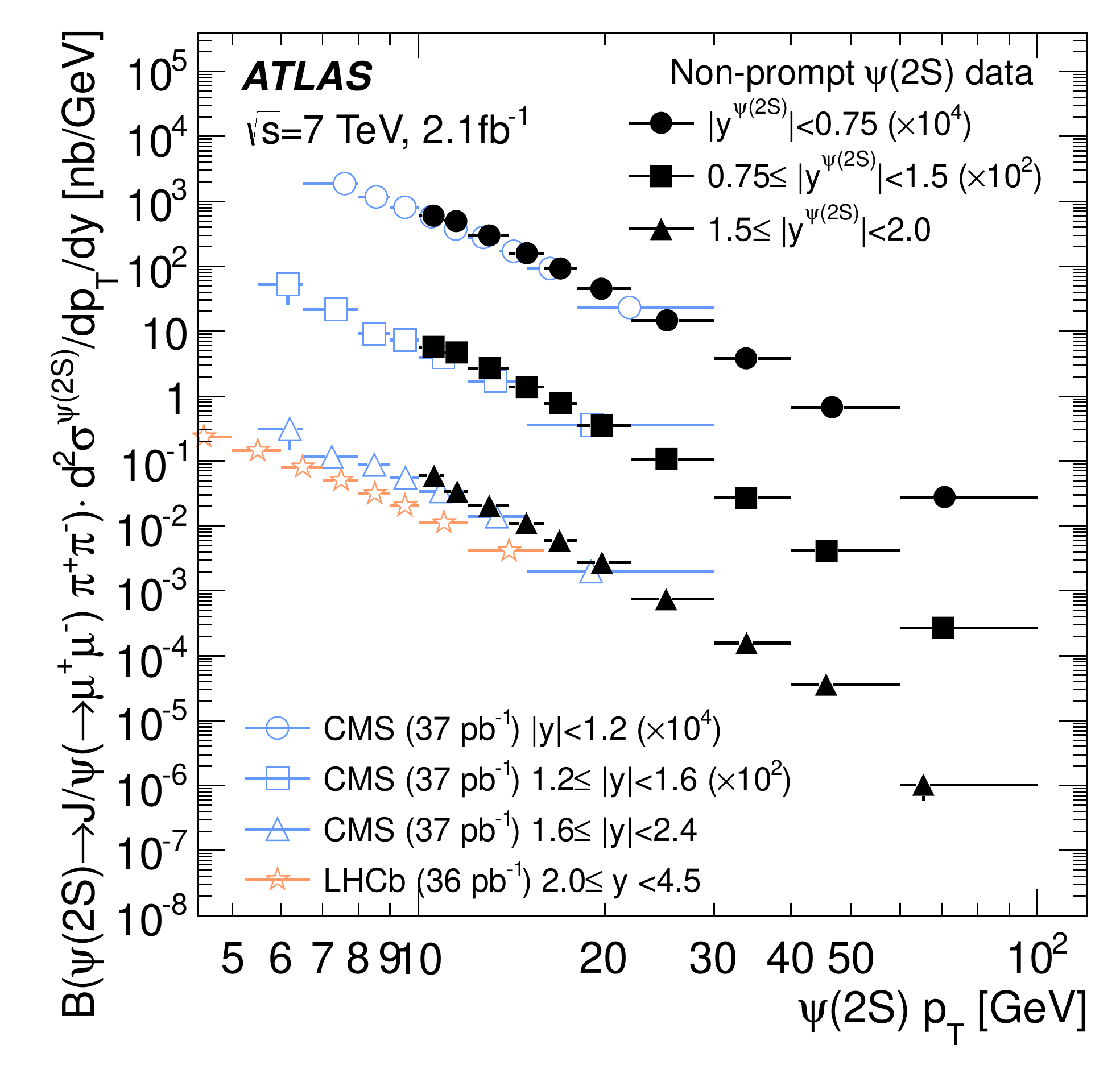}}
    \subfigure[Prompt production vs.\ $J/\psi$ $p_{\rm T}$\label{fig:promptDataJpsi}]{\includegraphics[width=0.49\textwidth]{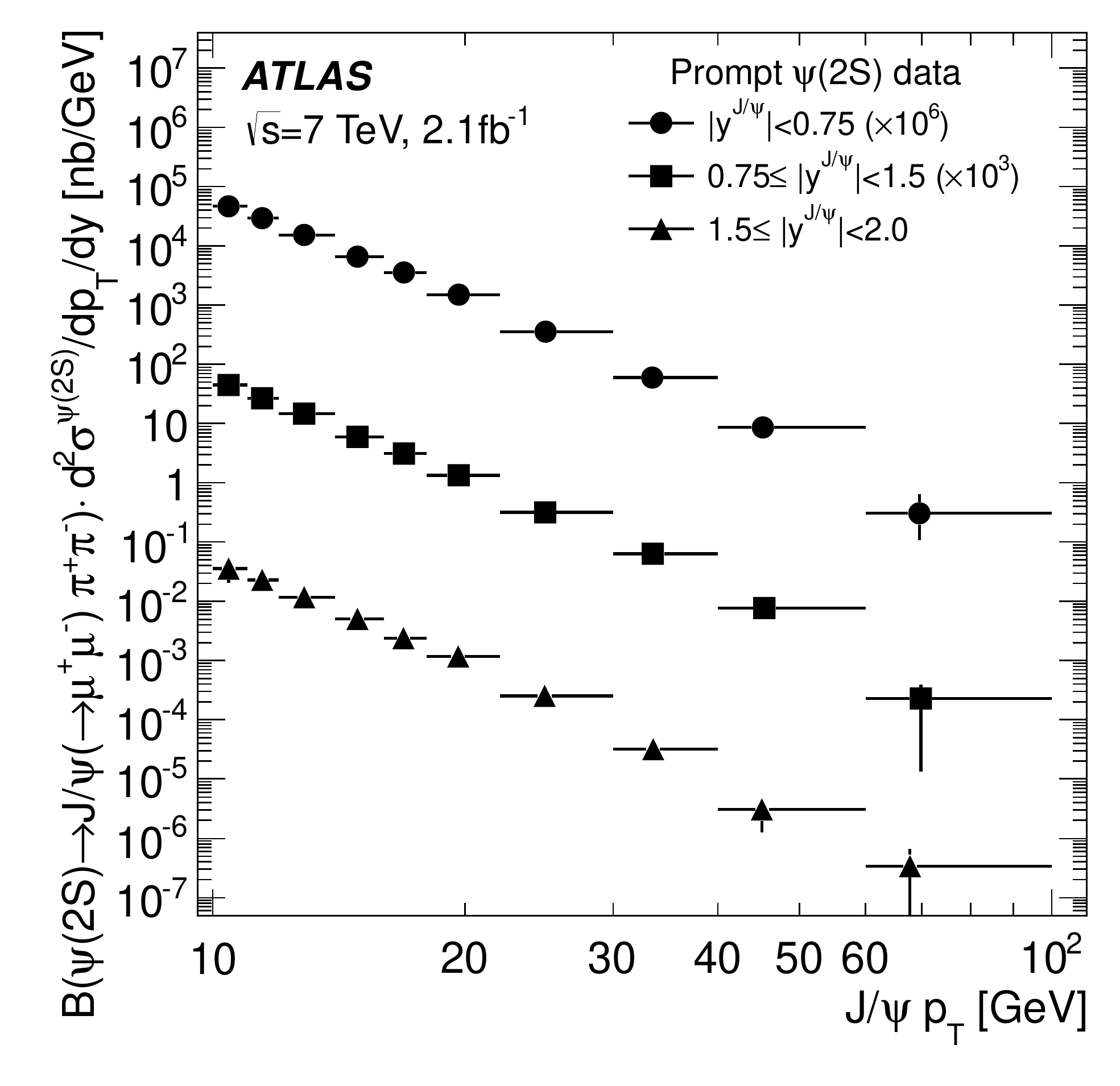}}
    \subfigure[Non-prompt production vs.\ $J/\psi$ $p_{\rm T}$\label{fig:nonpromptDataJpsi}]{\includegraphics[width=0.49\textwidth]{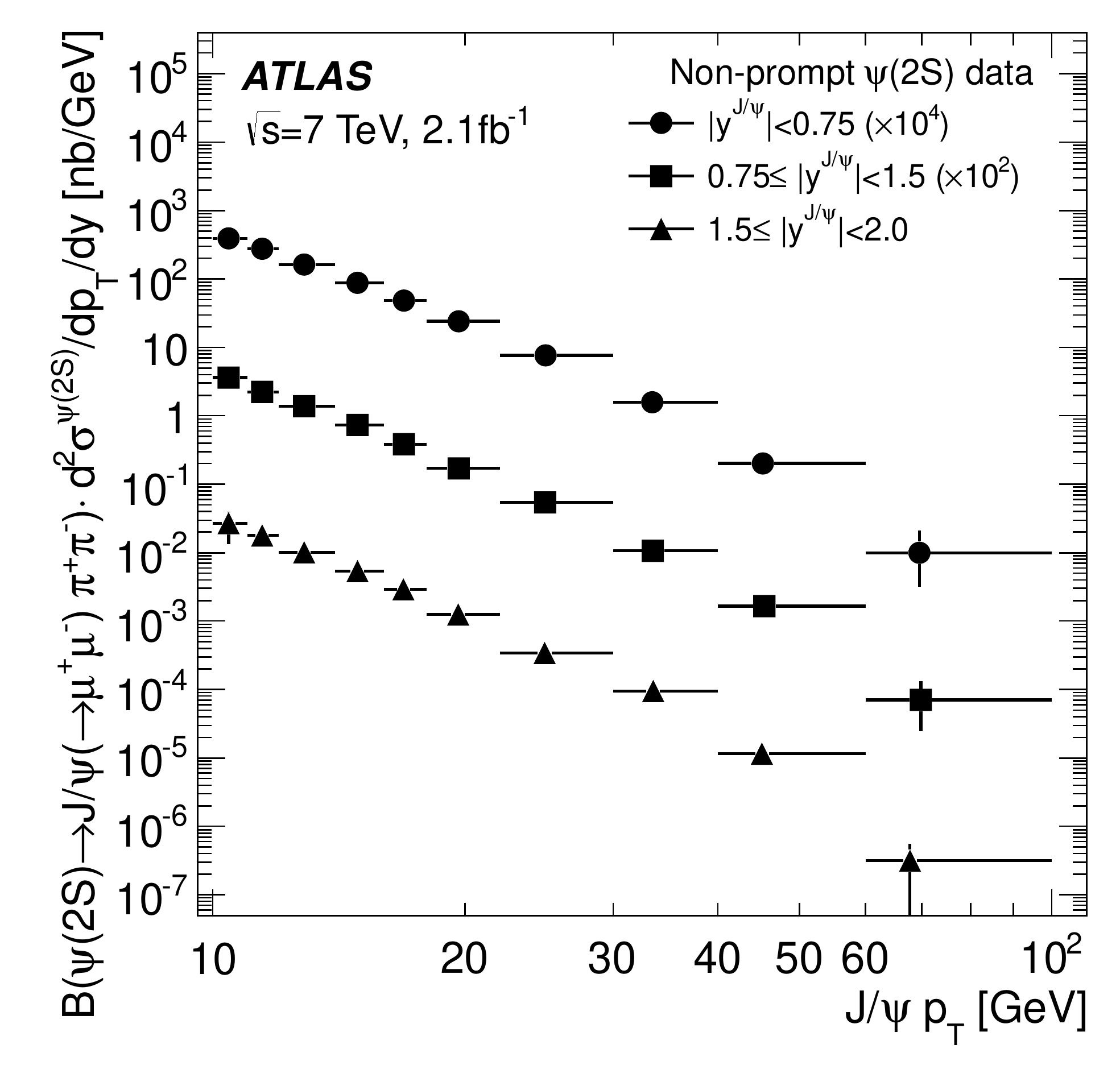}}
    \caption{Measured differential cross-sections for (a) prompt $\psi(2\mathrm{S})$ production and (b) non-prompt $\psi(2\mathrm{S})$ production
      as a function of $\psi(2\mathrm{S})$ transverse momentum for three $\psi(2\mathrm{S})$ rapidity intervals. 
      Also shown are (c) prompt and (d) non-prompt cross-sections expressed as a function of the transverse momentum of the $J/\psi$ from the $\psi(2\mathrm{S})\to\Jmumupipi$ decay for three $J/\psi$ rapidity intervals.
      The results in the various rapidity intervals are scaled by powers of ten for clarity of presentation. 
      The data points are at the mean of the efficiency and acceptance corrected $p_{\rm T}$ distribution in each $p_{\rm T}$ interval, indicated by the horizontal error bars, 
      and the vertical error bars represent the total statistical and systematic uncertainty (see figures~\ref{fig:Uncertainty:p} and~\ref{fig:Uncertainty:np}).
      Overlaid on the results presented as a function of $\psi(2\mathrm{S})$ $p_{\rm T}$ are measurements from the CMS and LHCb experiments.
      \label{fig:results_data}
    }
  \end{center}
\end{figure}

Fully corrected measurements of the differential prompt and non-prompt cross-sections as functions of $\psi(2\mathrm{S})$ $p_{\rm T}$ and rapidity
are presented in figures~\ref{fig:promptData} and~\ref{fig:nonpromptData} and are tabulated in table~\ref{tab:XS_psi2s_rap}.
These results are compared to results from CMS\,\cite{CMS} and LHCb\,\cite{LHCb} in similar or neighbouring rapidity intervals
(the LHCb and CMS data are also presented assuming isotropic $\psi(2\mathrm{S})$ production). 
The measured differential cross-sections of prompt and non-prompt production
of $\Jpsi$ mesons from $\psi(2\mathrm{S})\to \Jpipi$ decays are presented as functions of $J/\psi$ transverse momentum and rapidity
in figures~\ref{fig:promptDataJpsi} and~\ref{fig:nonpromptDataJpsi} and in table~\ref{tab:XS_jpsi_rap}.

The effects of the various polarisation scenarios described in section~\ref{sec:accept}
on the measured $\Jpsi$ cross-sections were also studied.
The corresponding correction factors for all $\Jpsi$ and $\psi(2\mathrm{S})$ $p_{\rm T}$---$|y|$ bins are tabulated in appendix~\ref{sec:acccorr}.

\begin{table}[htbp]
  \begin{center}
\footnotesize
\caption{\label{tab:XS_psi2s_rap}Prompt and non-prompt production cross-section times branching ratio as a function of $\psi(2\mathrm{S})$ $p_{\rm T}$ for three $\psi(2\mathrm{S})$ rapidity intervals.
The first uncertainty is statistical, the second is systematic. Spin-alignment and luminosity ($\pm 1.8\%$) uncertainties are not included.}
\vspace{2mm}
\begin{tabular}{cc r@{\hspace{1mm}$\pm$\hspace{1mm}}l@{\hspace{0.5mm}${}$\hspace{0.5mm}}l r@{\hspace{1mm}$\pm$\hspace{1mm}}l@{\hspace{0.5mm}${}$\hspace{0.5mm}}l}
\hline\hline
\multicolumn{8}{c}{$\mathcal{B}(\psi(2S)\rightarrow J/\psi(\rightarrow\mu^{+}\mu^{-}) \pi^{+}\pi^{-})\cdot \mathrm{d}^{2}\sigma^{\psi(2S)}/\mathrm{d}p_{\rm T}\,\mathrm{d}y $ } \\ 
\multicolumn{8}{c}{$0\leq |y|<0.75$} \\
$p_{\rm T}$ interval [GeV]  & $\langle p_{\rm T}\rangle$ [GeV]  & \multicolumn{3}{c}{Prompt [pb/GeV]} &  \multicolumn{3}{c}{Non-prompt [pb/GeV]} \\[-0.5ex]
$10.0$--$11.0$ & $ 10.6$ & $ 89$ & $7$ & ${}^{+10}_{-7}$ & $60.4$ & $4.8$ & ${}^{+4.6}_{-5.0}$ \\
$11.0$--$12.0$ & $ 11.5$ & $ 61.6$ & $2.1$ & ${}^{+7.1}_{-4.8}$ & $49.4$ & $1.7$ & ${}^{+4.8}_{-4.2}$ \\
$12.0$--$14.0$ & $ 13.0$ & $ 34.1$ & $0.7$ & ${}^{+3.8}_{-2.4}$ & $29.8$ & $0.6$ & ${}^{+2.2}_{-2.4}$ \\
$14.0$--$16.0$ & $ 15.0$ & $ 15.4$ & $0.3$ & ${}^{+1.6}_{-1.0}$ & $16.0$ & $0.3$ & ${}^{+1.2}_{-1.1}$ \\
$16.0$--$18.0$ & $ 17.0$ & $ 7.84$ & $0.19$ & ${}^{+0.79}_{-0.52}$ & $9.30$ & $0.20$ & ${}^{+0.65}_{-0.64}$ \\
$18.0$--$22.0$ & $ 19.8$ & $ 3.21$ & $0.07$ & ${}^{+0.32}_{-0.20}$ & $4.54$ & $0.08$ & ${}^{+0.30}_{-0.30}$ \\
$22.0$--$30.0$ & $ 25.2$ & $ 0.822$ & $0.024$ & ${}^{+0.086}_{-0.055}$ & $1.46$ & $0.03$ & ${}^{+0.10}_{-0.10}$ \\
$30.0$--$40.0$ & $ 33.8$ & $ 0.171$ & $0.009$ & ${}^{+0.019}_{-0.014}$ & $0.381$ & $0.012$ & ${}^{+0.030}_{-0.029}$ \\
$40.0$--$60.0$ & $ 46.6$ & $ 0.0241$ & $0.0026$ & ${}^{+0.0038}_{-0.0027}$ & $0.0670$ & $0.0040$ & ${}^{+0.0072}_{-0.0073}$ \\
$60.0$--$100.0$ & $ 70.8$ & $ 0.00106$ & $0.00034$ & ${}^{+0.00028}_{-0.00023}$ & $0.00279$ & $0.00047$ & ${}^{+0.00056}_{-0.00058}$ \\
\hline\hline
\multicolumn{8}{c}{$\mathcal{B}(\psi(2S)\rightarrow J/\psi(\rightarrow\mu^{+}\mu^{-}) \pi^{+}\pi^{-})\cdot \mathrm{d}^{2}\sigma^{\psi(2S)}/\mathrm{d}p_{\rm T}\,\mathrm{d}y $ }\\ 
\multicolumn{8}{c}{$0.75\leq |y|<1.5$} \\
$p_{\rm T}$ interval [GeV]  & $\langle p_{\rm T}\rangle$ [GeV]  & \multicolumn{3}{c}{Prompt [pb/GeV]} &  \multicolumn{3}{c}{Non-prompt [pb/GeV]} \\[-0.5ex]
$10.0$--$11.0$ & $ 10.6$ & $ 70$ & $11$ & ${}^{+8}_{-6}$ & $57$ & $9$ & ${}^{+6}_{-5}$ \\
$11.0$--$12.0$ & $ 11.5$ & $ 60.7$ & $4.1$ & ${}^{+6.4}_{-7.0}$ & $47.3$ & $3.4$ & ${}^{+4.8}_{-5.2}$ \\
$12.0$--$14.0$ & $ 13.0$ & $ 33.6$ & $1.1$ & ${}^{+3.3}_{-3.0}$ & $26.9$ & $0.9$ & ${}^{+2.5}_{-2.4}$ \\
$14.0$--$16.0$ & $ 15.0$ & $ 14.0$ & $0.5$ & ${}^{+1.4}_{-1.1}$ & $13.7$ & $0.4$ & ${}^{+1.2}_{-1.2}$ \\
$16.0$--$18.0$ & $ 16.9$ & $ 6.92$ & $0.25$ & ${}^{+0.64}_{-0.56}$ & $7.72$ & $0.25$ & ${}^{+0.62}_{-0.62}$ \\
$18.0$--$22.0$ & $ 19.8$ & $ 2.97$ & $0.09$ & ${}^{+0.30}_{-0.23}$ & $3.51$ & $0.10$ & ${}^{+0.30}_{-0.27}$ \\
$22.0$--$30.0$ & $ 25.2$ & $ 0.712$ & $0.028$ & ${}^{+0.073}_{-0.056}$ & $1.075$ & $0.031$ & ${}^{+0.094}_{-0.084}$ \\
$30.0$--$40.0$ & $ 33.8$ & $ 0.145$ & $0.010$ & ${}^{+0.016}_{-0.012}$ & $0.269$ & $0.013$ & ${}^{+0.024}_{-0.023}$ \\
$40.0$--$60.0$ & $ 45.6$ & $ 0.0259$ & $0.0027$ & ${}^{+0.0031}_{-0.0029}$ & $0.0412$ & $0.0035$ & ${}^{+0.0052}_{-0.0047}$ \\
$60.0$--$100.0$ & $ 70.4$ & $ 0.00068$ & $0.00032$ & ${}^{+0.00014}_{-0.00018}$ & $0.00269$ & $0.00050$ & ${}^{+0.00052}_{-0.00057}$ \\

\hline\hline
\multicolumn{8}{c}{$\mathcal{B}(\psi(2S)\rightarrow J/\psi(\rightarrow\mu^{+}\mu^{-}) \pi^{+}\pi^{-})\cdot \mathrm{d}^{2}\sigma^{\psi(2S)}/\mathrm{d}p_{\rm T}\,\mathrm{d}y $ }\\ 
\multicolumn{8}{c}{$1.5\leq |y|<2$} \\
$p_{\rm T}$ interval [GeV]  & $\langle p_{\rm T}\rangle$ [GeV]  & \multicolumn{3}{c}{Prompt [pb/GeV]} &  \multicolumn{3}{c}{Non-prompt [pb/GeV]} \\[-0.5ex]
$10.0$--$11.0$ & $ 10.6$ & $ 70$ & $16$ & ${}^{+11}_{-13}$ & $59$ & $10$ & ${}^{+9}_{-11}$ \\
$11.0$--$12.0$ & $ 11.5$ & $ 51.2$ & $9.1$ & ${}^{+4.9}_{-5.0}$ & $33.9$ & $6.4$ & ${}^{+3.3}_{-3.2}$ \\
$12.0$--$14.0$ & $ 13.0$ & $ 29.0$ & $1.5$ & ${}^{+3.0}_{-2.7}$ & $20.5$ & $1.1$ & ${}^{+2.2}_{-2.0}$ \\
$14.0$--$16.0$ & $ 14.9$ & $ 12.3$ & $0.6$ & ${}^{+1.2}_{-1.1}$ & $11.0$ & $0.5$ & ${}^{+1.1}_{-1.0}$ \\
$16.0$--$18.0$ & $ 16.9$ & $ 6.23$ & $0.36$ & ${}^{+0.67}_{-0.59}$ & $5.94$ & $0.35$ & ${}^{+0.63}_{-0.56}$ \\
$18.0$--$22.0$ & $ 19.8$ & $ 2.35$ & $0.13$ & ${}^{+0.27}_{-0.21}$ & $2.73$ & $0.14$ & ${}^{+0.27}_{-0.26}$ \\
$22.0$--$30.0$ & $ 25.1$ & $ 0.636$ & $0.042$ & ${}^{+0.078}_{-0.074}$ & $0.74$ & $0.05$ & ${}^{+0.09}_{-0.07}$ \\
$30.0$--$40.0$ & $ 33.9$ & $ 0.108$ & $0.012$ & ${}^{+0.013}_{-0.012}$ & $0.157$ & $0.015$ & ${}^{+0.018}_{-0.017}$ \\
$40.0$--$60.0$ & $ 45.5$ & $ 0.0095$ & $0.0035$ & ${}^{+0.0014}_{-0.0014}$ & $0.0358$ & $0.0038$ & ${}^{+0.0049}_{-0.0046}$ \\
$60.0$--$100.0$ & $ 65.3$ & $ 0.00072$ & $0.00031$ & ${}^{+0.00023}_{-0.00016}$ & $0.00103$ & $0.00037$ & ${}^{+0.00024}_{-0.00023}$ \\
\hline\hline
\end{tabular}
  \end{center}
\end{table}

\begin{table}[htbp]
  \begin{center}
\footnotesize
\caption{\label{tab:XS_jpsi_rap}Prompt and non-prompt production cross-section times branching ratio as a function of $J/\psi$ $p_{\rm T}$ for three $J/\psi$ rapidity intervals.
  The first uncertainty is statistical, the second is systematic. Spin-alignment and luminosity ($\pm 1.8\%$) uncertainties are not included.}
\vspace{2mm}
\begin{tabular}{cc r@{\hspace{1mm}$\pm$\hspace{1mm}}l@{\hspace{0.5mm}${}$\hspace{0.5mm}}l r@{\hspace{1mm}$\pm$\hspace{1mm}}l@{\hspace{0.5mm}${}$\hspace{0.5mm}}l}
\hline\hline
\multicolumn{8}{c}{$\mathcal{B}(\psi(2S)\rightarrow J/\psi(\rightarrow\mu^{+}\mu^{-}) \pi^{+}\pi^{-})\cdot \mathrm{d}^{2}\sigma^{\psi(2S)}/\mathrm{d}p_{\rm T}\,\mathrm{d}y $ }\\ 
\multicolumn{8}{c}{$0\leq |y|<0.75$} \\
$p_{\rm T}$ interval [GeV]  & $\langle p_{\rm T}\rangle$ [GeV]  & \multicolumn{3}{c}{Prompt [pb/GeV]} &  \multicolumn{3}{c}{Non-prompt [pb/GeV]} \\[-0.5ex]
$10.0$--$11.0$ & $ 10.6$ & $ 46.5$ & $1.7$ & ${}^{+6.1}_{-3.5}$ & $39.4$ & $1.5$ & ${}^{+3.2}_{-3.6}$ \\
$11.0$--$12.0$ & $ 11.5$ & $ 29.3$ & $0.9$ & ${}^{+3.4}_{-2.1}$ & $27.4$ & $0.8$ & ${}^{+2.2}_{-2.2}$ \\
$12.0$--$14.0$ & $ 13.0$ & $ 15.4$ & $0.3$ & ${}^{+1.6}_{-1.0}$ & $16.3$ & $0.3$ & ${}^{+1.2}_{-1.2}$ \\
$14.0$--$16.0$ & $ 15.0$ & $ 6.59$ & $0.18$ & ${}^{+0.65}_{-0.43}$ & $8.79$ & $0.19$ & ${}^{+0.61}_{-0.58}$ \\
$16.0$--$18.0$ & $ 17.0$ & $ 3.54$ & $0.11$ & ${}^{+0.36}_{-0.23}$ & $4.80$ & $0.13$ & ${}^{+0.32}_{-0.32}$ \\
$18.0$--$22.0$ & $ 19.8$ & $ 1.50$ & $0.05$ & ${}^{+0.15}_{-0.10}$ & $2.40$ & $0.06$ & ${}^{+0.16}_{-0.16}$ \\
$22.0$--$30.0$ & $ 25.2$ & $ 0.353$ & $0.015$ & ${}^{+0.039}_{-0.025}$ & $0.762$ & $0.019$ & ${}^{+0.054}_{-0.053}$ \\
$30.0$--$40.0$ & $ 33.8$ & $ 0.0602$ & $0.0054$ & ${}^{+0.0095}_{-0.0052}$ & $0.157$ & $0.008$ & ${}^{+0.013}_{-0.013}$ \\
$40.0$--$60.0$ & $ 46.6$ & $ 0.0086$ & $0.0015$ & ${}^{+0.0018}_{-0.0014}$ & $0.0203$ & $0.0020$ & ${}^{+0.0028}_{-0.0030}$ \\
$60.0$--$100.0$ & $ 70.8$ & $ 0.00030$ & $0.00017$ & ${}^{+0.00030}_{-0.00010}$ & $0.0010$ & $0.0006$ & ${}^{+0.0009}_{-0.0003}$ \\
\hline\hline
\multicolumn{8}{c}{$\mathcal{B}(\psi(2S)\rightarrow J/\psi(\rightarrow\mu^{+}\mu^{-}) \pi^{+}\pi^{-})\cdot \mathrm{d}^{2}\sigma^{\psi(2S)}/\mathrm{d}p_{\rm T}\,\mathrm{d}y $ } \\ 
\multicolumn{8}{c}{$0.75\leq |y|<1.5$} \\
$p_{\rm T}$ interval [GeV]  & $\langle p_{\rm T}\rangle$ [GeV]  & \multicolumn{3}{c}{Prompt [pb/GeV]} &  \multicolumn{3}{c}{Non-prompt [pb/GeV]} \\[-0.5ex]
$10.0$--$11.0$ & $ 10.6$ & $ 44.5$ & $2.2$ & ${}^{+4.9}_{-4.2}$ & $36.3$ & $1.9$ & ${}^{+3.8}_{-3.5}$ \\
$11.0$--$12.0$ & $ 11.5$ & $ 27.0$ & $1.2$ & ${}^{+2.7}_{-2.5}$ & $22.4$ & $1.2$ & ${}^{+2.3}_{-2.1}$ \\
$12.0$--$14.0$ & $ 13.0$ & $ 14.8$ & $0.5$ & ${}^{+1.5}_{-1.3}$ & $13.8$ & $0.5$ & ${}^{+1.2}_{-1.2}$ \\
$14.0$--$16.0$ & $ 15.0$ & $ 5.95$ & $0.23$ & ${}^{+0.62}_{-0.48}$ & $7.32$ & $0.25$ & ${}^{+0.63}_{-0.60}$ \\
$16.0$--$18.0$ & $ 16.9$ & $ 3.14$ & $0.16$ & ${}^{+0.30}_{-0.24}$ & $3.84$ & $0.17$ & ${}^{+0.33}_{-0.29}$ \\
$18.0$--$22.0$ & $ 19.8$ & $ 1.34$ & $0.06$ & ${}^{+0.12}_{-0.11}$ & $1.71$ & $0.06$ & ${}^{+0.15}_{-0.14}$ \\
$22.0$--$30.0$ & $ 25.2$ & $ 0.316$ & $0.017$ & ${}^{+0.033}_{-0.026}$ & $0.544$ & $0.022$ & ${}^{+0.049}_{-0.046}$ \\
$30.0$--$40.0$ & $ 33.8$ & $ 0.0636$ & $0.0060$ & ${}^{+0.0073}_{-0.0061}$ & $0.107$ & $0.008$ & ${}^{+0.010}_{-0.010}$ \\
$40.0$--$60.0$ & $ 45.6$ & $ 0.0078$ & $0.0015$ & ${}^{+0.0011}_{-0.0010}$ & $0.0165$ & $0.0020$ & ${}^{+0.0024}_{-0.0023}$ \\
$60.0$--$100.0$ & $ 70.4$ & $ 0.00023$ & $0.00014$ & ${}^{+0.00010}_{-0.00017}$ & $0.00070$ & $0.00042$ & ${}^{+0.00046}_{-0.00017}$ \\

\hline\hline
\multicolumn{8}{c}{$\mathcal{B}(\psi(2S)\rightarrow J/\psi(\rightarrow\mu^{+}\mu^{-}) \pi^{+}\pi^{-})\cdot \mathrm{d}^{2}\sigma^{\psi(2S)}/\mathrm{d}p_{\rm T}\,\mathrm{d}y $ } \\ 
\multicolumn{8}{c}{$1.5\leq |y|<2$} \\
$p_{\rm T}$ interval [GeV]  & $\langle p_{\rm T}\rangle$ [GeV]  & \multicolumn{3}{c}{Prompt [pb/GeV]} &  \multicolumn{3}{c}{Non-prompt [pb/GeV]} \\[-0.5ex]
$10.0$--$11.0$ & $ 10.6$ & $ 36$ & $14$ & ${}^{+5}_{-6}$ & $27$ & $12$ & ${}^{+4}_{-5}$ \\
$11.0$--$12.0$ & $ 11.5$ & $ 22.9$ & $2.2$ & ${}^{+3.4}_{-4.2}$ & $17.7$ & $1.8$ & ${}^{+2.5}_{-3.0}$ \\
$12.0$--$14.0$ & $ 13.0$ & $ 11.7$ & $0.7$ & ${}^{+1.6}_{-2.1}$ & $10.2$ & $0.6$ & ${}^{+1.3}_{-1.8}$ \\
$14.0$--$16.0$ & $ 14.9$ & $ 5.01$ & $0.28$ & ${}^{+0.61}_{-0.77}$ & $5.37$ & $0.28$ & ${}^{+0.67}_{-0.80}$ \\
$16.0$--$18.0$ & $ 16.9$ & $ 2.38$ & $0.19$ & ${}^{+0.26}_{-0.37}$ & $2.93$ & $0.21$ & ${}^{+0.32}_{-0.45}$ \\
$18.0$--$22.0$ & $ 19.8$ & $ 1.18$ & $0.09$ & ${}^{+0.15}_{-0.17}$ & $1.25$ & $0.09$ & ${}^{+0.15}_{-0.17}$ \\
$22.0$--$30.0$ & $ 25.1$ & $ 0.251$ & $0.024$ & ${}^{+0.027}_{-0.024}$ & $0.343$ & $0.029$ & ${}^{+0.036}_{-0.033}$ \\
$30.0$--$40.0$ & $ 33.9$ & $ 0.0318$ & $0.0004$ & ${}^{+0.0040}_{-0.0036}$ & $0.095$ & $0.001$ & ${}^{+0.012}_{-0.011}$ \\
$40.0$--$60.0$ & $ 45.5$ & $ 0.0031$ & $0.0006$ & ${}^{+0.0011}_{-0.0018}$ & $0.0115$ & $0.0008$ & ${}^{+0.0016}_{-0.0026}$ \\
$60.0$--$100.0$ & $ 65.3$ & $ 0.00034$ & $0.00032$ & ${}^{+0.00009}_{-0.00015}$ & $0.00031$ & $0.00023$ & ${}^{+0.00008}_{-0.00015}$ \\
\hline\hline
\end{tabular}
  \end{center}
\end{table}

\FloatBarrier

\subsection*{Prompt cross-section measurement versus theory}

\begin{figure}[htbp]
  \begin{center}
      \includegraphics[width=0.85\textwidth]{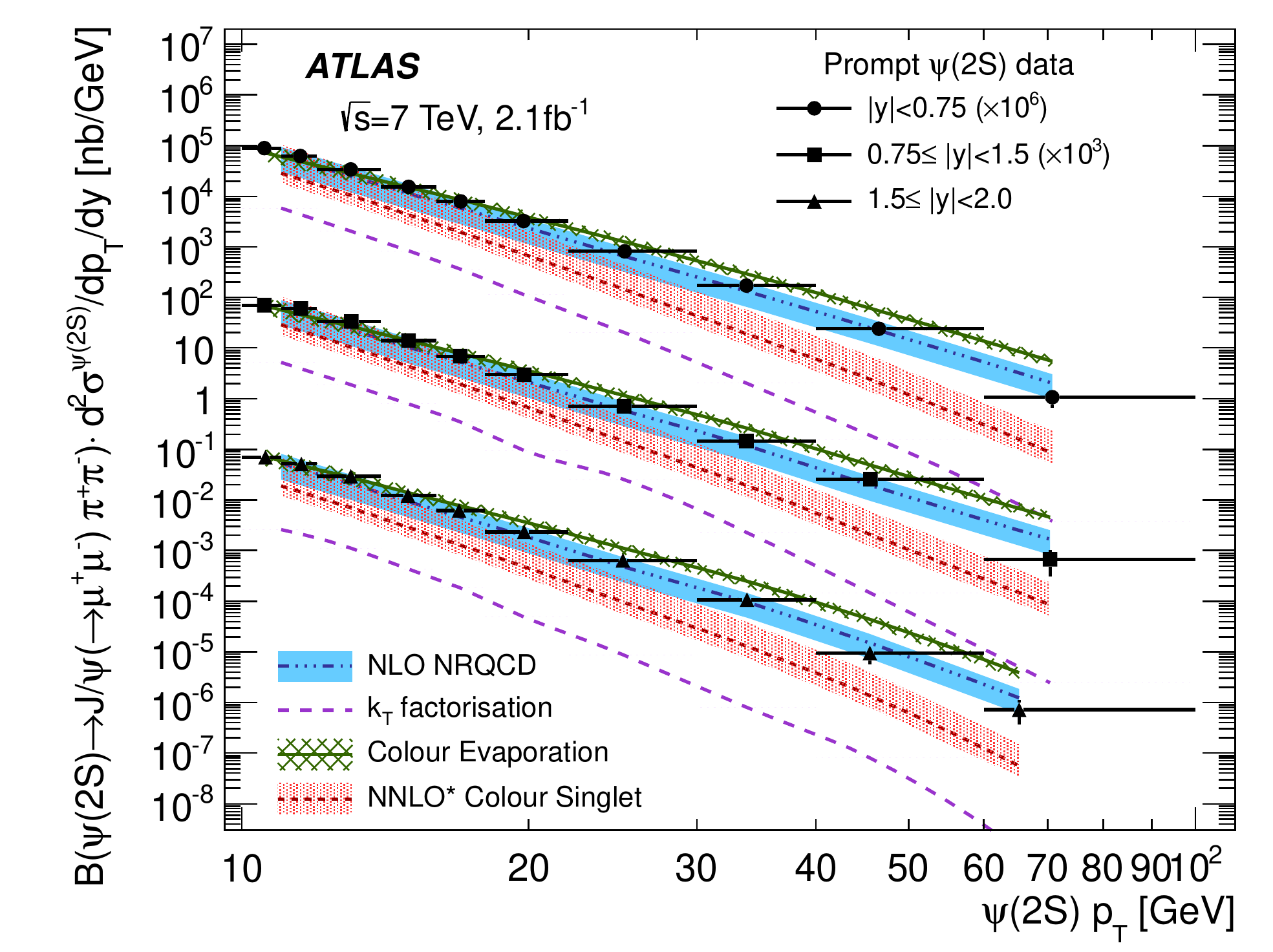}
      \includegraphics[width=0.85\textwidth]{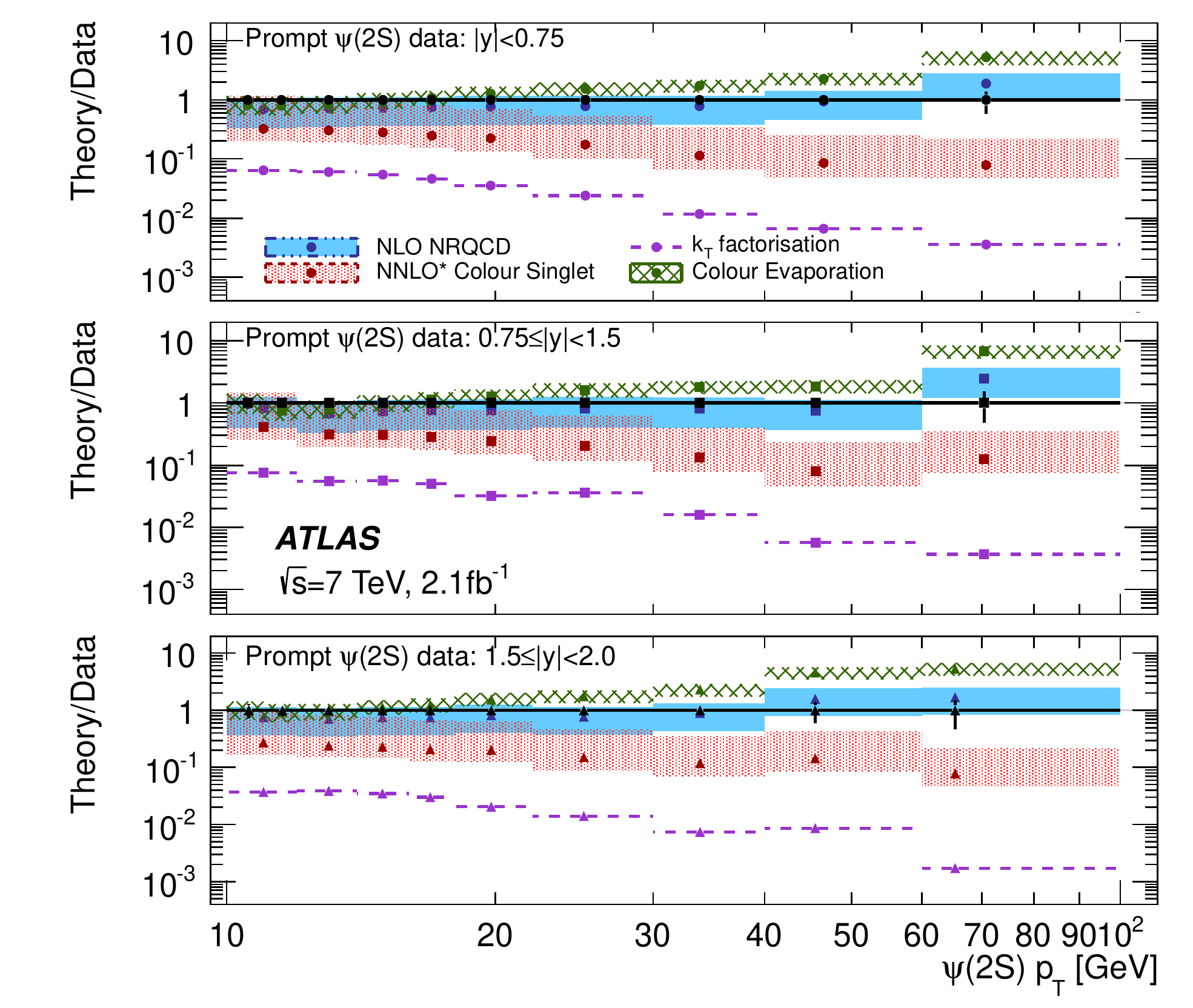}
    \caption{Measured differential cross-sections (top)
      and ratios of the predicted to measured differential cross-sections (bottom) for prompt $\psi(2\mathrm{S})$ production 
      as a function of $\psi(2\mathrm{S})$ transverse momentum for three $\psi(2\mathrm{S})$ rapidity intervals with comparison to theoretical predictions in the ATLAS fiducial region.
      The data points are at the mean of the efficiency and acceptance corrected $p_{\rm T}$ distribution in each $p_{\rm T}$ interval, indicated by the horizontal error bars, 
      and the vertical error bars represent the total statistical and systematic uncertainty (see figure~\ref{fig:Uncertainty:p}).
      \label{fig:results_prompt}
    }
  \end{center}
\end{figure}

In figure~\ref{fig:results_prompt}, the measured prompt production cross-sections are compared to predictions from 
colour-singlet\,\cite{CSM1,CSM2,CSM3,CSM4,CSM5,CSM6,CSM7} perturbative QCD calculations at partial next-to-next-to-leading-order (NNLO*)\,\cite{Lansberg:2008gk} using the CTEQ6M\,\cite{CTEQ}
parton distribution function set, 
leading-order (LO) and next-to-leading-order (NLO) non-relativistic QCD (NRQCD)\,\cite{NRQCD} (or `colour-octet' approach), the
colour evaporation model\,\cite{CEM1,CEM2,CEM3}, and a $k_{\rm T}$-factorisation approach\,\cite{kTFactor}.

The colour-singlet NNLO* predictions have no free parameters constrained from experimental data.
Uncertainties in these
predictions are assessed by variation of renormalisation and factorisation scales (which dominate the total uncertainty), 
and the charm quark mass used in the calculation as discussed in ref.~\cite{Lansberg:2008gk}. 
The central values of the NNLO* predictions underestimate the observed cross-sections 
by a factor of five, significantly outside the variation permitted by the associated scale uncertainties. Deviations from the data are enhanced at high $p_{\rm T}$
pointing to the need for further large singlet corrections or a sizeable colour octet contribution at these momenta.

The NRQCD predictions presented here are derived using HELAC-ONIA\,\cite{HELAConia,HELAC1,HELAC2,HELAC3},
an automatic matrix-element generator for the calculation of the heavy quarkonium helicity amplitudes in the framework of NRQCD factorisation.
Uncertainties in the predictions come from the uncertainties due to the choice of scale, charm quark mass and long-distance matrix elements (LDME) as discussed in ref.~\cite{HELAC1}.
NLO colour-octet LDME values from ref.~\cite{NRQCD} are used.
NLO predictions do well in describing the shape and normalisation of prompt production data over the full range of transverse momenta probed, with the agreement particularly notable
at large $p_{\rm T}$ where prior constraints on the LDME were not available.
The ratio of theory to data is also shown in figure~\ref{fig:results_prompt}.
 
Uncertainties in the colour evaporation model (CEM)\,\cite{Schuler:1996ku,Amundson:1995em,Amundson:1996qr} 
predictions from factorisation and renormalisation scale dependencies are estimated according to the prescription discussed in
ref.~\cite{Nelson:2012bc}, using a central value for the charm quark mass of $1.27$~GeV.
The predictions of the CEM are found to describe $\psi(2\mathrm{S})$ production well, and tend to follow the same behaviour as the NLO NRQCD predictions, but
at the highest $p_{\rm T}$ probed, there is a tendency for CEM to predict a somewhat harder spectrum than is observed in the data.

Parameter settings for the predictions of the $k_{\rm T}$-factorisation approach shown here are described in ref.~\cite{kTFactor}, 
take a parton-level cross-section prediction from the colour-singlet model\,\cite{CSM4,Baier:1981uk,CSM5} 
and make use of the CCFM A0 unintegrated gluon parameterisation\,\cite{CCFM_A0} that incorporates initial-state radiation dependencies.
Comparison with data shows that the $k_{\rm T}$-factorisation approach significantly underestimates the prompt $\psi(2\mathrm{S})$ production rate.
The theory-to-data ratio in figure~\ref{fig:results_prompt} highlights that this underestimation also has a $p_{\rm T}$-dependent shape.
This underestimation may be related to the observation\,\cite{ATLAS_chic} that the same model overestimates the production of $C$-even ($\chi_c$) charmonium states.
\afterpage{\clearpage}
Regarding the impact of possible spin-alignment variation on the prompt cross-section extracted
(see figure~\ref{fig:polRatio} and appendix~\ref{sec:acccorr}), it is clear that even in the most extreme cases disfavoured by available data\,\cite{CMSPol,Aaij:2014qea},
the maximum impact on the total reported cross-section is $(+62\%,-32\%)$ at a $p_{\rm T}$ of 10~GeV
and drops to $(+8\%,-12\%)$ at high $p_{\rm T}$. This range of variation is significantly smaller than the observed differences between some theories and data.


\subsection*{Non-prompt cross-section measurement versus theory}
For non-prompt production, comparison is made to theoretical predictions from fixed-order next-to-leading-logarithm (FONLL) calculations\,\cite{FONLL_2001,FONLL_2012}, 
which have been successful in describing $J/\psi$\,\cite{JPsi} and $B$-meson production\,\cite{ATLAS:2013cia} at the LHC,
and NLO predictions in the general-mass variable-flavour-number scheme (GM-VFNS), which have also proved reliable at describing production of non-prompt $J/\psi$ 
at low $p_{\rm T}$ and central rapidities\,\cite{Bolzoni:2013tca}.

Comparison of the non-prompt spectra is made to FONLL predictions obtained by first
determining the $b$-hadron production spectrum from a next-to-leading order QCD calculation matched with an all-order resummation to next-to-leading-logarithm accuracy
in the limit where the transverse momentum of the $b$-quark is much larger than its mass. This distribution is then convolved
with a phenomenological spectrum, obtained from experimental data that describe the momentum distribution of the $\psi(2\mathrm{S})$ in $B$-meson decays.
The parton distribution function set CTEQ~6.6\,\cite{CTEQ} is used and the renormalisation and factorisation scales are chosen to be $\mu=\sqrt{m^2+p^2_{T}}$, where $m$ and $p_{\rm T}$
refer to the mass and transverse momentum of the $b$-quark, where a $b$-quark mass of $4.75$~GeV is used.
The Kartvelishvili--Likhoded--Petrov fragmentation function parameterisation\,\cite{Kartvelishvili:1977pi} is used for determination of the $b$-quark fragmentation distribution.
Uncertainties on the predictions are assessed by varying the $b$-quark mass (by $\pm0.25$ GeV), evaluating the parton distribution function uncertainties and varying the renormalisation and factorisation scales 
independently up and down by a factor of two from their nominal values, with the additional constraint that the ratio of two scales must be in the range $0.5$--$2.0$.

NLO GM-VFNS predictions also use the CTEQ~6.6 parton distribution function set, the same choice of renormalisation and factorisation scales and variation as for FONLL, with a $c$-quark mass of $1.3$~GeV
and $b$-quark mass of $4.5$~GeV. The NLO predictions make use of a fragmentation function derived from NLO fits to LEP data\,\cite{Kniehl:2008zza}. 

\begin{figure}[htbp]
  \begin{center}
      \includegraphics[width=0.85\textwidth]{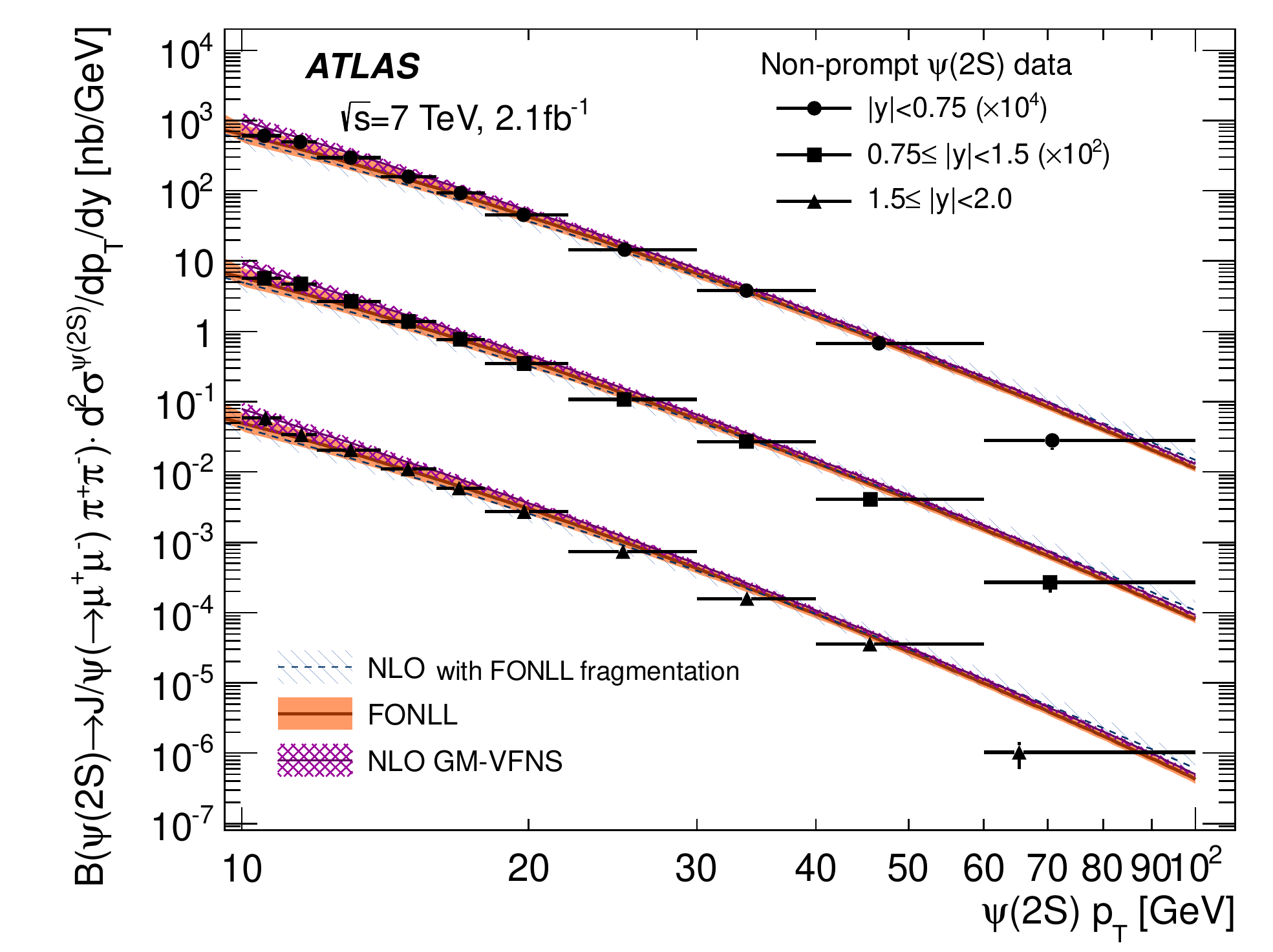}
      \includegraphics[width=0.85\textwidth]{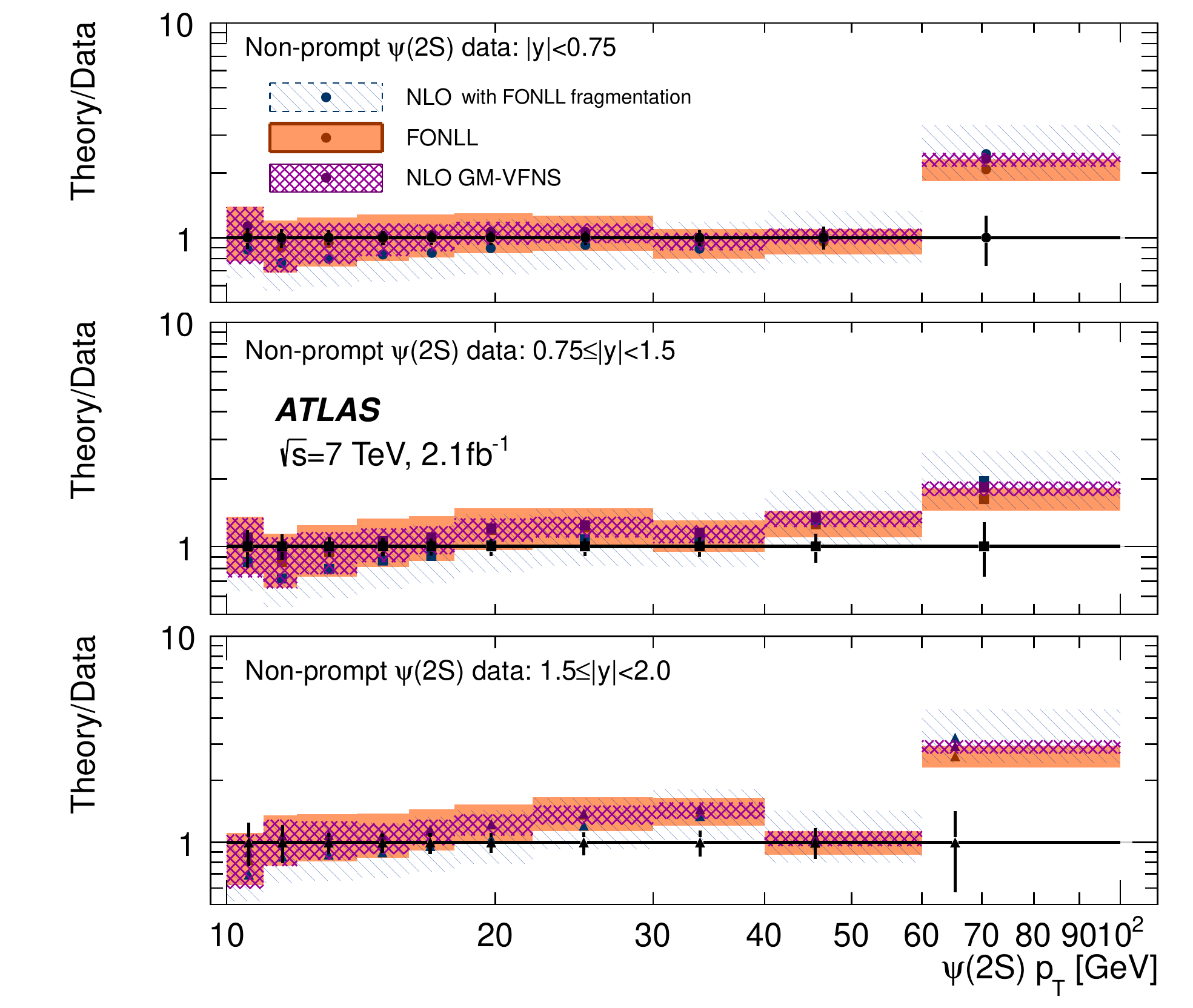}
    \caption{Measured differential cross-sections (top)
      and ratios of the predicted to measured differential cross-sections (bottom) for non-prompt $\psi(2\mathrm{S})$ production 
      as a function of $\psi(2\mathrm{S})$ transverse momentum for three $\psi(2\mathrm{S})$ rapidity intervals with comparison to theoretical predictions in the ATLAS fiducial region.
      The data points are at the mean of the efficiency and acceptance corrected $p_{\rm T}$ distribution in each $p_{\rm T}$ interval, indicated by the horizontal error bars, 
      and the vertical error bars represent the total statistical and systematic uncertainty (see figure~\ref{fig:Uncertainty:np}).
      \label{fig:results_nonprompt}
    }
  \end{center}
\end{figure}

Figure~\ref{fig:results_nonprompt} shows a comparison of FONLL and NLO GM-VFNS predictions to the non-prompt experimental data.
Also shown is a comparison of NLO predictions using the FONLL fragmentation functions.
At small and moderate transverse momenta, near and not significantly larger than the $b$-quark mass, NLO approaches are expected to do well, and scale uncertainties from the GM-VFNS approach
are smaller than those from FONLL.

\afterpage{\clearpage}
Both the FONLL and NLO GM-VFNS predictions describe the data well over the transverse momentum range studied but tend to predict a slightly harder $p_{\rm T}$ spectrum than observed in the data.
This tendency is more noticeable in NLO predictions using the FONLL fragmentation functions.
The differences observed between data and theoretical expectations are significantly larger than can be expected from any modification to the acceptance of $\psi(2\mathrm{S})$
due to a non-isotropic spin-alignment.
Our data supports hints of a similar trend observed in CMS data\,\cite{CMS}, and extends the comparison with theory to higher momenta.
Given that FONLL is able to describe reasonably well the production of fully reconstructed
charged $B$ mesons  in a similar range of transverse momenta\,\cite{ATLAS:2013cia}, the deviation
observed in this measurement seems to point towards possible mismodelling in $b$-hadron composition and decay kinematics, rather than in the $b$-quark fragmentation.

\section{Conclusions}

The prompt and non-prompt production cross-sections and the non-prompt production fraction of the $\psi(2\mathrm{S})$ decaying into $\Jmumupipi$
were measured
in the rapidity range $|y|<2.0$ for transverse momenta between $10$ and $100$~GeV. 
This measurement was carried out using 2.1~fb$^{-1}$ of $pp$ collision data at a centre-of-mass energy of 7~TeV recorded by the ATLAS experiment at the LHC.
The results presented here significantly extend the range of the measurement to higher transverse momenta with increased precision over previous measurements.

Theoretical models of prompt $\psi(2\mathrm{S})$ production vary significantly in their predictions of overall rate and kinematic dependence.
NLO NRQCD predictions were found to describe the data satisfactorily across the full range of transverse momentum studied.
Predictions from the colour evaporation model were able to describe all but the 
the highest $p_{\rm T}$ region, where the production rates were significantly overestimated.
NNLO* colour-singlet calculations, in contrast, undershoot the data by an order of magnitude at the highest $p_{\rm T}$ studied.
The addition of further large corrections to NNLO* colour-singlet calculations, or a significant colour-octet contribution at high transverse momentum is needed to describe the data.
Predictions of the $k_{\rm T}$-factorisation model exhibit a softer $p_{\rm T}$ spectrum than observed and clearly undershoot the data in overall rate. 
Together with the recent observation\,\cite{ATLAS_chic} of an overestimate of the production 
rate of $C$-even $\chi_c$ charmonium states in the $k_{\rm T}$-factorisation approach, these measurements provide coherent input to improve the $k_{\rm T}$-dependent approach.

In non-prompt $\psi(2\mathrm{S})$  production, both NLO GM-VFNS and FONLL calculations describe the data well, but a tendency is observed for 
the theory to predict a slightly harder $p_{\rm T}$ spectrum than is measured in data.
This supports trends previously observed in CMS data at lower $p_{\rm T}$, with the ATLAS and CMS data consistent in the region of overlap.

\FloatBarrier

\acknowledgments

\input{Acknowledgements}




\bibliographystyle{atlasBibStyleWithTitle} 
\bibliography{jpsipipi_JHEP}

\appendix
\section{Acceptance correction factors}
\label{sec:acccorr}

Tables~\ref{tab:tableSAratio} and~\ref{tab:tableSAratioJpsi} document correction factors that can be used to correct measured prompt $\psi(2\mathrm{S})$ production cross-sections from 
isotropic production cross-sections presented in the main text to an alternative spin-alignment scenario.

\begin{sidewaystable}[tb]
\begin{center}
\footnotesize
\caption{Multiplicative factors to correct measured production cross-sections measured in $\psi(2\mathrm{S})$ $p_{\rm T}$ and $|y|$ from isotropic production to an alternative spin-alignment scenario.}
\vspace{2mm}
\begin{tabular}[h]{crcccccccccc}
\hline \hline
 $p_{\rm T}$ [GeV] & 10--11 & 11--12 & 12--14 & 14--16 & 16--18 & 18--22 & 22--30 & 30--40 & 40--60 & 60--100 \\
\hline\hline
Longitudinal\\ \hline
$|y|<$ 0.75 & 0.68 & 0.68 & 0.69 & 0.70 & 0.72 & 0.74 & 0.77 & 0.81 & 0.85 & 0.89 \\
0.75$\leq |y|<$ 1.50 & 0.69 & 0.70 & 0.70 & 0.71 & 0.72 & 0.74 & 0.77 & 0.81 & 0.85 & 0.89 \\
1.50$\leq |y|<$ 2.00 & 0.70 & 0.70 & 0.71 & 0.72 & 0.73 & 0.74 & 0.77 & 0.81 & 0.85 & 0.88 \\ \hline
Transverse zero\\ \hline
$|y|<$ 0.75 & 1.32 & 1.30 & 1.29 & 1.27 & 1.25 & 1.22 & 1.18 & 1.14 & 1.10 & 1.06 \\
0.75$\leq |y|<$ 1.50 & 1.29 & 1.28 & 1.27 & 1.25 & 1.24 & 1.21 & 1.18 & 1.13 & 1.10 & 1.07 \\
1.50$\leq |y|<$ 2.00 & 1.27 & 1.27 & 1.26 & 1.25 & 1.23 & 1.21 & 1.17 & 1.13 & 1.10 & 1.08 \\ \hline
Transverse positive\\ \hline
$|y|<$ 0.75 & 1.61 & 1.44 & 1.35 & 1.30 & 1.27 & 1.23 & 1.18 & 1.14 & 1.10 & 1.07 \\
0.75$\leq |y|<$ 1.50 & 1.62 & 1.44 & 1.36 & 1.30 & 1.27 & 1.23 & 1.19 & 1.14 & 1.10 & 1.07 \\
1.50$\leq |y|<$ 2.00 & 1.62 & 1.42 & 1.36 & 1.30 & 1.27 & 1.23 & 1.19 & 1.14 & 1.10 & 1.08 \\ \hline
Transverse negative\\ \hline
$|y|<$ 0.75 & 1.11 & 1.20 & 1.22 & 1.23 & 1.22 & 1.21 & 1.17 & 1.13 & 1.10 & 1.07 \\
0.75$\leq |y|<$ 1.50 & 1.07 & 1.16 & 1.19 & 1.21 & 1.21 & 1.19 & 1.17 & 1.13 & 1.10 & 1.06 \\
1.50$\leq |y|<$ 2.00 & 1.05 & 1.14 & 1.18 & 1.20 & 1.20 & 1.19 & 1.16 & 1.13 & 1.10 & 1.07 \\ \hline
Off-plane positive\\ \hline
$|y|<$ 0.75 & 1.06 & 1.05 & 1.05 & 1.04 & 1.03 & 1.02 & 1.02 & 1.01 & 1.01 & 1.00 \\
0.75$\leq |y|<$ 1.50 & 1.12 & 1.12 & 1.10 & 1.08 & 1.07 & 1.05 & 1.04 & 1.02 & 1.01 & 1.01 \\
1.50$\leq |y|<$ 2.00 & 1.14 & 1.15 & 1.12 & 1.10 & 1.08 & 1.07 & 1.04 & 1.02 & 1.01 & 1.01 \\ \hline
Off-plane negative\\ \hline
$|y|<$ 0.75 & 0.96 & 0.95 & 0.96 & 0.96 & 0.97 & 0.98 & 0.98 & 0.99 & 0.99 & 1.00 \\
0.75$\leq |y|<$ 1.50 & 0.91 & 0.91 & 0.92 & 0.93 & 0.94 & 0.95 & 0.97 & 0.98 & 0.99 & 0.99 \\
1.50$\leq |y|<$ 2.00 & 0.90 & 0.89 & 0.90 & 0.91 & 0.93 & 0.94 & 0.96 & 0.97 & 0.99 & 0.99 \\ \hline
\hline \hline
\end{tabular}
\label{tab:tableSAratio}
\end{center}
\end{sidewaystable}

\begin{sidewaystable}[tb]
\begin{center}
\footnotesize
\caption{Multiplicative factors to correct measured production cross-sections measured in $p_{\rm T}$ and $|y|$ for $J/\psi$ in the $\psi(2\mathrm{S})\to\Jmumupipi$ decay from isotropic production to an alternative spin-alignment scenario.}
\vspace{2mm}
\begin{tabular}[h]{crcccccccccc}
\hline \hline
$p_{\rm T}$ [GeV] & 10--11 & 11--12 & 12--14 & 14--16 & 16--18 & 18--22 & 22--30 & 30--40 & 40--60 & 60--100\\ \hline \hline
Longitudinal\\ \hline
$|y|<$ 0.75 & 0.69 & 0.69 & 0.71 & 0.72 & 0.74 & 0.76 & 0.79 & 0.83 & 0.86 & 0.91 \\
0.75$\leq |y|<$ 1.50 & 0.70 & 0.71 & 0.71 & 0.73 & 0.74 & 0.76 & 0.79 & 0.83 & 0.86 & 0.90 \\
1.50$\leq |y|<$ 2.00 & 0.71 & 0.71 & 0.72 & 0.73 & 0.75 & 0.76 & 0.79 & 0.83 & 0.86 & 0.87 \\ \hline
Transverse zero\\ \hline
$|y|<$ 0.75 & 1.29 & 1.28 & 1.26 & 1.24 & 1.22 & 1.19 & 1.15 & 1.12 & 1.09 & 1.06 \\
0.75$\leq |y|<$ 1.50 & 1.28 & 1.27 & 1.25 & 1.23 & 1.21 & 1.19 & 1.15 & 1.11 & 1.09 & 1.06 \\
1.50$\leq |y|<$ 2.00 & 1.26 & 1.26 & 1.24 & 1.23 & 1.21 & 1.19 & 1.15 & 1.11 & 1.09 & 1.04 \\ \hline
Transverse positive\\ \hline
$|y|<$ 0.75 & 1.38 & 1.33 & 1.30 & 1.26 & 1.23 & 1.20 & 1.16 & 1.12 & 1.09 & 1.06 \\
0.75$\leq |y|<$ 1.50 & 1.39 & 1.34 & 1.30 & 1.26 & 1.23 & 1.20 & 1.16 & 1.12 & 1.09 & 1.06 \\
1.50$\leq |y|<$ 2.00 & 1.38 & 1.34 & 1.30 & 1.26 & 1.23 & 1.20 & 1.16 & 1.12 & 1.09 & 1.04 \\ \hline
Transverse negative\\ \hline
$|y|<$ 0.75 & 1.22 & 1.23 & 1.23 & 1.22 & 1.20 & 1.18 & 1.15 & 1.11 & 1.09 & 1.05 \\
0.75$\leq |y|<$ 1.50 & 1.19 & 1.20 & 1.21 & 1.20 & 1.19 & 1.17 & 1.15 & 1.11 & 1.09 & 1.05 \\
1.50$\leq |y|<$ 2.00 & 1.16 & 1.19 & 1.20 & 1.20 & 1.19 & 1.17 & 1.15 & 1.11 & 1.09 & 1.07 \\ \hline
Off-plane positive\\ \hline
$|y|<$ 0.75 & 1.05 & 1.04 & 1.04 & 1.03 & 1.02 & 1.02 & 1.01 & 1.01 & 1.01 & 1.00 \\
0.75$\leq |y|<$ 1.50 & 1.11 & 1.09 & 1.08 & 1.07 & 1.05 & 1.04 & 1.03 & 1.02 & 1.01 & 1.01 \\
1.50$\leq |y|<$ 2.00 & 1.13 & 1.12 & 1.10 & 1.08 & 1.06 & 1.05 & 1.03 & 1.02 & 1.01 & 1.00 \\ \hline
Off-plane negative\\ \hline
$|y|<$ 0.75 & 0.95 & 0.96 & 0.97 & 0.97 & 0.98 & 0.98 & 0.99 & 0.99 & 1.00 & 1.00 \\
0.75$\leq |y|<$ 1.50 & 0.91 & 0.92 & 0.93 & 0.94 & 0.95 & 0.96 & 0.97 & 0.99 & 0.99 & 0.99 \\
1.50$\leq |y|<$ 2.00 & 0.90 & 0.91 & 0.92 & 0.93 & 0.94 & 0.95 & 0.97 & 0.98 & 0.99 & 0.98 \\ \hline
\hline \hline
\end{tabular}
\label{tab:tableSAratioJpsi}
\end{center}
\end{sidewaystable}

\onecolumn
\clearpage 
\include{atlas_authlist}

\end{document}

%% file: Acknowledgements.tex


We thank Carlos Louren\c{c}o and Hermine W\"{o}hri for pointing out an issue with the preliminary version of the data presented in this paper.

We thank CERN for the very successful operation of the LHC, as well as the
support staff from our institutions without whom ATLAS could not be
operated efficiently.

We acknowledge the support of ANPCyT, Argentina; YerPhI, Armenia; ARC,
Australia; BMWF and FWF, Austria; ANAS, Azerbaijan; SSTC, Belarus; CNPq and FAPESP,
Brazil; NSERC, NRC and CFI, Canada; CERN; CONICYT, Chile; CAS, MOST and NSFC,
China; COLCIENCIAS, Colombia; MSMT CR, MPO CR and VSC CR, Czech Republic;
DNRF, DNSRC and Lundbeck Foundation, Denmark; EPLANET, ERC and NSRF, European Union;
IN2P3-CNRS, CEA-DSM/IRFU, France; GNSF, Georgia; BMBF, DFG, HGF, MPG and AvH
Foundation, Germany; GSRT and NSRF, Greece; ISF, MINERVA, GIF, I-CORE and Benoziyo Center,
Israel; INFN, Italy; MEXT and JSPS, Japan; CNRST, Morocco; FOM and NWO,
Netherlands; BRF and RCN, Norway; MNiSW and NCN, Poland; GRICES and FCT, Portugal; MNE/IFA, Romania; MES of Russia and ROSATOM, Russian Federation; JINR; MSTD,
Serbia; MSSR, Slovakia; ARRS and MIZ\v{S}, Slovenia; DST/NRF, South Africa;
MINECO, Spain; SRC and Wallenberg Foundation, Sweden; SER, SNSF and Cantons of
Bern and Geneva, Switzerland; NSC, Taiwan; TAEK, Turkey; STFC, the Royal
Society and Leverhulme Trust, United Kingdom; DOE and NSF, United States of
America.

The crucial computing support from all WLCG partners is acknowledged
gratefully, in particular from CERN and the ATLAS Tier-1 facilities at
TRIUMF (Canada), NDGF (Denmark, Norway, Sweden), CC-IN2P3 (France),
KIT/GridKA (Germany), INFN-CNAF (Italy), NL-T1 (Netherlands), PIC (Spain),
ASGC (Taiwan), RAL (UK) and BNL (USA) and in the Tier-2 facilities
worldwide.

%% file: atlas_authlist.tex
\begin{flushleft}
{\Large The ATLAS Collaboration}

\bigskip

G.~Aad$^{\rm 84}$,
B.~Abbott$^{\rm 112}$,
J.~Abdallah$^{\rm 152}$,
S.~Abdel~Khalek$^{\rm 116}$,
O.~Abdinov$^{\rm 11}$,
R.~Aben$^{\rm 106}$,
B.~Abi$^{\rm 113}$,
M.~Abolins$^{\rm 89}$,
O.S.~AbouZeid$^{\rm 159}$,
H.~Abramowicz$^{\rm 154}$
% ...
\bigskip
\\
% ...
$^{11}$ Institute of Physics, Azerbaijan Academy of Sciences, Baku, Azerbaijan\\
% ...
$^{84}$ CPPM, Aix-Marseille Universit{\'e} and CNRS/IN2P3, Marseille, France\\
% ...
$^{89}$ Department of Physics and Astronomy, Michigan State University, East Lansing MI, United States of America\\
% ...
$^{106}$ Nikhef National Institute for Subatomic Physics and University of Amsterdam, Amsterdam, Netherlands\\
% ...
$^{112}$ Homer L. Dodge Department of Physics and Astronomy, University of Oklahoma, Norman OK, United States of America\\
$^{113}$ Department of Physics, Oklahoma State University, Stillwater OK, United States of America\\
% ...
$^{116}$ LAL, Universit{\'e} Paris-Sud and CNRS/IN2P3, Orsay, France\\
% ...
$^{152}$ Institute of Physics, Academia Sinica, Taipei, Taiwan\\
% ...
$^{154}$ Raymond and Beverly Sackler School of Physics and Astronomy, Tel Aviv University, Tel Aviv, Israel\\
% ...
$^{159}$ Department of Physics, University of Toronto, Toronto ON, Canada\\
% ...
\end{flushleft}
